\documentclass[twocolumns,10pt,final]{IEEEtran}
\usepackage{latexsym}
\usepackage{cite}
\usepackage{amssymb}
\usepackage{bbm}
\usepackage{bm}
\usepackage[usenames,dvipsnames]{color}
\usepackage{amsmath, amssymb}
\usepackage{dsfont}
\usepackage[dvips]{graphics}
\DeclareMathOperator{\supp}{supp}
\usepackage{graphicx}
\usepackage{lipsum}
\usepackage{algorithmicx}
\usepackage{algorithm} 
\usepackage[noend]{algpseudocode} 
\usepackage{psfrag}
\usepackage{units}
\usepackage{comment}
\usepackage{subfig}
\usepackage{setspace}
\usepackage{algorithm}
\usepackage{algpseudocode}
\usepackage{theorem}
\usepackage{float}
\usepackage{mathtools}
 \allowdisplaybreaks

\newtheorem{definition}{Definition}

\newtheorem{theorem}{Theorem}
\newtheorem{lemma}{Lemma}
\newtheorem{remark}{Remark}

\newtheorem{corollary}{Corollary}
\newtheorem{proposition}{Proposition}

\DeclareMathOperator{\lcm}{lcm}
\newcommand\numberthis{\addtocounter{equation}{1}\tag{\theequation}}

{ 
\bibliographystyle{unsrt}
\bibliographystyle{IEEEtran}
\begin{document}

\makeatletter
\newcommand{\vasti}{\bBigg@{3}}
\newcommand{\vastii}{\bBigg@{3.5}}
\newcommand{\vast}{\bBigg@{4}}
\newcommand{\Vast}{\bBigg@{5}}
\makeatother
\newcommand{\be}{\begin{equation}}
\newcommand{\ee}{\end{equation}}
\newcommand{\ba}{\begin{align}}
\newcommand{\ea}{\end{align}}
\newcommand{\baa}{\begin{align*}}
\newcommand{\eaa}{\end{align*}}
\newcommand{\bea}{\begin{eqnarray}}
\newcommand{\eea}{\end{eqnarray}}
\newcommand{\beaa}{\begin{eqnarray*}}
\newcommand{\eeaa}{\end{eqnarray*}}
\newcommand{\p}[1]{\left(#1\right)}
\newcommand{\pp}[1]{\left[#1\right]}
\newcommand{\ppp}[1]{\left\{#1\right\}}
\newcommand{\ber}{$\ \mbox{Bernoulli}$}

\title{Arbitrarily Varying Wiretap Channels with Type Constrained States}

\author{Ziv Goldfeld, \emph{Student Member, IEEE}, Paul Cuff, \emph{Member, IEEE}, and Haim H. Permuter, \emph{Senior Member, IEEE}
\thanks{Z. Goldfeld and H. H. Permuter were supported in part by the Cyber Security Research Center within the Ben-Gurion University of the Negev, in part by the European Research Council under the European Union's Seventh Framework Programme (FP7/2007-2013)/ERC grant agreement n$^\circ$337752 and in part by the Israel Science Foundation. P. Cuff was supported in part by the National Science Foundation under Grant CCF-1350595 and in part by the Air Force Office of Scientific Research under Grant FA9550-15-1-0180. 
\newline This paper was presented in part at the 2016 International Conference on the Science of Electrical Engineers (ICSEE), Eilat, Israel.
\newline Z. Goldfeld and H. H. Permuter are with the Department of Electrical and Computer Engineering, Ben-Gurion University of the Negev, Beer-Sheva, Israel (gziv@post.bgu.ac.il, haimp@bgu.ac.il). Paul Cuff is with the Department of Electrical Engineering, Princeton University, Princeton, NJ 08544 USA (e-mail: cuff@princeton.edu).}}
\maketitle

\begin{abstract}
Determining a single-letter secrecy-capacity formula for the arbitrarily varying wiretap channel (AVWTC) is an open problem largely because of two main challenges. Not only does it capture the difficulty of the compound wiretap channel (another open problem), it also requires that secrecy is ensured with respect to exponentially many possible channel state sequences. By extending the strong soft-covering lemma, recently derived by the authors, to the heterogeneous scenario, this work accounts for the exponential number of secrecy constraints while only imposing single-letter constraints on the communication rate. Through this approach we derive a single-letter characterization of the correlated-random (CR) assisted semantic-security (SS) capacity of an AVWTC with a type constraint on the allowed state sequences. The allowed state sequences are the ones in a typical set around a single constraining type. The stringent SS requirement is established by showing that the mutual information between the message and the eavesdropper's observations is negligible even when maximized over all message distributions, choices of state sequences and realizations of the CR-code.

Both the achievability and the converse proofs of the type constrained coding theorem rely on stronger claims than actually required. The direct part establishes a novel single-letter lower bound on the CR-assisted SS-capacity of an AVWTC with state sequences constrained by any convex and closed set of state probability mass functions. This bound achieves the best known single-letter secrecy rates for a corresponding compound wiretap channel over the same constraint set. In contrast to other single-letter results in the AVWTC literature, this work does not assume the existence of a best channel to the eavesdropper. Instead, SS follows by leveraging the heterogeneous version of the strong soft-covering lemma and a CR-code reduction argument. Optimality is a consequence of a max-inf upper bound on the CR-assisted SS-capacity of an AVWTC with state sequences constrained to any collection of type-classes. When adjusted to the aforementioned compound WTC, the upper bound simplifies to a max-min structure, thus strengthening the previously best known single-letter upper bound by Liang \emph{et al.} that has a min-max form. The proof of the upper bound uses a novel distribution coupling argument. The capacity formula shows that the legitimate users effectively see an averaged main channel, while security must be ensured versus an eavesdropper with perfect channel state information. An example visualizes our single-letter results, and their relation to the past multi-letter secrecy-capacity characterization of the AVWTC is highlighted.
\end{abstract}



\section{introduction}

Modern communication systems usually present an architectural separation between error correction and data encryption. The former is typically realized at the physical layer by transforming the noisy communication channel into a reliable ``bit pipe''. The data encryption is implemented on top of that by applying cryptographic principles. The cryptographic approach assumes no knowledge on the quality of the eavesdropper's channel and relies solely on restricting the computational power of the eavesdropper. The looming prospect of quantum computers (QCs) (some companies have recently reported a working prototype of a QC with over than 1000 qbits \cite{DWAVE1,DWAVE2,DWAVE3,DWAVE4,DWAVE5}), however, would boost computational abilities, rendering some critical cryptosystems insecure and weakening others.\footnote{More specifically, asymmetric ciphers that rely on the hardness of integer factorization or discrete logarithms can be completely broken using QCs via Shor's algorithm (or a variant thereof) \cite{Shor_Polynomial1999,Bernstein_postQCcrypto2009}. Symmetric encryption, on the other hand, would be weakened by QC attacks but could regain its strength by increasing the size of the key \cite{Perlner_Quantum2009}. This essentially follows since a QC can search through a space of size $2^n$ in time $2^{\frac{n}{2}}$, so by doubling the size of the key a symmetric cryptosystem would offer the same protection versus a QC attack, as the original system did versus a classic attack.} Post-QC cryptography offers partial solutions that rely on larger keys, but even now considerable efforts are made to save this expensive resource. Nonetheless, cryptography remains the main practical tool for protecting data, at least for the time being.


Physical Layer Security, rooted in information-theoretic principles, is an alternative approach to
provably secure communication that dates back to Wyner's celebrated paper on the wiretap channel (WTC) \cite{Wyner_Wiretap1975}. Essentially, Wyner's main idea was to exploit the noise of the communication channel along with proper physical layer coding to guarantee secrecy against a computationally-unlimited eavesdropper. Protection against such an eavesdropper, however, comes at a price of assuming that the eavesdropper's channel is perfectly known to the legitimate parties and stays fixed during the transmission. Many of the information-theoretic secrecy results that followed relied on extending Wyner's ideas, and therefore, are derived under the same hypothesis. Much of the critique by the cryptographic community towards information-theoretic security is aimed exactly at that assumption.

Practical systems suffer from limited channel state information (CSI) due to inaccuracies in the channel's estimation process and imperfect feedback. Furthermore, adversarial eavesdroppers will refrain from providing the legitimate parties with any information about their channels to make securing the data even harder. Accordingly, limited CSI (especially about the eavesdropper's channel) must be assumed to successfully model a practical communication system. The model of an arbitrarily varying WTC (AVWTC), that is the focus of this work, does just that. The AVWTC combines the WTC \cite{Wyner_Wiretap1975,Csiszar_Korner_BCconfidential1978} and the arbitrarily varying channel (AVC) \cite{blackwell1960capacities,Ahlsweded_Elimination1978,Jahn_multiuser_AVC1981,Csiszar_Narayan_CR_AVC1988,Csiszar_Narayan_det_AVC1988}. It consists of a collection of discrete-memoryless WTCs indexed by elements in a finite state space. The state at each time instance is chosen in an arbitrary manner and is unknown to the legitimate parties. Being aware of the state space, however, the legitimate users can place the actual channel realization within a certain uncertainty set, which models their limited eavesdropper's CSI. A relaxed scenario where the main channel is fixed in time but the eavesdropper's channel is varying and unknown was studied in \cite{Yener_MIMO_AVWTC2014}. The authors of \cite{Yener_MIMO_AVWTC2014} considered the MIMO Gaussian case and proved the existence of a universal secure coding scheme.

Inspired by wiretap channel instances that involve active eavesdroppers \cite{Mihaljevic_WTCII1994,Luo_WTCII2005,Liu_Nested_WTCII2007,Calderbank_active_WTCII2009}, the AVWTC was first introduced in \cite{MolavianJazi_AVWTC_thesis2009}. The author of \cite{MolavianJazi_AVWTC_thesis2009} derived single-letter lower and upper bounds on the correlated-random (CR) assisted weak secrecy-capacity of the AVWTC. The relation between the CR-assisted weak secrecy-capacity and the uncorrelated weak secrecy-capacity was also established in Theorem 5 of that work (see also \cite{Boche_superactive_AVWTC2016} for similar results for strong secrecy). It turns out that it makes a difference whether CR codes or their uncorrelated counterparts are used. In particular, the uncorrelated secrecy-capacity may be zero (if the main channel is symmetrizable \cite[Definition 2]{Csiszar_AVWTC_constrained_det1988}), while the CR-assisted secrecy-capacity is positive. Thus, viewing CR as an additional resource for communication, this resource can make communication possible where it is impossible without, as long as the choice of the state sequence is independent of the realization of the CR. On the other hand, CR should not be viewed as a cryptographic key to be exploited for secrecy, and therefore, it is assumed to be known to the eavesdropper. A single-letter characterization of the CR-assisted secrecy-capacity remains an open problem and only a multi-letter description has been established \cite{Boche_AVWTC_multiletterISIT2015,Boche_AVWTC_multiletter_capacity2015}. Despite the computational infeasibility of that formula, it was used in \cite{Boche_AVWTC_multiletter_capacity2015} to prove that the CR-assisted secrecy-capacity of the AVWTC is a continuous function of the uncertainty set. In contrast, the work of  \cite{Boche_AVWTC_continuity2015} showed that the same in not true for uncorrelated secrecy-capacity, by exemplifying a discontinuity point.


The challenge presented by the AVWTC is twofold. First, it subsumes the difficulty of the compound WTC (where the channel's state is constant in time), for which a single-letter secrecy-capacity characterization is also an open problem \cite{Blackwell1959capacity,Wolfowitz1959simultaneous,Liang_compoundWTC_Journal2007,Ekrem_compundWTC2010,Khitsi_MISO_compoundWTC2010,Khitsi_compoundWTC2011,Boche_compoundWTC2013,Schaefer_compound_WTC2013}. While a multi-letter description of the compound WTC's secrecy-capacity was found in \cite{Boche_compoundWTC2013}, it is currently unknown how to single-letterize this expression. The underlying gap is that while reliability must be ensured with respect to the worst main channel, security is measured under the best eavesdropper channel; a single channel state under which these extremes simultaneously materialize, however, does not necessarily exist. The second difficulty concerning AVWTCs is that security must be ensured under all possible state sequences, whose number grows exponentially with the blocklength. To get single-letter results, the latter is usually dealt with by assuming the existence of \emph{a best channel to the eavesdropper} and establishing secrecy with respect to that channel only (see, e.g., \cite{MolavianJazi_AVWTC_thesis2009,Boche_AVWTC_book2013}). Yet, the only single-letter secrecy-capacity characterization for an AVWTC that the authors are aware of assumes even more \cite[Theorem 4]{MolavianJazi_AVWTC_thesis2009}. On top of the existence of such a best channel, the derivation of \cite[Theorem 4]{MolavianJazi_AVWTC_thesis2009} also relies on the AVWTC being strongly-degraded and having independent (main channel and eavesdropper channel) states. A related setting for which a single-letter formula is known is an AVWTC where the CR is used as a secret key and there is a sufficient amount thereof \cite{Boche_superactive_AVWTC2016}. Although such a model slightly deviates from the operational meaning of CR as considered in this work, formally, it can be viewed as another scenario for which the multi-letter formula from \cite{Boche_AVWTC_multiletter_capacity2015} is single-letterizable.



We consider a \emph{general} AVWTC with a type constraint on the allowed state sequences, and establish in Theorem~\ref{TM:AVWTC_CR_capacity} a single-letter characterization of its CR-assisted semantic-security (SS) capacity. Our approach relies neither on the existence of a best channel to the eavesdropper nor on the benefit of secret CR. Instead, we show that the exponential number of security requirements are satisfied, even while using a random codebook construction under single-letter constraints on the communication rate, via a finer analysis that uses a heterogeneous strong soft-covering lemma, on which we expand the discussion subsequently. A full characterization of both the CR-assisted and the uncorrelated capacities of the classic AVC with a pair of linear constraints on the state and input sequences is due to Csisz{\'a}r and Narayan \cite{Csiszar_AVWTC_constrained_random1988,Csiszar_AVWTC_constrained_det1988}. The extension of this setting to the AVWTC scenario is the focus of \cite{Boche_constrained_AVWTC2015}, where a multi-letter description of the CR-assisted secrecy-capacity is given. In our case, the type constraint essentially means that the viable state sequences are only the ones of the prescribed type. However, since a fixed distribution (even if rational) is not a valid type for all blocklengths, we define achievability by allowing the empirical distribution of the state sequences to be within a small gap from the type. By doing so, the type constrained AVWTC is well defined for all blocklengths. As a consequence, our uncertainty set is a typical set around the allowed type, which still contains exponentially many state sequences. The structure of the CR-assisted SS-capacity formula suggests that the legitimate users effectively see the averaged channel (i.e., the expectation of the main channels with respect to the type) while security must be ensured versus an eavesdropper with perfect CSI. A specific instance of a type constrained AVWTC that is related to binary symmetric - binary erasure (BS-BE) WTC that was studied in \cite{Ulukus_BSCBEC_WTC2011} is used to visualize the result.

The results are derived while adopting the prescription of \cite{Vardy_Semantic_WTC2012} to replace the commonly used strong secrecy metric with the stricter SS metric \cite{Goldwasser_Semantic_Security1984,Bellare_Semantic_Security1997}. The authors of \cite{Vardy_Semantic_WTC2012} advocate SS as the new standard for information-theoretic security, because from a cryptographic point of view, strong secrecy is insufficient to provide security of applications. Its main drawback lies in the assumption that the message is random and uniformly distributed, as real-life messages are neither (messages may be files, votes or any type of structured data, often with low entropy). In turn, the uniformly distributed message makes the strong secrecy metric an average quantity, that might converge even when many\footnote{The number of unsecured messages may even grow exponentially with the blocklength, while still having a converging strong secrecy metric.} of the messages are actually not secured. Furthermore, to eliminate the benefit of CR for secrecy purposes, we demand that SS holds for each realization of the CR (a similar approach was taken in \cite{Boche_AVWTC_multiletterISIT2015,Boche_AVWTC_multiletter_capacity2015} with respect to the strong secrecy metric). This essentially means that the transmission is semantically-secure even if the choice of the state sequence depends on the realization of the~CR. 

In Lemma~\ref{LEMMA:soft_covering} we develop a heterogeneous soft-covering analysis tool that is key in ensuring SS under the exponentially many state sequences of the AVWTC. By means of the Chernoff bound, the lemma guarantees a double-exponential decay of the probability that soft-covering fails to occur under the relative entropy metric. The probability is taken with respect to a random codebook and the convergence occurs as long as the rate of the codebook is greater than the mutual information between the channel's input and output random variables. In turn, this allows us to furnish a single-letter achievability result without assuming that a best channel to the eavesdropper exists. Doubly-exponentially decaying probabilities coming from the Chernoff bound were previously used in the context of secrecy in, e.g., \cite{Boche_AVWTC_book2013,Boche_AVWTC_multiletter_capacity2015,Boche_compoundWTC2013,Boche_AVWTC_multiletterISIT2015,Csiszar_almost_independent1996,Boche_superactive_AVWTC2016}. In particular, claims similar to these presented in this work (but under the total variation metric) were used in \cite{Boche_AVWTC_multiletter_capacity2015,Boche_AVWTC_multiletterISIT2015,Boche_superactive_AVWTC2016} for the security analysis under the AVWTC scenario. Similar concentration results also previously appeared in quantum information theory sources \cite{Devetak_Stronger_SCL2003,Wilde_Book2013}. Nevertheless, we emphasize the significance of the strong soft-covering lemma as a stand-alone claim because of the effectiveness of soft-covering in proofs of secrecy, resolvability \cite{Bloch_Resolvability_Secrecy2013,Kramer_EffectiveSecrecy2014}, and channel synthesis \cite{Cuff_Synthesis2013}. Furthermore, the convergence to 0 of the relative entropy implied by Lemma~\ref{LEMMA:soft_covering} naturally relates to the definition of SS that uses mutual information.


To prove our coding theorem for the type constrained AVWTC (i.e., the main result in Theorem~\ref{TM:AVWTC_CR_capacity}), we provide both a stronger achievability and a stronger converse than is actually required. The broader achievability claim, found in Theorem~\ref{TM:AVWTC_CR_achievability_general}, is a lower bound on the CR-assisted SS-capacity of an AVWTC with state sequences constrained by any convex and closed set of state PMFs. This bound shows that the best known achievable single-letter secrecy rates for a similarly constrained compound WTC \cite{Liang_compoundWTC_Journal2007,Boche_compoundWTC2013} can be achieved also in the AVWTC.\footnote{This connection is expounded in Remark~\ref{R:AVWTC and CWTC}.} The lower bound is derived by first generating a CR-code over a large family of uncorrelated codes, whose size grows doubly-exponentially with the blocklength, and establish reliability by arguments similar to those used for the classic AVC with constrained states \cite{Csiszar_AVWTC_constrained_random1988}. Then, we invoke a Chernoff bound to show that a uniform CR-code over a family that is no more than polynomial in size is sufficient. Having this, SS follows via the union bound and the strong soft-covering lemma, because the combined number of codes, state sequences and messages grows only exponentially with the blocklenght. The lemma is still sharp enough to imply that the probability of a random codebook violating security is doubly-exponentially small. The obtained single-letter achievability formula is shown to be recoverable from the multi-letter CR-assisted secrecy-capacity description from \cite{Boche_AVWTC_multiletterISIT2015,Boche_AVWTC_multiletter_capacity2015} when specialized to the unconstrained states scenario. 


The polynomial size of the reduced CR-code is of consequence for the uncorrelated scenario as well. Provided that the uncorrelated SS-capacity is strictly positive, the relatively small CR-code allows to replace the shared randomness between the legitimate parties (used for selecting a code from the family) with a local randomness at the transmitter. The transmitter may select the code and inform the receiver which code is in use by sending its index as a short prefix. The positivity of the uncorrelated capacity is essential to allow the reliable transmission of the short prefix with a vanishing rate. Thus, the missing piece for establishing the uncorrelated SS-capacity of the type constrained AVWTC is a dichotomy result (in the spirit of \cite{Ahlsweded_Elimination1978,Csiszar_Narayan_det_AVC1988}) based on a condition that distinguishes weather its uncorrelated and CR-assisted secrecy-capacities are equal or not. Such results are available for AVWTCs without constraints on the state space \cite{MolavianJazi_AVWTC_thesis2009,Boche_AVWTC_book2013,Boche_superactive_AVWTC2016}. CR SS-capacity being the focus of this work, we pose the dichotomy result and the corresponding threshold property for the constrained states scenario as questions for future research.

To prove the converse part of the main result, we claim in Theorem~\ref{TM:AVWTC_CR_converse_general} an upper bound on CR-assisted SS-capacity of an AVWTC with state sequences from any collection of type-classes. The upper bound is of a max-inf form, i.e., first an infimum over the constraint set is taken, and then the result is maximized over the input distributions. When specializing the result to the aforementioned compound WTC, it produces an upper bound that improves upon the previously best known single-letter upper bound for this setting \cite[Theorem 2]{Liang_compoundWTC_Journal2007}. The latter result has a min-max structure, while our upper bound has a max-min form. This strengthening is due to a derivation that is uniform over the constraint set. The analysis is preformed per each type in the set and shows that reliability and SS under states from even a single type-class imply similar performance limits as the same channel but where the state sequence is independently and identically distributed (i.i.d.) according to the type. The main challenge is in upper bounding the normalized equivocation of the message given an output sequence that is generated by the average channel. This step relies on the equivocation being continuous in the set of viable state sequences. A novel distribution coupling argument is used to establish this desired property.



This paper is organized as follows. Section \ref{SEC:preliminaries} provides definitions and basic properties. In Section \ref{SEC:soft_covering} we state and prove the heterogeneous strong soft-covering lemma. The AVWTC with type constrained states is studied in Section \ref{SEC:AVWTC}, where the setup is defined and the CR-assisted SS-capacity is characterized and proven. Section \ref{SEC:AVWTC} also states the lower and upper bounds for the more general setup, relates the achievability result to past multi-letter descriptions of the CR-assisted secrecy-capacity and provides an example. The lower and upper bounds are proven in Sections \ref{SEC:AVWTC_CR_achievability_general_proof} and \ref{SEC:AVWTC_CR_converse_general_proof}, respectively. Finally, Section \ref{SEC:summary} summarizes the main achievements and insights of this work.


\section{Notations and Preliminaries}\label{SEC:preliminaries}

\par We use the following notations. As customary $\mathbb{N}$ is the set of natural numbers (which does not include 0), $\mathbb{Q}$ denotes the rational numbers, while  $\mathbb{R}$ are the reals. We further define $\mathbb{R}_+=\{x\in\mathbb{R}|x\geq 0\}$ and $\mathbb{R}_{++}=\{x\in\mathbb{R}|x> 0\}$. Given two real numbers $a,b$, we denote by $[a\mspace{-3mu}:\mspace{-3mu}b]$ the set of integers $\big\{n\in\mathbb{N}\big| \lceil a\rceil\leq n \leq\lfloor b \rfloor\big\}$. Calligraphic letters denote sets, e.g., $\mathcal{X}$, the complement of $\mathcal{X}$ is denoted by $\mathcal{X}^c$, while $|\mathcal{X}|$ stands for its cardinality. $\mathcal{X}^n$ denotes the $n$-fold Cartesian product of $\mathcal{X}$. An element of $\mathcal{X}^n$ is denoted by $x^n=(x_1,x_2,\ldots,x_n)$; whenever the dimension $n$ is clear from the context, vectors (or sequences) are denoted by boldface letters, e.g., $\mathbf{x}$. For any $\mathcal{S}\subseteq[1:n]$, we use $\mathbf{x}^\mathcal{S}=(x_i)_{i\in\mathcal{S}}$ to denote the substring of $x^n$ defined by $\mathcal{S}$, with respect to the natural ordering of $\mathcal{S}$. For instance, if $\mathcal{S}=[i:j]$, where $1\leq i< j\leq n$, then $\mathbf{x}^\mathcal{S}=(x_i,x_{i+1},\ldots,x_j)$. For $\mathcal{S}=[i:j]$ as before we sometimes write $x_i^j$ instead of $\mathbf{x}^\mathcal{S}$; when $i=1$, the subscript is omitted. We also use $x^{n\backslash i}$ instead $\mathbf{x}^\mathcal{S}$, when $\mathcal{S}=[1:i-1]\cup[i+1:n]$, for $1\leq i\leq n$.


Let $\big(\mathcal{X},\mathcal{F},\mathbb{P}\big)$ be a probability space, where $\mathcal{X}$ is the sample space, $\mathcal{F}$ is the $\sigma$-algebra and $\mathbb{P}$ is the probability measure. Random variables over $\big(\mathcal{X},\mathcal{F},\mathbb{P}\big)$ are denoted by uppercase letters, e.g., $X$, with conventions for random vectors similar to those for deterministic sequences. The probability of an event $\mathcal{A}\in\mathcal{F}$ is denoted by $\mathbb{P}(\mathcal{A})$, while $\mathbb{P}(\mathcal{A}\big|\mathcal{B}\mspace{2mu})$ denotes conditional probability of $\mathcal{A}$ given $\mathcal{B}$. We use $\mathds{1}_\mathcal{A}$ to denote the indicator function of $\mathcal{A}$. The set of all probability mass functions (PMFs) on a finite set $\mathcal{X}$ is denoted by $\mathcal{P}(\mathcal{X})$, i.e., 
\begin{equation}
    \mathcal{P}(\mathcal{X})=\left\{P:\mathcal{X}\to[0,1]\Bigg| \sum_{x\in\mathcal{X}}P(x)=1]\right\}.
\end{equation}
PMFs are denoted by the uppercase letters such as $P$ or $Q$, with a subscript that identifies the random variable and its possible conditioning. For example, for a discrete probability space $\big(\mathcal{X},\mathcal{F},\mathbb{P}\big)$ and two correlated random variables $X$ and $Y$ over that space, we use $P_X$, $P_{X,Y}$ and $P_{X|Y}$ to denote, respectively, the marginal PMF of $X$, the joint PMF of $(X,Y)$ and the conditional PMF of $X$ given $Y$. In particular, $P_{X|Y}$ represents the stochastic matrix whose elements are given by $P_{X|Y}(x|y)=\mathbb{P}\big(X=x|Y=y\big)$. Expressions such as $P_{X,Y}=P_XP_{Y|X}$ are to be understood as $P_{X,Y}(x,y)=P_X(x)P_{Y|X}(y|x)$, for all $(x,y)\in\mathcal{X}\times\mathcal{Y}$. Accordingly, when three random variables $X$, $Y$ and $Z$ satisfy $P_{X|Y,Z}=P_{X|Y}$, they form a Markov chain, which we denote by $X-Y-Z$. We omit subscripts if the arguments of a PMF are lowercase versions of the random variables. The support of a PMF $P$ and the expectation of a real-valued random variable $X$ are denoted by $\supp(P)$ and $\mathbb{E}\big[X\big]$, respectively. If $X\sim P$, we emphasize that an expectation is taken with respect to the distribution on $X$ by writing $\mathbb{E}_X$ or $\mathbb{E}_P$ (choosing the simpler of the two). Similarly, we use $H_P$ and $I_P$ to indicate that an entropy or a mutual information term are calculated with respect to a PMF~$P$.

For a discrete measurable space $(\mathcal{X},\mathcal{F})$, a PMF $Q\in\mathcal{P}(\mathcal{X})$ gives rise to a probability measure on $(\mathcal{X},\mathcal{F})$, which we denote by $\mathbb{P}_Q$; accordingly, $\mathbb{P}_Q\big(\mathcal{A})=\sum_{x\in\mathcal{A}}Q(x)$, for every $\mathcal{A}\in\mathcal{F}$. For a sequence of random variable $X^n$, if the entries of $X^n$ are drawn in an independent and identically distributed (i.i.d.) manner according to $P_X$, then for every $\mathbf{x}\in\mathcal{X}^n$ we have $P_{X^n}(\mathbf{x})=\prod_{i=1}^nP_X(x_i)$ and we write $P_{X^n}(\mathbf{x})=P_X^n(\mathbf{x})$. Similarly, if for every $(\mathbf{x},\mathbf{y})\in\mathcal{X}^n\times\mathcal{Y}^n$ we have $P_{Y^n|X^n}(\mathbf{y}|\mathbf{x})=\prod_{i=1}^nP_{Y|X}(y_i|x_i)$, then we write $P_{Y^n|X^n}(\mathbf{y}|\mathbf{x})=P_{Y|X}^n(\mathbf{y}|\mathbf{x})$. We often use $Q_X^n$ or $Q_{Y|X}^n$ when referring to an i.i.d. sequence of random variables. The conditional product PMF $P_{Y|X}^n$ given a specific sequence $\mathbf{x}\in\mathcal{X}^n$ is denoted by $P_{Y|X=\mathbf{x}}^n$.

The type $\nu_{\mathbf{x}}$ of a sequence $\mathbf{x}\in\mathcal{X}^n$ is
\begin{equation}
\nu_{\mathbf{x}}(x)\triangleq\frac{N(x|\mathbf{x})}{n},\label{EQ:empirical_PMF}
\end{equation}
where $N(x|\mathbf{x})=\sum_{i=1}^n\mathds{1}_{\{x_i=x\}}$. The subset of $\mathcal{P}(\mathcal{X})$ that contains all possible types
of sequences $\mathbf{x}\in\mathcal{X}^n$ is denoted by $\mathcal{P}_n(\mathcal{X})$. By \cite[Lemma II.1]{Csiszar_MOT1998}, 
\begin{equation}
\big|\mathcal{P}_n(\mathcal{X})\big|=\binom {n+|\mathcal{X}|-1} {|\mathcal{X}|-1}\leq (n+1)^{|\mathcal{X}|}.\label{EQ:set_of_types_cardinality}
\end{equation}
For $P\in\mathcal{P}_n(\mathcal{X})$, the type-class $\big\{\mathbf{x}\in\mathcal{X}^n\big|\nu_\mathbf{x}=P\big\}$ is denoted by $\mathcal{T}^n_P$. We use $\mathcal{T}_\epsilon^{n}(P)$ to denote the set of letter-typical sequences with respect to the PMF $P\in\mathcal{P}(\mathcal{X})$ and the non-negative number $\epsilon$ defined by
\begin{equation}
\mathcal{T}_\epsilon^{n}(P)=\Big\{\mathbf{x}\in\mathcal{X}^n\Big|
    \big|\nu_{\mathbf{x}}(x)-P(x)\big|\leq\frac{\epsilon}{|\mathcal{X}|}\mathds{1}_{\{P(x)>0\}}\Big\}.\label{EQ:letter_typical_set}
\end{equation}
This definition of the letter-typical set resembles this from \cite[Chapter 2]{Csiszar_Korner_Book2011}, with the only difference being the normalization of $\epsilon$ by $|\mathcal{X}|$. This gives rise to an upper bound on the probability of an i.i.d. sequence being atypical that is uniform in the underlying i.i.d. distribution. Namely, by a simple adaptation of \cite[Lemma 2.12]{Csiszar_Korner_Book2011}, if $X^n$ is i.i.d. according to $P\in\mathcal{P}(\mathcal{X})$, then
\begin{equation}
\mathbb{P}_{P^n}\Big(X^n\notin\mathcal{T}_\epsilon^n(P)\Big)\leq 2|\mathcal{X}|e^{-2n\frac{\epsilon^2}{|\mathcal{X}|^2}}.\label{EQ:uniform_atypical_bound}
\end{equation}
This uniform bound plays an important role in the proof of Theorem \ref{TM:AVWTC_CR_converse_general}, where an upper bound on the CR-assisted SS-capacity of the AVWTC is established.

\begin{definition}[Relative Entropy]
	Let $(\mathcal{X},\mathcal{F})$ be a measurable space and let $P$ and $Q$ be two probability measures on $\mathcal{F}$, with $P\ll Q$ (i.e., $P$ is absolutely continuous with respect to $Q$). The relative entropy between $P$ and $Q$ is
	\begin{equation}
		D(P||Q)=\int_\mathcal{X} dP\log\left(\frac{dP}{dQ}\right),\label{EQ:relative_entropy_def}
	\end{equation}
	where $\frac{dP}{dQ}$ denotes the Radon-Nikodym derivative between $P$ and $Q$. If the sample space $\mathcal{X}$ is countable, \eqref{EQ:relative_entropy_def} reduces to
	\begin{equation}
		D(P||Q)=\sum_{x\in\supp(P)}P(x)\log\left(\frac{P(x)}{Q(x)}\right)\label{EQ:relative_entropy_def_discrete}.
	\end{equation}
\end{definition}

%
\begin{definition}[Total Variation]
	Let $(\mathcal{X},\mathcal{F})$ be a measurable and $P$ and $Q$ be two probability measures on $\mathcal{F}$. The total variation between $P$ and $Q$ is
	\begin{equation}
		||P-Q||_{\mathrm{TV}}=\sup_{\mathcal{A}\in\mathcal{F}}\big|P(\mathcal{A})-Q(\mathcal{A})\big|.\label{EQ:total_variation_def}
	\end{equation}
	If the sample space $\mathcal{X}$ is countable, \eqref{EQ:total_variation_def} reduces to
	\begin{equation}
		||P-Q||_{\mathrm{TV}}=\frac{1}{2}\sum_{x\in\mathcal{X}}\big|P(x)-Q(x)\big|.\label{EQ:total_variation_def_discrete}
	\end{equation}
\end{definition}

\section{Heterogeneous Strong Soft-Covering Lemma}\label{SEC:soft_covering}

We present here a generalization of the original \emph{strong soft-covering lemma} first established by the authors in \cite[Lemma 1]{Goldfeld_WTCII_semantic2015}. The lemma in this work is a heterogeneous version on the original homogeneous claim. Lemma 1 from \cite{Goldfeld_WTCII_semantic2015} considers a discrete-memoryless channel (DMC) that does not change throughout the block transmission, while here the memoryless channel may vary from symbol to symbol. This variation is modeled as an $n$-fold state-dependent channel $Q^n_{V|U,S}$ over which a codeword $\mathbf{u}\in\mathcal{U}^n$ is transmitted under a state sequence $\mathbf{s}\in\mathcal{S}^n$. Thus, in each time instance $i\in[1:n]$, the $i$-th symbol of $\mathbf{u}$ is transmitted over the channel $Q_{V|U,S=s_i}$.

\begin{figure}[t!]
    \begin{center}
        \begin{psfrags}
            \psfragscanon
            \psfrag{A}[][][1]{$\mspace{103mu}W\sim\mbox{Unif}(\mathcal{W}_n)$}
            \psfrag{B}[][][1]{\ \ \ \ \ \ \ \ \ \ \ \ $\mathcal{B}_n\mspace{-5mu}=\mspace{-5mu}\big\{\mspace{-2mu}\mathbf{U}(w)\mspace{-2mu}\big\}$}
            \psfrag{C}[][][1]{\ \ \ \ \ \ \ \ \ $\mathbf{U}(W)$}
            \psfrag{D}[][][1]{\ \ \ \ \ \ \ \ \ \ \ \ \ $Q^n_{V|U,S=\mathbf{s}}$}
            \psfrag{E}[][][1]{\ \ \ \ \ \ \ \ \ \ \ \ \ \ \ $\mathbf{V}\sim P^{(\mathbf{s})}_{\mathbf{V}|\mathsf{B}_n=\mathcal{B}_n}$}
            \hspace{-10mm}\includegraphics[scale = .44]{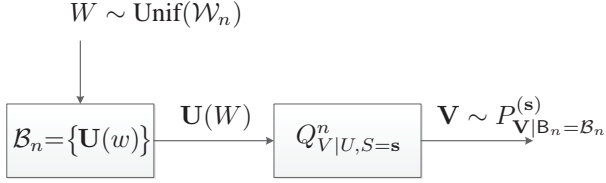}
             \caption{Coding problem with the goal of making $P^{(\mathbf{s})}_{\mathbf{V}|\mathsf{B}_n=\mathcal{B}_n}$ resemble $Q_{V|S=\mathbf{s}}^n$.} \label{FIG:soft_covering}
            \psfragscanoff
        \end{psfrags}
     \end{center}
 \end{figure}

Let $\mathcal{B}_n$ be a randomly generated codebook of $u$-sequence, one of which is selected uniformly at random and passed through the channel $Q^n_{V|U,S=\mathbf{s}}$. Lemma \ref{LEMMA:soft_covering} gives a sufficient condition for the induced conditional distribution of the channel output given the state to result in a good approximation of $Q^n_{V|S=\mathbf{s}}$ in the limit of large $n$, for any $\mathbf{s}\in\mathcal{S}^n$ (Fig.~\ref{FIG:soft_covering}). The proximity between the induced and the desired distributions is measured in terms of relative entropy. Specifically, we show that as long as the codebook is of size $|\mathcal{B}_n|=2^{nR}$ with $R>I(U;V|S)$, where the mutual information is calculated with respect to the empirical PMF $\nu_{\mathbf{s}}$ of the state sequence, the relative entropy vanishes exponentially quickly with the blocklength $n$, with high probability with respect to the random codebook. Via the Chenoff bound, the negligible probability of the random set not producing this desired result is doubly-exponentially small.

The heterogeneous strong soft-covering lemma is subsequently invoked for the SS analysis of the AVWTC, where the double-exponential decay it provides plays a key role. Similar claims that use total variation were previously made in the context of AVWTCs in \cite{Boche_AVWTC_multiletter_capacity2015,Boche_AVWTC_multiletterISIT2015,Boche_superactive_AVWTC2016} (in particular, see \cite[Lemma 1]{Boche_superactive_AVWTC2016}), though the codebook design was slightly different in those works. The stronger notion of soft-covering was also previously observed in works on  quantum information theory \cite{Devetak_Stronger_SCL2003,Wilde_Book2013}. We emphasize Lemma \ref{LEMMA:soft_covering} as a stand-alone tool due to its simplicity, and consequently, the prospect of it coming in handy for other proofs of secrecy, channel resolvability, channel synthesis, etc.



\subsection{Soft-Covering Setup and Result}

Let $\mathcal{S}$ be a finite set and let $\mathbf{s}\in\mathcal{S}^n$ be a sequence with an empirical PMF $\nu_\mathbf{s}$. Let $\mathsf{B}_n = \big\{\mathbf{U}(w)\big\}_{w\in\mathcal{W}_n}$, where\footnote{To simplify notation, we assume that $2^{nR}$ is an integer, for all $n\in\mathbb{N}$. Otherwise, simple modifications of some of the subsequent expressions using floor operations are needed.} $\mathcal{W}_n=[1:2^{nR}]$ with $R\in\mathbb{R}_+$, be a set of random vectors that are i.i.d. according to $Q_{U|S=\mathbf{s}}^n$, where $Q_{U|S}:\mathcal{S}\to\mathcal{P}(\mathcal{U})$. We refer to $\mathsf{B}_n$ as the random codebook, denote by $\mathfrak{B}_n$ the set of all its possible realizations, while a specific realization is denoted by $\mathcal{B}_n=\big\{\mathbf{u}(w)\big\}_{w\in\mathcal{W}_n}$. For every $\mathcal{B}_n\in\mathfrak{B}_n$, a sequence $\mathbf{u}(w)$ is randomly and uniformly selected and passed through a memoryless non-stationary channel $Q^n_{V|U,S=\mathbf{s}}$. For each $\mathbf{s}\in\mathcal{S}^n$, the induced joint distribution of $\mathsf{B}_n$, $W$ and $\mathbf{V}$ is
\begin{align*}
&P^{(\mathbf{s})}_{\mathsf{B}_n,W,\mathbf{V}}(\mathcal{B}_n,w,\mathbf{v})\\
&=\left[\prod_{w'\in\mathcal{W}_n}Q^n_{U|S}\big(\mathbf{u}(w')\big|\mathbf{s}\big)\right]2^{-nR}\cdot Q^n_{V|U,S}\big(\mathbf{v}\big|\mathbf{u}(w),\mathbf{s}\big)\numberthis\label{EQ:soft_covering_induced_jointPMF},
\end{align*}
which gives rise to a probability measure denoted by $\mathbb{P}$.\footnote{Since $\mathbf{s}\in\mathcal{S}^n$ stays fixed throughout this section, the notation $\mathbb{P}$ omits the dependence of the state sequence.} When switching to other probability measures, we do so in accordance to the notations defined in Section \ref{SEC:preliminaries}. On account of \eqref{EQ:soft_covering_induced_jointPMF}, the induced output distribution conditioned on a codebook $\mathcal{B}_n\in\mathfrak{B}_n$ for each state sequence  $\mathbf{s}\in\mathcal{S}^n$ is:
\begin{equation}
    P^{(\mathbf{s})}_{\mathbf{V}|\mathsf{B}_n}(\mathbf{v}|\mathcal{B}_n)=2^{-nR} \sum_{w\in\mathcal{W}_n} Q_{V|U,S}^n\big(\mathbf{v}\big|\mathbf{u}(w),\mathbf{s}\big).\label{EQ:soft_covering_induced_PMF}
\end{equation}
The following lemma states that as long as the codebook is of size $2^{nR}$, with\footnote{The subscript $\nu_\mathbf{s} Q$ indicates that the mutual information is calculated with respect to $\nu_\mathbf{s}Q_{U|S}Q_{V|U,S}$} $R>I_{\nu_\mathbf{s} Q}(U;V|S)$, the induced output PMF constitutes a good approximation of $Q_{V|S=\mathbf{s}}^n$ in the limit of large $n$, with high probability. Namely, the probability that the relative entropy between the induced PMF and product PMF vanishes exponentially quickly with the blocklength $n$, is double exponentially close to 1.

\begin{lemma}[Heterogeneous Strong Soft-Covering Lemma]\label{LEMMA:soft_covering}
For $\mathbf{s}\in\mathcal{S}^n$ with empirical PMF $\nu_\mathbf{s}$, and any $\zeta>0$, $Q_{U,V|S}:\mathcal{S}\to\mathcal{P}(\mathcal{U}\times\mathcal{V})$, and $R>I_{\nu_\mathbf{s}Q}(U;V|S)+\zeta$, where $|\mathcal{S}|,|\mathcal{V}|<\infty$, there exist $\gamma_1,\gamma_2>0$, such that for $n$ large enough
\begin{equation}
        \mathbb{P}\bigg(D\Big(P^{(\mathbf{s})}_{\mathbf{V}|\mathsf{B}_n}\Big|\Big|Q_{V|S=\mathbf{s}}^n\Big)> e^{-n\gamma_1}\bigg)\leq e^{- e^{n\gamma_2}}.\label{EQ:soft_covering}
\end{equation}
More precisely, for any $n \in \mathbb{N}$ and $\delta \in \big(0,R-I_{\nu_\mathbf{s} Q}(U;V|S)\big)$
\begin{equation}
\mathbb{P}\bigg(D\Big(P^{(\mathbf{s})}_{\mathbf{V}|\mathsf{B}_n}\Big|\Big|Q_{V|S=\mathbf{s}}^n\Big)> c_\delta n2^{-n\gamma_{\delta}} \bigg)\leq \big( 1 + |{\cal V}|^n \big) e^{-\frac{1}{3} 2^{n\delta}},\label{EQ:soft_covering_precise}
\end{equation}
where
\begin{subequations}
\begin{align}
    \gamma_{\delta}\mspace{-2mu} &= \mspace{-2mu}\sup_{\eta > 1} \frac{\eta\mspace{-2mu}-\mspace{-2mu}1}{2\eta\mspace{-2mu}-\mspace{-2mu} 1}\mspace{-4.5mu} \left(\mspace{-4.5mu}R\mspace{-2mu}-\mspace{-2mu}\delta\mspace{-2mu}-\mspace{-2mu}\max_{s\in\mathcal{S}}d_{\eta} (Q_{U,V|S=s},\mspace{-2mu}Q_{U|S=s} Q_{V|S=s})\mspace{-4.5mu}\right) \label{EQ:soft_covering_exponent}\\
    c_\delta &= 3\log e+\gamma_\delta2\log2+2 \log \left(\max_{\substack{(s,v)\in\mathcal{S}\times\mathcal{V}:\\Q_{V|S}(v|s)>0}} \frac{1}{Q_{V|S}(v|s)} \right),\label{EQ:soft_covering_coefficient1}
\end{align}
and $d_{\eta}(\mu,\nu)=\frac{1}{\eta-1}\log_2\int d\mspace{2mu}\mu\left(\frac{d\mspace{2mu}\mu}{d\mspace{2mu}\nu}\right)^{1-\eta}$ is the R\'{e}nyi divergence of order $\eta$.
\end{subequations}
\end{lemma}

The proof of Lemma \ref{LEMMA:soft_covering} bears close resemblance to the proof of the homogeneous version of the strong soft-covering lemma from \cite[Lemma 1]{Goldfeld_WTCII_semantic2015}. The main difference, is in the bound on the expected value of the probability of atypical sequences. To avoid verbatim repetition of the arguments from \cite{Goldfeld_WTCII_semantic2015}, we summarize (most of) the technical parts from the proof of the homogeneous case
in our Lemma \ref{LEMMA:technical} and invoke it to establish Lemma \ref{LEMMA:soft_covering}.


The important quantity in the lemma above is $\gamma_{\delta}$, which is the exponent that soft-covering achieves. We see in \eqref{EQ:soft_covering_precise} that the double-exponential convergence of probability occurs with exponent $\delta>0$. Thus, the best soft-covering exponent that the lemma achieves with confidence, over all $\delta>0$, is
\begin{align*}
    \gamma^*&=\sup_{\delta>0}\gamma_{\delta}\\
            &=\gamma_0\\
            &=\sup_{\eta > 1} \frac{\eta\mspace{-2mu}-\mspace{-2mu}1}{2\eta\mspace{-2mu}-\mspace{-2mu} 1}\mspace{-3mu} \left(\mspace{-3mu}R\mspace{-2mu}-\mspace{-2mu}\max_{s\in\mathcal{S}}d_{\eta} (Q_{U,V|S=s},\mspace{-2mu}Q_{U|S=s} Q_{V|S=s})\mspace{-3mu}\right).\numberthis
\end{align*}
The double-exponential confidence rate $\delta$ acts as a reduction in codebook rate $R$ in the definition of $\gamma_{\delta}$. Consequently, $\gamma_{\delta}=0$ for $\delta \geq R-I_{\nu_\mathbf{s} Q}(U;V|S)$.

\begin{remark} The role of $\zeta$ in the statement of Lemma \ref{LEMMA:soft_covering} is merely to ensure that a fixed $\delta\in\big(0,R-I_{\nu_\mathbf{s} Q}(U;V|S)\big)$ can be found for all $n$. Since the mutual information is calculated with respect to $\nu_\mathbf{s}$, its value may vary with $n$. Taking $R>I_{\nu_\mathbf{s} Q}(U;V|S)+\zeta$ implies that $(0,\zeta)\subseteq\big(0,R-I_{\nu_\mathbf{s} Q}(U;V|S)\big)$, and therefore, a fixed value of $\delta$ as needed exists.

\end{remark}

\begin{remark}[Total Variation Exponent of Decay]
The strong soft-covering lemma can be reproduced while replacing the relative entropy with total variation. Although, relative entropy can be used to bound total variation via Pinsker's inequality, this approach causes a loss of a factor of 2 in the exponent of decay. Alternatively, the proof of Lemma \ref{LEMMA:soft_covering} can be modified to produce the bound on the total variation instead of the relative entropy. This direct method keeps the error exponents the same for the total variation case as it is for relative entropy.
\end{remark}

\begin{IEEEproof}[Proof of Lemma \ref{LEMMA:soft_covering}]
We state the proof in terms of arbitrary distributions $Q_{U|S}$ and $Q_{V|U,S}$ (not necessarily discrete). We assume $|\mathcal{S}|<\infty$, and will specialize to a finite output alphabet $\mathcal{V}$ only when needed. 

First, define conditional information density $i_{Q_{U,V|S=s}}$, which is a function on the space $\mathcal{U}\times\mathcal{V}$ specified by
\begin{equation}
  i_{Q_{U,V|S=s}}(u,v)\triangleq\log\left(\frac{dQ_{V|U=u,S=s}}{dQ_{V|S=s}}(v)\right).\label{EQ:information_density}
\end{equation}
In \eqref{EQ:information_density}, the argument of the logarithm is the Radon-Nikodym derivative between $Q_{V|U=u,S=s}$ and $Q_{V|S=s}$. Let $\epsilon\geq0$ be arbitrary, and define the conditional jointly typical set of $u$- and $v$-sequences given $\mathbf{s}$ as
\begin{equation}
        \mathcal{A}_{\epsilon}(\mathbf{s})\mspace{-3mu}\triangleq\mspace{-3mu}\left\{\mspace{-3mu}(\mathbf{u},\mspace{-1.5mu}\mathbf{v})\mspace{-3mu}\in\mspace{-2mu}\mathcal{U}^n\mspace{-4mu}\times\mspace{-4mu}\mathcal{V}^n\bigg|\frac{1}{n}i_{Q^n_{U,V|S=\mathbf{s}}}\mspace{-3mu}(\mathbf{u},\mspace{-1.5mu}\mathbf{v})\mspace{-3mu}<\mspace{-3mu}I(U;V|S)\mspace{-2mu}+\mspace{-2mu}\epsilon\mspace{-1mu}\right\}\label{EQ:typical_set}
    \end{equation}
and note that
\begin{equation}
  i_{Q^n_{U,V|S=\mathbf{s}}}(\mathbf{u},\mathbf{v})=\sum_{t=1}^ni_{Q_{U,V|S=s_t}}(u_t,v_t).\label{EQ:information_density_product}
\end{equation}
For brevity, in \eqref{EQ:typical_set} and henceforth, we use $I(U;V|S)$ instead of $I_{\nu_\mathbf{s} Q}(U;V|S)$. 

Next, for every $\mathcal{B}_n\in\mathfrak{B}_n$, we split $P^{(\mathbf{s})}_{\mathbf{V}|\mathsf{B}_n=\mathcal{B}_n}$ into two parts, making use of the indicator function. For every $\mathbf{v}\in\mathcal{V}^n$, define
\begin{subequations}
    \begin{align}
        P_{\mathcal{B}_n,\mathbf{s}}^{(1)}(\mathbf{v})\mspace{-1.5mu} &\triangleq\mspace{-1.5mu} 2^{-nR}\mspace{-5mu} \sum_{w\in\mathcal{W}_n}\mspace{-2.5mu} Q_{V|U,S}^n\big(\mathbf{v}\big|\mathbf{u}(w),\mathbf{s}\big)\mathbf{1}_{\big\{\mspace{-1.5mu}\big(\mathbf{u}(w),\mathbf{v}\big)\in\mathcal{A}_{\epsilon}(\mathbf{s})\mspace{-1.5mu}\big\}}\\
        P_{\mathcal{B}_n,\mathbf{s}}^{(2)}(\mathbf{v})\mspace{-1.5mu} &\triangleq\mspace{-1.5mu} 2^{-nR}\mspace{-5mu} \sum_{w\in\mathcal{W}_n}\mspace{-2.5mu} Q_{V|U,S}^n\big(\mathbf{v}\big|\mathbf{u}(w),\mathbf{s}\big)\mathbf{1}_{\big\{\mspace{-1.5mu}\big(\mathbf{u}(w),\mathbf{v}\big)\notin\mathcal{A}_{\epsilon}(\mathbf{s})\mspace{-1.5mu}\big\}}\mspace{-2mu}.
    \end{align}
\end{subequations}
The measures $P_{\mathcal{B}_n,\mathbf{s}}^{(1)}$ and $P_{\mathcal{B}_n,\mathbf{s}}^{(2)}$ on the space $\mathcal{V}^n$ are not probability measures, but $P_{\mathcal{B}_n,\mathbf{s}}^{(1)}+P_{\mathcal{B}_n,\mathbf{s}}^{(2)}=P^{(\mathbf{s})}_{\mathbf{V}|\mathsf{B}_n=\mathcal{B}_n}$ for each codebook $\mathcal{B}_n\in\mathfrak{B}_n$. For every $\mathbf{v}\in\mathcal{V}^n$, we define
\begin{equation}
\Delta_{\mathcal{B}_n,\mathbf{s}}(\mathbf{v})=\frac{dP^{(\mathbf{s})}_{\mathbf{V}|\mathsf{B}_n=\mathcal{B}_n}}{dQ_{V|S=\mathbf{s}}^n}(\mathbf{v}),\label{EQ:soft_covering_Radon}
\end{equation}
where the right-hand side (RHS) is the Radon-Nikodym derivative between $P^{(\mathbf{s})}_{\mathbf{V}|\mathsf{B}_n=\mathcal{B}_n}$ and $Q_{V|S=\mathbf{s}}^n$, and also split it into $\Delta_{\mathcal{B}_n,\mathbf{s}}(\mathbf{v})=\Delta_{\mathcal{B}_n,\mathbf{s}}^{(1)}(\mathbf{v})+\Delta_{\mathcal{B}_n,\mathbf{s}}^{(2)}(\mathbf{v})$, where
\begin{subequations}
\begin{align}
    \Delta_{\mathcal{B}_n,\mathbf{s}}^{(1)}(\mathbf{v})&\triangleq\frac{dP_{\mathcal{B}_n,\mathbf{s}}^{(1)}}{dQ_{V|S=\mathbf{s}}^n}(\mathbf{v})\\
    \Delta_{\mathcal{B}_n,\mathbf{s}}^{(2)}(\mathbf{v})&\triangleq\frac{dP_{\mathcal{B}_n,\mathbf{s}}^{(2)}}{dQ_{V|S=\mathbf{s}}^n}(\mathbf{v}).
\end{align}\label{EQ:delta_def}%
\end{subequations}

To see that the RHS of \eqref{EQ:soft_covering_Radon} indeed exists, on could easily verify that 
$P^{(\mathbf{s})}_{\mathbf{V}|\mathsf{B}_n=\mathcal{B}_n}$ is absolutely continuous with respect to $Q_{V|S=\mathbf{s}}^n$. This essentially follows since if $Q_{V|S=\mathbf{s}}^n(\mathcal{A})=0$, for some $\mathcal{A}\subseteq\mathcal{V}^n$, then $Q^n_{V|U=\mathbf{u},S=\mathbf{s}}(\mathcal{A})=0$, for every $\mathbf{u}\in\supp\left(Q^n_{U|S=\mathbf{s}}\right)$. The structure of $P^{(\mathbf{s})}_{\mathbf{V}|\mathsf{B}_n=\mathcal{B}_n}$ given in \eqref{EQ:soft_covering_induced_PMF} then implies that $P^{(\mathbf{s})}_{\mathbf{V}|\mathsf{B}_n=\mathcal{B}_n}(\mathcal{A})=0$.

Note that $\int dP_{\mathsf{B}_n,\mathbf{s}}^{(2)}$ is an average of exponentially many i.i.d. random variables bounded between 0 and 1, given by  
\begin{align*}
    \int& dP_{\mathsf{B}_n,\mathbf{s}}^{(2)}\\
    &=\sum_{w\in\mathcal{W}_n} 2^{-nR}\cdot 
    \mathbb{P}_{Q_{V|U,S=\mathbf{s}}^n}\Big(\big(\mathbf{U}(w),\mathbf{V}\big)\notin\mathcal{A}_\epsilon(\mathbf{s})\Big|\mathbf{U}(w)\Big).\numberthis\label{EQ:chernoff_rvs1}
\end{align*}
With respect to \eqref{EQ:chernoff_rvs1} and the above definitions, the heterogeneous strong soft-covering lemma is established by the following technical lemma.
\begin{lemma}\label{LEMMA:technical}
Let $|\mathcal{S}|,|\mathcal{V}|<\infty$ and $\epsilon\geq 0$. If there exist $\alpha,\beta_\epsilon>0$ such that
\begin{subequations}
\begin{align}
&\mathbb{E}_{\mathsf{B}_n}\mathbb{P}_{Q_{V|U,S=\mathbf{s}}^n}\Big(\big(\mathbf{U}(w),\mathbf{V}\big)\notin\mathcal{A}_\epsilon(\mathbf{s})\Big|\mathbf{U}(w)\Big)\leq 2^{-n\beta_\epsilon}\label{EQ:technical_assumption1}\\
&\mathbb{P}\left(\Delta^{(2)}_{\mathsf{B}_n,\mathbf{s}}(\mathbf{v})\leq \alpha^n\right)=1,\quad\forall\mspace{3mu}\mathbf{v}\in\mathcal{V}^n,\label{EQ:technical_assumption2}
\end{align}\label{EQ:technical_assumptions}%
\end{subequations}
then
\begin{align*}
\mathbb{P}\bigg(D\Big(&P^{(\mathbf{s})}_{\mathbf{V}|\mathsf{B}_n}\Big|\Big|Q_{V|S=\mathbf{s}}^n\Big)\geq c_{\beta,\epsilon,\alpha} n2^{-n\beta_\epsilon}\bigg)\\
&\leq e^{-\frac{1}{3}2^{n(R-\beta_\epsilon)}}+|\mathcal{V}|^ne^{-\frac{1}{3}2^{n\left(R-I(U;V|S)-\epsilon-2\beta_\epsilon\right)}},\numberthis\label{EQ:technical_established}
\end{align*}
where $c_{\beta,\epsilon,\alpha}= 3\log e+2\beta_\epsilon\log2+2 \log\alpha$.
\end{lemma}
Lemma \ref{LEMMA:technical} essentially follows from the proof of Lemma 1 from \cite{Goldfeld_WTCII_semantic2015}. More specifically, the derivation repeats the steps between Equations (18) and (40) in the proof of \cite[Lemma 1]{Goldfeld_WTCII_semantic2015} and is, therefore, omitted. Having this, it remains to be shown that \eqref{EQ:technical_assumptions} holds for certain positive $\beta_\epsilon$ and $\alpha$. For \eqref{EQ:technical_assumption1}, observe that
\begin{align*}
    &\mathbb{E}_{\mathsf{B}_n}\mathbb{P}_{Q_{V|U,S=\mathbf{s}}^n}\Big(\big(\mathbf{U}(w),\mathbf{V}\big)\notin\mathcal{A}_\epsilon(\mathbf{s})\Big|\mathbf{U}(w)\Big)\\
    &=\mathbb{P}_{Q_{U,V|S=\mathbf{s}}^n}\Big(\big(\mathbf{U},\mathbf{V}\big)\notin\mathcal{A}_\epsilon(\mathbf{s})\Big)\\
    &= \mathbb{P}_{Q_{U,V|S=\mathbf{s}}^n} \left( \sum_{t=1}^n i_{Q_{U,V|S=s_t}}(U_t,V_t) \geq n \big(I(U;V|S) + \epsilon\big)\right) \\
    &\stackrel{(a)}= \mathbb{P}_{Q_{U,V|S=\mathbf{s}}^n} \left( 2^{\lambda\sum_{t=1}^n i_{Q_{U,V|S=s_t}}(U_t,V_t)} \geq 2^{n\lambda(I(U;V|S) + \epsilon)}\right) \\
    &\stackrel{(b)}\leq \frac{\mathbb{E}_{Q_{U,V|S=\mathbf{s}}^n} 2^{\lambda \sum_{t=1}^ni_{Q_{U,V|S=s_t}}(U_t,V_t)}}{2^{n\lambda (I(U;V|S) + \epsilon)}},\numberthis\label{EQ:atypical_expectation_UB1}
\end{align*}
where (a) is true for any $\lambda\geq0$ and (b) is Markov's inequality. For the numerator on the RHS of \eqref{EQ:atypical_expectation_UB1}, we have
\begin{align*}
\mathbb{E}_{Q_{U,V|S=\mathbf{s}}^n} &2^{\lambda \sum_{t=1}^n i_{Q_{U,V|S=s_t}}(U_t,V_t)}\\
&\stackrel{(a)}=\prod_{t=1}^n\mathbb{E}_{Q_{U,V|S=s_t}} 2^{\lambda i_{Q_{U,V|S=s_t}}(U_t,V_t)}\\
&\leq\left(\max_{t\in[1:n]}\mathbb{E}_{Q_{U,V|S=s_t}} 2^{\lambda i_{Q_{U,V|S=s_t}}(U_t,V_t)}\right)^n\\
&\stackrel{(b)}\leq\left(\max_{s\in\mathcal{S}}\mathbb{E}_{Q_{U,V|S=s}} 2^{\lambda i_{Q_{U,V|S=s}}(U_s,V_s)}\right)^n,\numberthis\label{EQ:atypical_expectation_numerator_UB}
\end{align*}
where (a) uses the independence across time, while (b) follows because $|\mathcal{S}|<\infty$ and by defining $(U_s,V_s)\sim Q_{U,V|S=s}$. Plugging \eqref{EQ:atypical_expectation_numerator_UB} back into \eqref{EQ:atypical_expectation_UB1}, gives
\begin{align*}
    &\mathbb{E}_{\mathsf{B}_n}\mathbb{P}_{Q_{V|U,S=\mathbf{s}}^n}\Big(\big(\mathbf{U}(w),\mathbf{V}\big)\notin\mathcal{A}_\epsilon(\mathbf{s})\Big|\mathbf{U}(w)\Big)\\
    &\leq\left( \frac{\max_{s\in\mathcal{S}}\mathbb{E}_{Q_{U,V|S=s}} 2^{\lambda i_{Q_{U,V|S=s}}(U_s,V_s)}}{2^{\lambda (I(U;V|S) + \epsilon)}} \right)^n \\
    &=\left( \frac{2^{\log_2\left(\max\limits_{s\in\mathcal{S}}\mathbb{E}_{Q_{U,V|S=s}} 2^{\lambda i_{Q_{U,V|S=s}}(U_s,V_s)}\right)}}{2^{\lambda (I(U;V|S) + \epsilon)}} \right)^n \\
    &\stackrel{(a)}=\left( \frac{2^{\lambda \max\limits_{s\in\mathcal{S}}\frac{1}{\lambda}\log_2\left(\mathbb{E}_{Q_{U,V|S=s}} 2^{\lambda i_{Q_{U,V|S=s}}(U_s,V_s)}\right)}}{2^{\lambda (I(U;V|S) + \epsilon)}} \right)^n \\
    &= 2^{n \lambda \left(\max\limits_{s\in\mathcal{S}} \frac{1}{\lambda} \log_2 \mathbb{E}_{Q_{U,V|S=s}} \big[2^{\lambda i_{Q_{U,V|S=s}}(U_s,V_s)}\big] - I(U;V|S) - \epsilon \right)} \\
    &\stackrel{(b)}= 2^{n \lambda \big( \max\limits_{s\in\mathcal{S}}d_{\lambda+1}(Q_{U,V|S=s},Q_{U|S=s} Q_{V|S=s}) - I(U;V|S) - \epsilon \big)},\numberthis\label{EQ:atypical_expectation_UB2}
\end{align*}
where (a) is because the logarithm is non-decreasing and by restricting $\lambda$ to be strictly positive, while (b) is from the definition of the R\'{e}nyi divergence of order $\lambda+1$. Substituting $\eta = \lambda+1$ into \eqref{EQ:atypical_expectation_UB2} yields
\begin{equation}
    \mathbb{E}_{\mathsf{B}_n}\mathbb{P}_{Q_{V|U,S=\mathbf{s}}^n}\Big(\big(\mathbf{U}(w),\mathbf{V}\big)\notin\mathcal{A}_\epsilon(\mathbf{s})\Big|\mathbf{U}(w)\Big)\leq 2^{-n\beta_{\eta,\epsilon}},\label{EQ:atypical probability expectation}
\end{equation}
where
\begin{align*}
     \beta_{\eta,\epsilon}=(\eta - 1) \bigg(I(&U;V|S) + \epsilon\\
     &-\max_{s\in\mathcal{S}}d_{\eta}(Q_{U,V|S=s}, Q_{U|S=s} Q_{V|S=s}) \bigg),\numberthis\label{EQ:chernoff_rvs1_prop_beta}
\end{align*}
for every $\eta>1$ and $\epsilon\geq 0$. Thus \eqref{EQ:technical_assumption1} holds with $\beta_{\eta,\epsilon}$ in the role of $\beta_\epsilon$.

The relation in  \eqref{EQ:technical_assumption2} follows by noting that $|\mathcal{S}|,|\mathcal{V}|<\infty$ implies that for any $\mathbf{v}\in\mathcal{V}^n$ and $\mathcal{B}_n\in\mathfrak{B}_n$, we have
\begin{equation}
    \Delta_{\mathcal{B}_n,\mathbf{s}}^{(2)}(\mathbf{v}) \leq \left( \max_{\substack{(s,v)\in\mathcal{S}\times\mathcal{V}:\\Q_{V|S}(v|s)>0}} \frac{1}{Q_{V|S}(v|s)} \right)^n.\label{EQ:atypical divergence bound}
\end{equation}
Notice that the maximum is only over the pair $(s,v)$ for which $Q_{V|S}(v|s)>0$, which makes this bound finite. The underlying reason for this restriction is that with probability one a conditional distribution is absolutely continuous with respect to its associated marginal distribution. By \eqref{EQ:atypical divergence bound}, we obtain
\begin{equation}
\mathbb{P}\left(\Delta^{(2)}_{\mathsf{B}_n,\mathbf{s}}(\mathbf{v})\leq\alpha^n\right)=1,\quad\forall\mspace{3mu}\mathbf{v}\in\mathcal{V}^n,
\end{equation}
with 
\begin{equation}
\alpha=\max_{\substack{(s,v)\in\mathcal{S}\times\mathcal{V}:\\Q_{V|S}(v|s)>0}} \frac{1}{Q_{V|S}(v|s)}.\label{EQ:technical_alpha}
\end{equation}

Thus, by Lemma \ref{LEMMA:technical} we have \eqref{EQ:technical_established} with $\beta_{\eta,\epsilon}$ and $\alpha$ from \eqref{EQ:chernoff_rvs1_prop_beta} and \eqref{EQ:technical_alpha}, respectively. Recalling that we may optimize over $\eta>1$ and $\epsilon\geq 0$, we fix $\delta \in \big(0,R-I(U;V|S)\big)$ and set
\begin{align*}
    \mspace{-10mu}\epsilon_{\eta,\delta}&\mspace{-1.5mu}=\mspace{-1.5mu}\frac{ \frac{1}{2} (\mspace{-1mu}R\mspace{-2mu}-\mspace{-2mu}\delta)\mspace{-2.5mu}+\mspace{-2.5mu}(\eta\mspace{-2mu}-\mspace{-2mu}1\mspace{-1.5mu}) \max\limits_{s\in\mathcal{S}}d_{\eta}(\mspace{-1mu}Q_{U,V|S=s},\mspace{-1.5mu}Q_{U|S=s} Q_{V|S=s}\mspace{-1mu})}{ \frac{1}{2} + (\eta- 1) }\\
    &\mspace{270mu}-I (U;V|S).\numberthis\label{EQ:optimized_epsilon}
\end{align*}
Substituting into $\beta_{\eta,\epsilon}$ gives
\begin{align*}
    \beta_{\eta,\delta}&\triangleq\beta_{\eta,\epsilon_{\eta,\delta}}\\
                       &=\mspace{-2mu}\frac{\eta\mspace{-2mu}-\mspace{-2mu}1}{2\eta\mspace{-2mu}-\mspace{-2mu}1}\mspace{-3mu} \left(\mspace{-3mu}R\mspace{-2mu}-\mspace{-2mu}\delta\mspace{-2mu}-\mspace{-2mu}\max_{s\in\mathcal{S}}d_{\eta}(Q_{U,V|S=s},\mspace{-1mu}Q_{U|S=s}Q_{V|S=s})\mspace{-2mu}\right)\mspace{-3mu}.\numberthis\label{EQ:beta_epsilon}
\end{align*}
Plugging $\alpha$ and $\beta_{\eta,\delta}$ into $c_{\beta,\epsilon,\alpha}$, which we relabel as $c_{\eta,\delta}$, we have
\begin{equation}
    c_{\eta,\delta}= 3\log e+2\beta_{\eta,\delta} \log2+2 \log \left(\max_{\substack{(s,v)\in\mathcal{S}\times\mathcal{V}:\\Q_{V|S}(v|s)>0}} \frac{1}{Q_{V|S}(v|s)}\right)\mspace{-3mu}.\label{EQ:c_etadelta}
\end{equation}
Observe that $\epsilon_{\eta,\delta}$ in \eqref{EQ:optimized_epsilon} is non-negative under the assumption that $R-\delta> I(U;V|S)$, because $\eta>1$ and
\begin{align*}
\max_{s\in\mathcal{S}}d_{\eta}(Q_{U,V|S=s}&, Q_{U|S=s} Q_{V|S=s})\\
&\geq \max_{s\in\mathcal{S}}d_1(Q_{U,V|S=s}, Q_{U|S=s} Q_{V|S=s})\\
&\geq I(U;V|S).\numberthis
\end{align*}

Reevaluating \eqref{EQ:technical_established} based on \eqref{EQ:optimized_epsilon}-\eqref{EQ:c_etadelta} gives
\begin{align*}
\mathbb{P}\bigg(D\Big(&P^{(\mathbf{s})}_{\mathbf{V}|\mathsf{B}_n}\Big|\Big|Q_{V|S=\mathbf{s}}^n\Big)\geq c_{\eta,\delta} n2^{-n\beta_{\eta,\delta}}\bigg)\\
&\leq e^{-\frac{1}{3}2^{n\left(R-\beta_{\eta,\epsilon}\right)}}+|\mathcal{V}|^ne^{-\frac{1}{3}2^{n\left(R-I(U;V|S)-\epsilon_{\eta,\delta}-2\beta_{\eta,\delta}\right)}}\\
&= e^{-\frac{1}{3}2^{n\left(R-\beta_{\eta,\delta}\right)}} + |\mathcal{V}|^n\cdot e^{-\frac{1}{3} 2^{n\delta}}\\
&\stackrel{(a)}\leq \left(1+|\mathcal{V}|^n\right)e^{-\frac{1}{3} 2^{n\delta}},\numberthis\label{EQ:bad_divergence_final_UB}
\end{align*}
where (a) is because $\beta_{\eta,\delta} \leq \frac{1}{2}(R-\delta)$. Denoting $c_\delta\triangleq \sup_{\eta> 1}c_{\eta,\delta}$, \eqref{EQ:bad_divergence_final_UB} further gives
\begin{equation}
\mathbb{P}\bigg(\mspace{-2mu}D\Big(P^{(\mathbf{s})}_{\mathbf{V}|\mathsf{B}_n}\Big|\Big|Q_{V|S=\mathbf{s}}^n\Big)\mspace{-1.5mu}\geq\mspace{-1.5mu} c_\delta n2^{-n\beta_{\eta,\delta}}\Big)\mspace{-1.5mu}\leq\mspace{-1.5mu} \left(1+|\mathcal{V}|^n\right)e^{-\frac{1}{3} 2^{n\delta}}\mspace{-3mu}.\label{EQ:bad_divergence_final_UB_final}
\end{equation}
Since \eqref{EQ:bad_divergence_final_UB_final} is true for all $\eta>1$, it must also be true, with strict inequality in the LHS, when replacing $\beta_{\eta,\delta}$ with
\begin{align*}
  \gamma_{\delta}&\triangleq\sup_{\eta> 1}\beta_{\eta,\delta}\\
                 &=\mspace{-2mu}\sup_{\eta > 1} \frac{\eta\mspace{-2mu}-\mspace{-2mu}1}{2\eta\mspace{-2mu}-\mspace{-2mu} 1}\mspace{-4.5mu} \left(\mspace{-4.5mu}R\mspace{-2mu}-\mspace{-2mu}\delta\mspace{-2mu}-\mspace{-2mu}\max_{s\in\mathcal{S}}d_{\eta} (Q_{U,V|S=s},\mspace{-2mu}Q_{U|S=s} Q_{V|S=s})\mspace{-4.5mu}\right) \label{EQ:soft_covering_exponent_def}
\end{align*}
which is the exponential rate of convergence stated in \eqref{EQ:soft_covering_exponent} that we derive for the heterogeneous strong soft-covering lemma. This establishes the statement from \eqref{EQ:soft_covering_precise} and proves Lemma \ref{LEMMA:soft_covering}.

Concluding, if $R > I(U;V|S)+\zeta$, for any $\zeta>0$ arbitrarily small, then for any $\delta\in\big(0,R-I(U;V|S)\big)$, we get exponential convergence of the relative entropy at rate $O(2^{- n\gamma_\delta})$ with double-exponential certainty. Discarding the precise exponents of convergence and coefficients, we state that there exist $\gamma_1,\gamma_2>0$, such that for $n$ large enough
\begin{equation}
   \mathbb{P}\bigg(D\Big(P^{(\mathbf{s})}_{\mathbf{V}|\mathsf{B}_n}\Big|\Big|Q_{V|S=\mathbf{s}}^n\Big)>e^{-n\gamma_1}\bigg)\leq e^{-e^{n\gamma_2}}.
\end{equation}
\end{IEEEproof}


\section{Arbitrarily Varying Wiretap Channels with Type Constrained States}\label{SEC:AVWTC}


\subsection{Problem Setup and Definitions}\label{SUBSEC:AVWTC_setup}


\begin{figure}[t!]
    \begin{center}
        \begin{psfrags}
            \psfragscanon
            \psfrag{I}[][][1]{\ \ $m$}
            \psfrag{J}[][][1]{\ \ \ \ \ \ \ \ Enc $f_n$}
            \psfrag{X}[][][1]{\ \ $(\mathfrak{W},\mathfrak{V})$}
            \psfrag{K}[][][1]{\ \ \ \ \ $X^n$}
            \psfrag{S}[][][1]{\ \ \ \ \ \ \ \ \ \ AVWTC}
            \psfrag{M}[][][1]{\ \ \ $Y^n_\mathbf{s}$}
            \psfrag{N}[][][1]{\ \ \ $Z^n_\mathbf{s}$}
            \psfrag{O}[][][1]{\ \ \ \ \ \ \ Dec $\phi_n$}
            \psfrag{P}[][][0.8]{\ \ \ \ \ \ \ \ \ Eavesdropper}
            \psfrag{Q}[][][1]{\ \ $\hat{m}$}
            \psfrag{U}[][][1]{$\mspace{4mu}m$}
            \psfrag{T}[][][1]{\ \ \ \ \ \ \ \ \ \  $\mathbf{s}\in\mathcal{S}_\mathcal{Q}^n$}
            \hspace{15mm}\includegraphics[scale = .45]{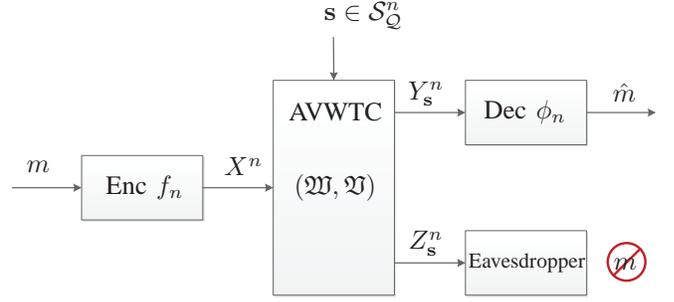}
            \caption{The AVWTC with $\mathcal{Q}$-constrained states, i.e., when the allowed state sequences have empirical PMFs that belong to~$\mathcal{Q}$.} \label{FIG:AVWTC}
            \psfragscanoff
        \end{psfrags}
     \end{center}
 \end{figure}


Let $\mathcal{X}$, $\mathcal{Y}$, $\mathcal{Z}$ and $\mathcal{S}$ be finite sets. A discrete-memoryless (DM) arbitrarily varying wiretap channel (AVWTC), as illustrated in Fig. \ref{FIG:AVWTC}, is defined by a pair $(\mathfrak{W},\mathfrak{V})$ of families of channels $\mathfrak{W}=\big\{W_s:\mathcal{X}\to\mathcal{P}(\mathcal{Y})\big|s\in\mathcal{S}\big\}$ and $\mathfrak{V}=\big\{V_s:\mathcal{X}\to\mathcal{P}(\mathcal{Z})\big|s\in\mathcal{S}\big\}$, from $\mathcal{X}$ to $\mathcal{Y}$ and $\mathcal{Z}$, respectively. Thus, $s\in\mathcal{S}$ denotes the state of the channels and can be interpreted as an index identifying a particular pair
$(W,V)\in\mathfrak{W}\times\mathfrak{V}$. 

The $n$-th extension of the channel laws for input $\mathbf{x}\in\mathcal{X}^n$ and outputs $\mathbf{y}\in\mathcal{Y}^n$ and $\mathbf{z}\in\mathcal{Z}^n$, under the state sequence $\mathbf{s}\in\mathcal{S}^n$ are
\begin{subequations}
\begin{align}
W^n_{\mathbf{s}}(\mathbf{y}|\mathbf{x})\triangleq\prod_{i=1}^nW_{s_i}(y_i|x_i)\\
V^n_{\mathbf{s}}(\mathbf{z}|\mathbf{x})\triangleq\prod_{i=1}^nV_{s_i}(z_i|x_i).
\end{align}
\end{subequations}
The families of channels $W_\mathbf{s}^n:\mathcal{X}^n\to\mathcal{P}(\mathcal{Y}^n)$ and $V_\mathbf{s}^n:\mathcal{X}^n\to\mathcal{P}(\mathcal{Z}^n)$, for $\mathbf{s}\in\mathcal{S}^n$, are denoted by $\mathfrak{W}^n$ and $\mathfrak{V}^n$, respectively, and $(\mathfrak{W}^n,\mathfrak{V}^n)$ is referred to as the ($n$-fold) AVWTC. The random variables representing the outputs of the AVWTC $(\mathfrak{W}^n,\mathfrak{V}^n)$ observed by the legitimate user and by the eavesdropper under the state sequence $\mathbf{s}\in\mathcal{S}^n$ are denoted by $Y^n_\mathbf{s}$ and $Z^n_\mathbf{s}$, respectively.

For any $\mathcal{Q}\subseteq\mathcal{P}(\mathcal{S})$ define
\begin{equation}
\mathcal{S}_\mathcal{Q}^n\triangleq\Big\{\mathbf{s}\in\mathcal{S}^n\Big|\nu_\mathbf{s}\in\mathcal{Q}\Big\}.\label{EQ:Q_PMFS}
\end{equation}
We impose a constraint $\mathcal{Q}$ on the allowed state sequences, i.e., only $\mathbf{s}\in\mathcal{S}_\mathcal{Q}^n$ are permitted. The triple $(\mathfrak{W}^n,\mathfrak{V}^n,\mathcal{Q})$ is referred to as the ($n$-fold) \emph{$\mathcal{Q}$-constrained AVWTC}. We subsequently focus on the \emph{type constrained AVWTC} formally defined in Definitions \ref{DEF:error_prob_SS_metric_type} and \ref{DEF:CR_achievability_type}. Intuitively, one may think of the type constrained scenario as corresponding to $\mathcal{Q}$ being a singleton, i.e., $\mathcal{Q}=\big\{Q_S\big\}$, for some $Q_S\in\mathcal{P}(\mathcal{S})$. However, such a setting would not be well-defined if $Q_S$ is not a rational distribution. Even if $Q_S$ is rational, it is not a valid type for all $n\in\mathbb{N}$, thus restricting the feasible blocklengths for the AVWTC. To circumvent these pathologies, in Definitions \ref{DEF:error_prob_SS_metric_type} and \ref{DEF:CR_achievability_type} we restrict the state sequences to have types that are close to $Q_S$, but not necessarily $Q_S$ itself.

\begin{remark}
One easily verifies that defining an AVWTC in terms of the pair $(\mathfrak{W}^n,\mathfrak{V}^n)$ is without loss of generality. In general, any state-input pair $(s,x)\in\mathcal{S}\times\mathcal{X}$ induces a joint conditional output PMF $U_s(\cdot,\cdot|x)\in\mathcal{P}(\mathcal{Y}\times\mathcal{Z})$. However, the performance of any of the codes defined below is measured with respect to the marginal output PMFs $W_s(\cdot|x)\in\mathcal{P}(\mathcal{Y})$ and $V_s(\cdot|x)\in\mathcal{P}(\mathcal{Z})$. Thus, under the framework presented here, all AVWTCs with the same marginals $W$ and $V$ are equivalent.
\end{remark}


\begin{definition}[Uncorrelated Code]\label{DEF:deterministic_code}
An uncorrelated $(n,M_n)$-code $c_n$ for the AVWTC $(\mathfrak{W}^n,\mathfrak{V}^n)$ has a message set $\mathcal{M}_n=[1:M_n]$, a stochastic encoder $f_n:\mathcal{M}_n\to\mathcal{P}(\mathcal{X}^n)$ and decoder $\phi_n:\mathcal{Y}^n\to\hat{\mathcal{M}}_n$, where $\hat{\mathcal{M}}_n\triangleq\mathcal{M}_n\cup\{e\}$ and $e\notin\mathcal{M}_n$ is an error symbol.
\end{definition}

For any uncorrelated $(n,M_n)$-code $c_n$ and state sequence $\mathbf{s}\in\mathcal{S}^n$, the induced joint PMF on $\mathcal{M}_n\times\mathcal{X}^n\times\mathcal{Y}^n\times\mathcal{Z}^n\times\hat{\mathcal{M}}$~is
\begin{align*}
&P^{(c_n,\mathbf{s})}_{M,\mathbf{X},\mathbf{Y}_\mathbf{s},\mathbf{Z}_\mathbf{s},\hat{M}}(m,\mathbf{x},\mathbf{y},\mathbf{z},\hat{m})\\
&\mspace{20mu}\triangleq P_M(m)f_n(\mathbf{x}|m)W^n_\mathbf{s}(\mathbf{y}|\mathbf{x})V^n_\mathbf{s}(\mathbf{z}|\mathbf{x})\mathds{1}_{\big\{\hat{m}=\phi_n(\mathbf{y})\big\}},\numberthis\label{EQ:AVWTC_induced_PMF}
\end{align*}
where $P_M\in\mathcal{P}(\mathcal{M}_n)$. The performance of $c_n$ on the type constrained AVWTC $(\mathfrak{W}^n,\mathfrak{V}^n,Q_S)$ is evaluated in terms of its rate $\frac{1}{n}\log M_n$, the maximal decoding error probability and the SS-metric. Reliability and security must be ensured with respect to every allowed constrained state sequence.

\begin{definition}[Message Error Probability]\label{DEF:WTCI_error_probability} Let $c_n$ be an uncorrelated $(n,M_n)$-code for the AVWTC $(\mathfrak{W}^n,\mathfrak{V}^n)$. For any $m\in\mathcal{M}_n$ and $\mathbf{s}\in\mathcal{S}^n$, let $e_m(W^n_\mathbf{s},c_n)$ be the error probability in decoding $m$ under the state sequence $\mathbf{s}$, given by
\begin{equation}
e_m(W_\mathbf{s}^n,c_n)=\sum_{\mathbf{x}\in\mathcal{X}^n}f_n(\mathbf{x}|m)\sum_{\substack{\mathbf{y}\in\mathcal{Y}^n:\\\phi_n(\mathbf{y})\neq m}}W_\mathbf{s}^n(\mathbf{y}|\mathbf{x}).\label{EQ:AVWTC_state_error_prob}
\end{equation}
\end{definition}
\begin{definition}[SS Metric] Let $c_n$ be an uncorrelated $(n,M_n)$-code for the AVWTC $(\mathfrak{W}^n,\mathfrak{V}^n)$. The information leakage to the eavesdropper under the state sequence $\mathbf{s}\in\mathcal{S}^n$ and the message PMF $P_M\in\mathcal{P}(\mathcal{M}_n)$ is
\begin{equation}
\ell(V_\mathbf{s}^n,P_M,c_n)=I_{c_n}(M;\mathbf{Z}_\mathbf{s}),\label{EQ:AVWTC_info_leakage}
\end{equation}
where the subscript $c_n$ denotes that the mutual information term is calculated with respect to the induced joint distribution $P^{(c_n,\mathbf{s})}_{M,Z^n_\mathbf{s}}$ from \eqref{EQ:AVWTC_induced_PMF}. 
For any $\mathcal{Q}\subseteq\mathcal{P}(\mathcal{S})$, the SS metric with respect to $c_n$ and the $\mathcal{Q}$-constrained AVWTC $(\mathfrak{W}^n,\mathfrak{V}^n,\mathcal{Q})$ is\footnote{$\ell_\mathrm{Sem}(\mathfrak{V}^n,\mathcal{Q},c_n)$ is actually the mutual-information-security (MIS) metric, which is equivalent to SS by \cite{Vardy_Semantic_WTC2012}. We use the representation in \eqref{EQ:AVWTC_SS_metric} rather than the formal definition of SS (see, e.g., \cite[Equation (4)]{Vardy_Semantic_WTC2012}) out of analytical convenience.}
\begin{equation}
\ell_\mathrm{Sem}(\mathfrak{V}^n,\mathcal{Q},c_n)=\max_{\substack{\mathbf{s}\in\mathcal{S}_\mathcal{Q}^n,\\P_M\in\mathcal{P}(\mathcal{M}_n)}}\ell(V_\mathbf{s}^n,P_M,c_n).\label{EQ:AVWTC_SS_metric}
\end{equation}
\end{definition}

\begin{remark}
We use the convention that the maximum over an empty set is $-\infty$. Accordingly if $\mathcal{Q}$ contains no rational distributions then $\ell_\mathrm{Sem}(\mathfrak{V}^n,\mathcal{Q},c_n)=-\infty$, for all $n\in\mathbb{N}$. Even when there exists $Q_S\in\mathcal{P}_n(\mathcal{S})$ such that $Q_S\in\mathcal{Q}$, there are blocklengths $n$ for which $\nu_\mathbf{s}\neq Q_S$ for every $\mathbf{s}\in\mathcal{S}^n$, and consequently, $\ell_\mathrm{Sem}(\mathfrak{V}^n,\mathcal{Q},c_n)=-\infty$ for these values of $n$ as well.
\end{remark}

\begin{remark}\label{REM:Message_PMF_function_of_code} SS requires that the uncorrelated code $c_n$ works well for all message PMFs. This means that the mutual information term in \eqref{EQ:AVWTC_SS_metric} is maximized over $P_M$ when $c_n$ is known. In other words, although not stated explicitly, the optimal $P_M$ is a function of $c_n$.
\end{remark}




We proceed with defining correlated random (CR) codes, their associated maximal error probability and SS-metric, CR-assisted achievability and CR-assisted secrecy-capacity.

\begin{definition}[CR Code, Error Probability and SS Metric]\label{DEF:CR_code}
A CR $(n,M_n,K_n)$-code $\mathsf{C}_n$ for the AVWTC $(\mathfrak{W}^n,\mathfrak{V}^n)$ is given by a family of uncorrelated $(n,M_n)$-codes $\mathcal{C}_n=\big\{c_n(\gamma)\big\}_{\gamma\in\Gamma_n}$, where $\Gamma_n=[1:K_n]$, and a PMF $\mu_n\in\mathcal{P}(\Gamma_n)$. For any $m\in\mathcal{M}_n$ and $\mathbf{s}\in\mathcal{S}^n$, the associated error probability with respect to $\mathsf{C}_n$ is
\begin{equation}
\mathcal{E}_m(W_\mathbf{s}^n,\mathsf{C}_n)=\sum_{\gamma\in\Gamma_n}\mu_n(\gamma)e_m\big(W_\mathbf{s}^n,c_n(\gamma)\big)\label{EQ:error_message_state_CR_codes}
\end{equation}
The maximal error probability and SS-metric of $\mathsf{C}_n$ for the $\mathcal{Q}$-constrained AVWTC $(\mathfrak{W}^n,\mathfrak{V}^n,\mathcal{Q})$ are defined as
\begin{subequations}
\begin{align}
\mathcal{E}(\mathfrak{W}^n,\mathcal{Q},\mathsf{C}_n)&=\max_{\substack{\mathbf{s}\in\mathcal{S}_\mathcal{Q}^n,\\m\in\mathcal{M}_n}}\mathcal{E}_m(W_\mathbf{s}^n,\mathsf{C}_n)\label{EQ:reliability_CR_codes}\\
\mathcal{L}_\mathrm{Sem}(\mathfrak{V}^n,\mathcal{Q},\mathsf{C}_n)&=\max_{\gamma\in\Gamma_n}\ell_\mathrm{Sem}\big(\mathfrak{V}^n,\mathcal{Q},c_n(\gamma)\big)\nonumber\\
&=\max_{\substack{\gamma\in\Gamma_n,\\\mathbf{s}\in\mathcal{S}_\mathcal{Q}^n,\\P_M\in\mathcal{P}(\mathcal{M}_n)}}\ell\big (V_\mathbf{s}^n,P_M,c_n(\gamma)\big).\label{EQ:security_CR_codes}
\end{align}\label{EQ:reliability_security_CR_codes}%
\end{subequations}
\end{definition}

\begin{remark}
The choice of encoder-decoder in a CR code is based on a realization of a random experiment that is available to the transmitted and the legitimate receiver. However, this CR the legitimate users share should not be viewed as a cryptographic key to be exploited for secrecy. This is accounted for in \eqref{EQ:security_CR_codes} by requiring that every uncorrelated code in the family $\mathcal{C}_n$ is semantically-secure. The choice of the state sequence, on the other hand, may depend on the family $\mathcal{C}_n$ and the PMF $\mu_n$, but not on the realization itself.
\end{remark}

\begin{definition}[CR-Assisted Achievability]\label{DEF:CR_achievability}
A number $R\in\mathbb{R}_+$ is called an achievable CR-assisted SS-rate for the $\mathcal{Q}$-constrained AVWTC $(\mathfrak{W}^n,\mathfrak{V}^n,\mathcal{Q})$, if for every $\epsilon>0$ and sufficiently large $n$, there exists a CR $(n,M_n,K_n)$-code $\mathsf{C}_n$ with
\begin{subequations}
\begin{align}
\frac{1}{n}\log M_n&> R-\epsilon\label{EQ:CR_achievability_rate}\\
\mathcal{E}(\mathfrak{W}^n,\mathcal{Q},\mathsf{C}_n)&\leq\epsilon\label{EQ:CR_achievability_reliability}\\
\mathcal{L}_\mathrm{Sem}(\mathfrak{V}^n,\mathcal{Q},\mathsf{C}_n)&\leq\epsilon.\label{EQ:CR_achievability_security}
\end{align}\label{EQ:CR_achievability}%
\end{subequations}
\end{definition}

\begin{remark}\label{REM:no_rational_PMF}
Note that if there are no types in $\mathcal{Q}$ then any rate is achievable. Consequently, if  $\mathcal{Q}_1\subseteq\mathcal{Q}_2\subseteq\mathcal{P}(\mathcal{S})$, then any $R$ that is achievable for the $\mathcal{Q}_2$-constrained AVWTC is also achievable for the $\mathcal{Q}_1$-constrained AVWTC. The achievable rates are therefore an increasing set as the constraint set decreases. When specializing to the type constrained AVWTC, we allow state sequences with types that are $\delta$-close to the constraining distribution (see Definition \ref{DEF:CR_achievability_type}). Consequently, the set of feasible state sequences is never empty, for large enough values of $n$. Nonetheless, the aforementioned monotonicity of the type constrained CR-assisted capacity still holds.

\end{remark}

\begin{remark}
Our achievability proof shows that $\mathcal{L}_\mathrm{Sem}(\mathfrak{V}^n,\mathcal{Q},\mathsf{C}_n)$ vanishes exponentially fast. This is a standard requirement in cryptography, commonly referred to as strong-SS (see, e.g., \cite[Section 3.2]{Vardy_Semantic_WTC2012}).
\end{remark}

\begin{definition}[CR-Assisted Capacity]\label{DEF:CR_capacity}
The CR-assisted SS-capacity $C_{\mathrm{R}}(\mathfrak{W},\mathfrak{V},\mathcal{Q})$ of the $\mathcal{Q}$-constrained AVWTC is the supremum of the set of achievable CR-assisted SS-rates.
\end{definition}


Our main goal is solving the type constrained AVWTC $(\mathfrak{W}^n,\mathfrak{V}^n,Q_S)$, for $Q_S\in\mathcal{P}_n(\mathcal{S})$. However, since a fixed rational distribution $Q_S$ is not a valid type for all blocklengths, the definitions of the type constrained performance metrics and its achievability use a relaxation parameter. For any $Q_S\in\mathcal{P}(\mathcal{S})$ and $\delta>0$, let $\mathcal{Q}_\delta(Q_S)\triangleq\Big\{\nu_\mathbf{s}\in\mathcal{P}_n(\mathcal{S})\Big| \mathbf{s}\in\mathcal{T}_\delta^n(Q_S)\Big\}$. The definitions of the error probability and the SS-metric for the type constrained AVWTC repeat those from Definition \ref{DEF:CR_code} with $\mathcal{Q}_\delta(Q_S)$ instead of $\mathcal{Q}$.

\begin{definition}[Type Constrained Error Probability \& SS]\label{DEF:error_prob_SS_metric_type}
For any $Q_S\in\mathcal{P}(\mathcal{S})$ and $\delta>0$, the maximal error probability and SS-metric of a CR $(n,M_n,K_n)$-code $\mathsf{C}_n$ for the type constrained AVWTC $(\mathfrak{W}^n,\mathfrak{V}^n,Q_S)$ with relaxation $\delta$ are $\mathcal{E}\big(\mathfrak{W}^n,\mathcal{Q}_\delta(Q_S),\mathsf{C}_n\big)$ and $\mathcal{L}_\mathrm{Sem}\big(\mathfrak{V}^n,\mathcal{Q}_\delta(Q_S),\mathsf{C}_n\big)$, respectively (see \eqref{EQ:reliability_security_CR_codes}).
\end{definition}



\begin{definition}[Type Constrained Achievability \& Capacity]\label{DEF:CR_achievability_type}
A number $R\in\mathbb{R}_+$ is called an achievable CR-assisted SS-rate for the type constrained AVWTC $(\mathfrak{W}^n,\mathfrak{V}^n,Q_S)$, if there exists a $\delta>0$ such that for every $\epsilon>0$ and sufficiently large $n$, there exists a CR $(n,M_n,K_n)$-code $\mathsf{C}_n$ with 
\begin{subequations}
\begin{align}
\frac{1}{n}\log M_n&> R-\epsilon\label{EQ:CR_achievability_rate_type}\\
\mathcal{E}\big(\mathfrak{W}^n,\mathcal{Q}_\delta(Q_S),\mathsf{C}_n\big)&\leq\epsilon\label{EQ:CR_achievability_reliability_type}\\
\mathcal{L}_\mathrm{Sem}\big(\mathfrak{V}^n,\mathcal{Q}_\delta(Q_S),\mathsf{C}_n\big)&\leq\epsilon.\label{EQ:CR_achievability_security_type}
\end{align}\label{EQ:CR_achievability_type}%
\end{subequations}
The CR-assisted SS-capacity $C_{\mathrm{R}}(\mathfrak{W},\mathfrak{V},Q_S)$ of the type constrained AVWTC $(\mathfrak{W}^n,\mathfrak{V}^n,Q_S)$ is the supremum of the set of achievable CR-assisted SS-rates.
\end{definition}



\begin{remark}
The definition of type constrained achievability allows the empirical PMFs of the state sequences to be within a $\delta>0$ gap from $Q_S$. By doing so, the type constrained AVWTC is well-defined for all sufficiently large blocklengths $n\in\mathbb{N}$, even when $Q_S$ is not actually a type but a PMF on $\mathcal{S}$.
\end{remark}

\begin{remark}\label{REM:smaller_delta_AVWTC_achievability}
If $R$ is an achievable CR-assisted SS-rate for the type constrained AVWTC $(\mathfrak{W}^n,\mathfrak{V}^n,Q_S)$ and $\delta>0$ is its corresponding parameter, then \eqref{EQ:CR_achievability_reliability_type}-\eqref{EQ:CR_achievability_security_type} also hold for any $\delta'\in(0,\delta)$. This implies that $C_{\mathrm{R}}(\mathfrak{W},\mathfrak{V},Q_S)=\sup\limits_{\delta>0}C_{\mathrm{R}}\big(\mathfrak{W},\mathfrak{V},\mathcal{Q}_\delta(Q_S)\big)$.
\end{remark}


\subsection{Single-Letter CR-Capacity Results}\label{SUBSEC:AVWTC_result}

Our main result is a single-letter characterization of the CR-assisted SS-capacity $C_{\mathrm{R}}(\mathfrak{W},\mathfrak{V},Q_S)$ of the type constrained AVWTC $(\mathfrak{W}^n,\mathfrak{V}^n,Q_S)$, for any $Q_S\in\mathcal{P}(\mathcal{S})$. A multi-letter characterization of the CR-assisted strong secrecy-capacity of the AVWTC without constraints on the state sequences was found in \cite{Boche_AVWTC_multiletter_capacity2015}. The uncorrelated secrecy-capacity was then derived in \cite{Boche_superactive_AVWTC2016} by relating it to the CR-assisted secrecy-capacity via the corresponding coding result for the classic AVC \cite{Csiszar_Narayan_det_AVC1988}. To the best of our knowledge, the only single-letter characterization of a secrecy-capacity of an AVWTC outside the current work \cite[Theorem 4]{MolavianJazi_AVWTC_thesis2009} is under the following assumptions: (i) security under the weak secrecy metric (as shown in \cite[Corollary 1]{Boche_AVWTC_multiletter_capacity2015} an upgrade to strong secrecy under the same conditions (ii)-(iv) is possible); (ii) the state space decomposes as $\mathcal{S}=\mathcal{S}_y\times\mathcal{S}_z$, where $s_y\in\mathcal{S}_y$ and $s_z\in\mathcal{S}_z$ are the states of the main AVC and of the AVC to the eavesdropper, respectively; (iii) the eavesdroppers output is a degraded version of the output of the main AVC under any pair of state, i.e., $X-Y_{s_y}-Z_{s_z}$ forms a Markov chain, for all $(s_y,s_z)\in\mathcal{S}_y\times\mathcal{S}_z$; (iv) there exists a best channel to the eavesdropper, i.e., the exists $s_z^\star\in\mathcal{S}_z$ such that $X-Z_{s^\star_z}-Z_{s_z}$ forms a Markov chain, for all $s_z\in\mathcal{S}_z$.\footnote{An even stronger version of assumptions (iii) and (iv) was used in \cite{MolavianJazi_AVWTC_thesis2009}. Specifically, the degraded property and the existence of a best channel to the eavesdropper were assumed to hold not only for every pair of original states, but also for any pair of averaged states (defined as convex combinations of the original ones).}

Our single-letter CR-capacity characterization is derived without assuming any of the above, while upgrading the secrecy metric to SS. Constrained state sequences were considered in the context of the classic point-to-point (PTP) AVC and the corresponding CR-assisted and uncorrelated capacities were derived in \cite{Csiszar_AVWTC_constrained_random1988} and \cite{Csiszar_AVWTC_constrained_det1988}, respectively. An AVWTC with linear peak constraints on the input and the state sequences was studied in \cite{Boche_constrained_AVWTC2015}, where a multi-letter description of its CR-assisted strong secrecy-capacity was found.

\begin{theorem}[AVWTC CR-Assisted SS-Capacity]\label{TM:AVWTC_CR_capacity}
For any $Q_S\in\mathcal{P}(\mathcal{S})$, the CR-assisted SS-capacity of the type constrained AVWTC $(\mathfrak{W}^n,\mathfrak{V}^n,Q_S)$ is
\begin{equation}
C_{\mathrm{R}}(\mathfrak{W},\mathfrak{V},Q_S)=\max_{Q_{U,X}}\Big[I(U;Y)-I(U;Z|S)\Big],\label{EQ:AVWTC_CR_capacity}
\end{equation}
where the mutual information terms are calculated with respect to a joint PMF $Q_{U,X}Q_SQ_{Y|X,S}Q_{Z|X,S}$ with $Q_{Y|X,S}(y|x,s)=W_s(y|x)$ and $Q_{Z|X,S}(z|x,s)=V_s(z|x)$, for all $(s,x,y,z)\in\mathcal{S}\times\mathcal{X}\times\mathcal{Y}\times\mathcal{Z}$, and $|\mathcal{U}|\leq|\mathcal{X}|$.
\end{theorem}

Theorem~\ref{TM:AVWTC_CR_capacity} is a consequence of two other stronger results that state a lower and upper bound on the CR-assisted SS-capacity of a general $\mathcal{Q}$-constrained AVWTC. These bounds match when specialized to the type constrained scenario. The lower and upper bounds are given in Theorem \ref{TM:AVWTC_CR_achievability_general} and \ref{TM:AVWTC_CR_converse_general}, respectively. Theorem \ref{TM:AVWTC_CR_capacity} is proven in Section \ref{SUBSEC:AVWTC_CR_capacity_proof}.

\begin{remark}[SS-Capacity Interpretation]\label{REM:CR_capacity_interpretation}
The characterization of the CR-assisted SS-capacity $C_{\mathrm{R}}(\mathfrak{W},\mathfrak{V},Q_S)$ in \eqref{EQ:AVWTC_CR_capacity} has the common structure of two subtracted mutual information terms. The first term, which corresponds to the capacity of the main channel, suggests that the legitimate users effectively see the averaged DMC $W_Q:\mathcal{X}\to\mathcal{P}(\mathcal{Y})$ defined by $W_Q(y|x)\triangleq\sum_{s\in\mathcal{S}}Q_S(s)W_s(y|x)$. In general, the capacity of the averaged channel is no larger than the capacities of the main channels $W_s$ associated with each $s\in\mathcal{S}$. Namely, denoting the capacity of a PTP channel $W:\mathcal{X}\to\mathcal{P}(\mathcal{Y})$ by $C(W)$, it holds that $C(W_Q)\leq\min_{s\in\mathcal{S}}C(W_s)$. This is due to the convexity of the mutual information in the conditional PMF (for a fixed marginal) and Jensen's inequality.

The second (subtracted) mutual information term is the loss in capacity induced by the secrecy requirement. The independence of $U$ and $S$ allows one to rewrite the conditional mutual information as $I(U;S,Z)$, which implies that security must be ensured versus an eavesdropper with perfect CSI. 
\end{remark}

\begin{remark}[Relation to the IID State Scenario]
The formula in \eqref{EQ:AVWTC_CR_capacity} can be viewed as the secrecy-capacity of the WTC with state variables that are i.i.d. according to $Q_S$, when no CSI is available to the legitimate users while the eavesdropper has full CSI. For simplicity, we outline a proof of this claim under the strong secrecy metric; an upgrade to SS is possible by means of the Lemma \ref{LEMMA:soft_covering} herein. The direct part follows by constructing a classic WTC code using i.i.d. samples of a random variable $U\sim Q_U$ that is independent of $S\sim Q_S$. Setting the total number of codewords just below $2^{nI(U;Y)}$ ensures reliable decoding, while strong secrecy follows by standard soft-covering arguments as long as each subcodebook has a rate that is at least $I(U;Z,S)$ (see, \cite[Corollay VII.5]{Cuff_Synthesis2013}). The converse essentially follows by repeating the steps between Equations \eqref{EQ:UB_summation}-\eqref{EQ:type_converse} from Section \ref{SEC:AVWTC_CR_converse_general_proof}, while omitting the conditioning on the random variable $C_n$ in \eqref{EQ:UB_summation}.
\end{remark}

\begin{remark}[Cardinality Bound]
The cardinality bound on $\mathcal{U}$ in Theorem \ref{TM:AVWTC_CR_capacity} is established using a standard application of the Eggleston-Fenchel-Carath{\'e}odory Theorem \cite[Theorem 18]{Eggleston_Convexity1958}. The details are omitted.
\end{remark}

We have the following lower bound on the CR-assisted SS-capacity of a $\mathcal{Q}$-constrained AVWTC.

\begin{theorem}[Achievability with $\mathcal{Q}$-constrained States]\label{TM:AVWTC_CR_achievability_general}
For any convex and closed $\mathcal{Q}\subseteq\mathcal{P}(\mathcal{S})$, the CR-assisted SS-capacity of the $\mathcal{Q}$-constrained AVWTC $(\mathfrak{W}^n,\mathfrak{V}^n,\mathcal{Q})$ is lower bounded as
\begin{equation}
C_{\mathrm{R}}(\mathfrak{W},\mathfrak{V},\mathcal{Q})\geq\max_{Q_{U,X}}\left[\min_{Q^{(1)}_S\in\mathcal{Q}}I(U;Y)\mspace{-2mu}-\mspace{-2mu}\max_{Q^{(2)}_S\in\mathcal{Q}}I(U;Z|S)\right]\mspace{-3mu},\label{EQ:AVWTC_CR_unconstrained_achievability}
\end{equation}
where the mutual information terms are calculated with respect to joint PMFs $Q_{U,X}Q^{(j)}_SQ_{Y|X,S}Q_{Z|X,S}$, for $j=1,2$, with $Q_{Y|X,S}(y|x,s)=W_s(y|x)$ and $Q_{Z|X,S}(z|x,s)=V_s(z|x)$, for all $(s,x,y,z)\in\mathcal{S}\times\mathcal{X}\times\mathcal{Y}\times\mathcal{Z}$, and $|\mathcal{U}|\leq|\mathcal{X}|$.
\end{theorem}

The proof of Theorem \ref{TM:AVWTC_CR_achievability_general} is given in Section \ref{SEC:AVWTC_CR_achievability_general_proof}. It relies on the approach from \cite{Csiszar_AVWTC_constrained_random1988} for the error probability analysis of a CR-code over a family of codes that grows doubly-exponentially with the blocklength. Since this family of codes is too large to establish SS in the sense of \eqref{EQ:security_CR_codes}, we use the Chernoff bound to show that a sub-family with no more than polynomially many codes is sufficient for reliability. Having that, the double-exponential decay that Lemma~\ref{LEMMA:soft_covering} provides is leveraged to establish SS over the reduced CR-code. The fact that reliability and security must hold with respect to the worst case choice in $\mathcal{Q}$ is expressed in the minimization of $I(U;Y)$ over all $Q_S^{(1)}$ PMFs and the maximization of $I(U;Z|S)$ over $Q_S^{(2)}$.

Although no converse proof accompanies Theorem~\ref{TM:AVWTC_CR_achievability_general}, the lower bound it states is stronger than existing single-letter achievability results in the literature as it assumes no `best channel to the eavesdropper', doesn't impose any specific structure on the state space, and ensures SS.


\begin{remark}[AVWTC Main Challenges] The difficulty in obtaining single-letter results for the AVWTC is twofold. First, the AVWTC must (in particular) satisfy all the performance requirements of the compound WTC (where the channel remains constant throughout the block transmission). The second difficulty is in ensuring security under exponentially many possible state sequences. Setting sights on single-letter results, the common workaround for the latter problem is to assume the existence of `a best channel to the eavesdropper'. Formally, it means that there exists a PMF $Q^\star\in\mathcal{P}(\mathcal{S})$, such that the averaged channel $\sum_{s\in\mathcal{S}}Q^\star(s)V_s(y|x)$ is better than all other averaged channels in the sense that the corresponding outputs form a Markov chain with the channel input $X$. Secrecy is then guaranteed with respect to this `best channel' only. As Theorem \ref{TM:AVWTC_CR_achievability_general} is derived without any assumptions on the AVWTC, it highlights the strength of Lemma \ref{LEMMA:soft_covering} in proving that exponentially many secrecy constraints (strongly related to the soft-covering phenomenon) are simultaneously satisfied, while only single-letter rate bounds are needed.
\end{remark}

\begin{remark}[Relation to Compound WTCs]\label{R:AVWTC and CWTC}
Theorem~\ref{TM:AVWTC_CR_achievability_general} establishes that the AVWTC is no worse than the best known single-letter secrecy rates for the compound WTC. Take the $\mathcal{Q}$-constrained AVWTC from Theorem~\ref{TM:AVWTC_CR_achievability_general} with some convex and closed $\mathcal{Q}\subseteq\mathcal{P}(\mathcal{S})$. Consider a compound WTC derived from this AVWTC. The state of the compound WTC is any point $Q_S \in \mathcal{Q}$. The compound WTC itself follows the probability law of the AVWTC, with the arbitrarily varying state $S^n$ replaced by an i.i.d. state according to $Q_S$ and the $S^n$ sequence included in the channel output to the eavesdropper. For this compound WTC, the RHS of \eqref{EQ:AVWTC_CR_unconstrained_achievability} coincides with the sharpest single-letter lower bound on the secrecy-capacity of the compound WTC in the literature (see \cite[Theorem 1]{Liang_compoundWTC_Journal2007} and \cite[Theorem 3.6]{Boche_compoundWTC2013}).
\end{remark}


A general upper bound on the CR-assisted SS-capacity of a $\mathcal{Q}$-constrained AVWTC is given next. To state the result, for any countable alphabet $\mathcal{X}$ we defined $\mathcal{P}_\mathbb{Q}(\mathcal{X})$ as the set of rational PMFs on $\mathcal{X}$. Namely,
\begin{equation}
\mathcal{P}_\mathbb{Q}(\mathcal{X})\triangleq\Big\{P\in\mathcal{P}(\mathcal{X})\Big|P(x)\in\mathbb{Q},\quad\forall x\in\mathcal{X}\Big\}.\label{EQ:rational_PMFs_set}
\end{equation}

\begin{theorem}[Upper Bound with $\mathcal{Q}$-constrained States]\label{TM:AVWTC_CR_converse_general}
For any $\mathcal{Q}\subseteq\mathcal{P}(\mathcal{S})$, the CR-assisted SS-capacity of the $\mathcal{Q}$-constrained AVWTC $(\mathfrak{W}^n,\mathfrak{V}^n,\mathcal{Q})$ is upper bounded as
\begin{align*}
C&_{\mathrm{R}}(\mathfrak{W},\mathfrak{V},\mathcal{Q})\\
               &\leq\max_{Q_{V,U,X}}\inf_{Q_S\in\mathcal{Q}\cap\mathcal{P}_\mathbb{Q}(\mathcal{S})}\Big[I(U;Y|V)-I(U;S,Z|V)\Big],\numberthis\label{EQ:AVWTC_CR_unconstrained_UB}
\end{align*}
where the mutual information terms are calculated with respect to a joint PMF $Q_{V,U,X}Q_SQ_{Y|X,S}Q_{Z|X,S}$ with $Q_{Y|X,S}(y|x,s)=W_s(y|x)$ and $Q_{Z|X,S}(z|x,s)=V_s(z|x)$, for all $(s,x,y,z)\in\mathcal{S}\times\mathcal{X}\times\mathcal{Y}\times\mathcal{Z}$. Furthermore, one may restrict $|\mathcal{U}|\leq|\mathcal{X}|$ and $|\mathcal{V}|\leq |\mathcal{X}|^2-1$.
\end{theorem}

The proof of Theorem \ref{TM:AVWTC_CR_converse_general} is given in Section \ref{SEC:AVWTC_CR_converse_general_proof}. The max-inf structure of the RHS of \eqref{EQ:AVWTC_CR_unconstrained_UB} calls for a derivation that is uniform in $Q_S\in\mathcal{Q}$. The infimum is taken over $\mathcal{Q}\cap\mathcal{P}_\mathbb{Q}(\mathcal{S})$ (rather than over $\mathcal{Q}$) because the proof effectively considers only the rational distributions in $\mathcal{Q}$ while leveraging the monotonicity of the CR-assisted SS-capacity with respect to $\mathcal{Q}$ (see Remark \ref{REM:no_rational_PMF}). The proof relies on a novel argument based on distribution coupling. We show that for each $Q_S\in\mathcal{Q}\cap\mathcal{P}_\mathbb{Q}(\mathcal{S})$, reliability and SS under a type constraint $Q_S$ imply similar performance for the same channel but where the state sequence is i.i.d. according to $Q_S$. The main difficulty is in showing that even when transmitting over a DMC obtained by averaging the $W_s\in\mathfrak{W}$ with respect to $Q_S$, the normalized equivocation of the message given the output sequence at the \emph{legitimate user} is still small.

Usually, Fano's inequality is sufficient for $\frac{1}{n}H(M|Y^n)$ to become arbitrarily small with $n$. This, however, is not the case here. The reliability criterion from \eqref{EQ:CR_achievability_reliability} and Fano's inequality imply that $\max_{\mathbf{s}\in\mathcal{S}^n_\mathcal{Q}}\frac{1}{n}H(M|Y^n_\mathbf{s})$ is small, but Theorem \ref{TM:AVWTC_CR_converse_general} needs this to hold for $\frac{1}{n}H(M|Y^n)$. In general, the former must not imply the latter because for any $\mathbf{s}\in\mathcal{S}^n_\mathcal{Q}$, the channel $W_\mathbf{s}\in\mathfrak{W}$ is at least as good as $W_Q$, meaning that the averaged channel induces a possibly larger equivocation. We establish the desired convergence of the equivocation at the legitimate receiver by a continuity property which we derive via the coupling idea.

\begin{remark}[Relation to the Converse of Theorem \ref{TM:AVWTC_CR_capacity}]
The converse for Theorem \ref{TM:AVWTC_CR_capacity} is derived based on Theorem \ref{TM:AVWTC_CR_converse_general}. Since the latter is valid for any $\mathcal{Q}\subset\mathcal{P}(\mathcal{S})$, combining it with basic continuity arguments implies the optimality of the RHS of \eqref{EQ:AVWTC_CR_capacity}. However, Theorem \ref{TM:AVWTC_CR_converse_general} encapsulates an even stronger claim: the RHS of \eqref{EQ:AVWTC_CR_capacity} is the best achievable CR-assisted SS-rate even if the state sequence is constrained to a single type (that potentially might allows higher rates), rather than a vanishing typical set.

Although clearly sufficient, the strong claim of Theorem \ref{TM:AVWTC_CR_converse_general} is not necessary for the converse of Theorem \ref{TM:AVWTC_CR_capacity}. In fact, one can circumvent the main difficulty in proving Theorem \ref{TM:AVWTC_CR_converse_general} (as described above), by establishing the optimality of the RHS of \eqref{EQ:AVWTC_CR_capacity} based on arguments similar to those used for the converse of the classic AVC with constrained states \cite[Lemma 3.2]{Csiszar_Narayan_CR_AVC1988}. Specifically, the probability of a decoding error under a random state sequence that is i.i.d. according to $Q_S$ can be shown to be small simply by splitting the analysis to typical and atypical state sequences. The definition of the type constrained CR-assisted achievability (Definition \ref{DEF:CR_achievability_type}), that accounts for a typical set around $Q_S$, takes care of the typical part, while the atypical part is bounded above by the exponentially decaying probability of the atypical set. Having that, a standard application of Fano's inequality implies that $\frac{1}{n}H(M|Y^n)$ converges to 0 with the blocklength, as required. 

This simple argument, however, does not apply when there is an actual type constraint (rather than a typical set constraint) on the state sequences. This is because for any $Q_S\in\mathcal{P}_n(\mathcal{S})$, the probability of an i.i.d. $Q_S$ sequence of states not being in $\mathcal{T}_{Q_S}^n$ approaches 1 when $n$ grows. As a consequence, the corresponding decoding error probability might not be small. The proof of Theorem \ref{TM:AVWTC_CR_converse_general} does not directly deals with the error probability. Instead, the aforementioned distribution coupling arguments and continuity of entropy are used to show that the normalized equivocation converges for state sequences in the entire typical set, as long as it converges for at least one specific type in that typical set. Although the proof Theorem \ref{TM:AVWTC_CR_converse_general} is cumbersome and requires several non-trivial steps, we take this route (rather than a converse tailored for Theorem \ref{TM:AVWTC_CR_capacity}) due to the stronger and insightful claim it produces.
\end{remark}

\begin{remark}[Time-Sharing Random Variable $\bm{V}$] The conditioning on $V$ in the RHS of \eqref{EQ:AVWTC_CR_unconstrained_UB} effectively allows the legitimate user to choose a random mixture of $Q_{U,X}$ distributions. The advantage in doing so is that there might not exist a single state distribution  that is bad for the whole mixture. This is reminiscent of a two-player zero-sum game, where the player who fixes the strategy first often benefits from a mixed strategy. When specializing to the type constrained scenario, however, the time-sharing random variable is removed. This is since when only one state distribution is allowed, the aforementioned distribution mixing outcomes with no gain. 
\end{remark}

\begin{remark}[Relation to Compound WTCs]\label{REM:AVWTC_CWTC_UB}
The best previously known single-letter upper bound on the secrecy-capacity of the compound WTC is due to Liang \emph{et al.} \cite[Theorem~2]{Liang_compoundWTC_Journal2007}. That upper bound has a min-max structure, and it is derived by claiming that the secrecy-capacity of the compound WTC is bounded above by that of the worst WTC in the set. This type of bounds are commonly related to knowledge of the channel's state at the transmitter (cf. e.g., \cite{Permuter_compound2009}). Indeed, as shown in \cite{Boche_compoundWTC2013}, the upper bound from \cite{Liang_compoundWTC_Journal2007} is tight for the compound WTC with encoder CSI.

Specializing the max-inf upper bound from Theorem \ref{TM:AVWTC_CR_converse_general} to the compound WTC described in Remark \ref{R:AVWTC and CWTC} (i.e., over an appropriate constraint set), results in a strengthening of the claim from \cite[Theorem~2]{Liang_compoundWTC_Journal2007}. The obtained bound first minimizes the difference of mutual information terms from the RHS of \eqref{EQ:AVWTC_CR_unconstrained_UB} over the constraint set, and then maximizes the outcome over the input distribution. It is easily observed the difference between the two bounds can be strict. In fact, for the special case of a PTP compound channel (i.e., without an eavesdropper) our bound is the actual capacity, while the bound from \cite{Liang_compoundWTC_Journal2007} is loose. A simple example is a channel that consists of two orthogonal binary channels: one is noise free while the other one is purely noise (i.e., binary symmetric channel with crossover probability $\frac{1}{2}$). The state determines which channel is noisy, and the transmitter selects a binary input, which is unknown to the receiver, specifying which channel to use (both channels give an output each time, with one being pure noise). For this instance, the compound capacity is $\frac{1}{2}$ [bit/use], but the looser min-max bound gives 1 [bit/use].

Theorem \ref{TM:AVWTC_CR_converse_general} essentially says that the $\mathcal{Q}$-constrained AVWTC is no better than the compound WTC under the corresponding constraints. Although this point seems intuitively obvious, it actually requires some careful attention. At first glance, the compound channel (on a constraint set) seems like an AVWTC with a stricter restriction on the eavesdropper, that now must choose an i.i.d. state sequence from the constraint set (rather than an arbitrary one). However, this perspective is misleading since the i.i.d. state sequence might not fall within the type constraint, especially when dealing with a single type.
\end{remark}

\begin{remark}[Cardinality Bound]
The cardinality bound on $\mathcal{U}$ and $\mathcal{V}$ in Theorem \ref{TM:AVWTC_CR_converse_general} follow by applying the 
Eggleston-Fenchel-Carath{\'e}odory Theorem \cite[Theorem 18]{Eggleston_Convexity1958} twice. The details are omitted.
\end{remark}

A simple consequence of Theorem \ref{TM:AVWTC_CR_converse_general} is the following.

\begin{corollary}[Upper Bound when $\bm{\mathcal{Q}}$ is Open]\label{COR:AVWTC_CR_converse_openset} If $\mathcal{Q}\subseteq\mathcal{P}(\mathcal{S})$ is an open set, then the CR-assisted SS-capacity of the $\mathcal{Q}$-constrained AVWTC $(\mathfrak{W}^n,\mathfrak{V}^n,\mathcal{Q})$ is upper bounded as
\begin{equation}
C_{\mathrm{R}}(\mathfrak{W},\mathfrak{V},\mathcal{Q})\leq\max_{Q_{V,U,X}}\inf_{Q_S\in\mathcal{Q}}\Big[I(U;Y|V)-I(U;Z|S,V)\Big],\label{EQ:AVWTC_CR_unconstrained_UB_CORR}
\end{equation}
where joint PMF is as described in Theorem \ref{TM:AVWTC_CR_converse_general}.
\end{corollary}

When $\mathcal{Q}$ is an open set, the domain of the infimum requires no intersection with $\mathcal{P}_\mathbb{Q}(\mathcal{S})$. This essentially follows because the rational numbers are dense in the reals and the mutual information is continuous in the underlying distribution. The full details are omitted. 

\subsection{An Example}

We give a simple numerical example that visualizes the SS-capacity result of Theorem \ref{TM:AVWTC_CR_capacity}. Let $\mathcal{X}=\mathcal{Y}=\{0,1\}$ and $\mathcal{Z}=\{0,1,?\}$, where $?$ is an erased symbol. Further assume that the state space $\mathcal{S}$ decomposes as $\mathcal{S}=\mathcal{S}_1\times\mathcal{S}_2$, where $\mathcal{S}_j=\{0,1\}$, for $j=1,2$. Let $(\mathfrak{W},\mathfrak{V})$ be an AVWTC, where the elements of $\mathfrak{W}$ and $\mathfrak{V}$ are indexed by $s_1\in\mathcal{S}_1$ and $s_2\in\mathcal{S}_2$, respectively. Define the main channel $W_{s_1}:\mathcal{X}\to\mathcal{P}(\mathcal{Y})$, for $s_1\in\mathcal{S}_1$, as $W_{s_1}(y|x)=\mathds{1}_{\{y=x\oplus s_1\}}$, where $\oplus$ denotes the modulo 2 addition. For the eavesdropper, let $V_0:\mathcal{X}\to\mathcal{P}(\mathcal{Z})$ be a noiseless channel, while $V_1:\mathcal{X}\to\mathcal{P}(\mathcal{Z})$ outputs the symbol $?$ with probability 1. Namely,
\begin{equation}
V_{s_2}(z|x)=\begin{cases}
\mathds{1}_{\{z=x\}},\quad s_2=0\\
\mathds{1}_{\{z=?\}},\quad s_2=1
\end{cases}.
\end{equation}
Finally, we introduce a type constraint $Q_{S_1,S_2}=Q_{S_1}Q_{S_2}$ on the state sequences, where $Q_{S_1}(1)=\epsilon$ and $Q_{S_2}(1)=\alpha$, for some $\epsilon\in\left[0,\frac{1}{2}\right]$ and $\alpha\in[0,1]$. Denote the CR-assisted SS-capacity of this AVWTC by $C_\mathrm{R}(\epsilon,\alpha)$.

By Theorem \ref{TM:AVWTC_CR_capacity}, The CR-assisted SS-capacity is
\begin{equation}
C_{\mathrm{R}}(\epsilon,\alpha)=\max_{Q_{U,X}}\Big[I(U;Y)-I(U;Z|S_2)\Big],\label{EQ:CR_capacity_simp1}
\end{equation}
where the mutual information terms are calculated with respect to the joint distribution $Q_{S_1}(s_1)Q_{S_2}(s_2)Q_{U,X}(u,x)W_{s_1}(y|x)V_{s_2}(z|x)$. 

Note that for any $Q_{U,X}\in\mathcal{P}(\mathcal{U}\times\mathcal{X})$, we have
\begin{align*}
I(U;&Z|S_2)\\
&=Q_{S_2}(0)I(U;Z|S_2=0)+Q_{S_2}(1)I(U;Z|S_2=1)\\
 &\stackrel{(a)}=(1-\alpha)I(U;X),\numberthis\label{EQ:CR_capacity_example_eave}
\end{align*}
where (a) is because $Z=?$ whenever $S_2=1$ (thus nullifying the second mutual information term), while given on $S_2=0$, we have $Z=X$ and the conditioning is removed due to the independence of $S_2$ and $(U,X)$. Consequently, \eqref{EQ:CR_capacity_simp1} reduces to
\begin{equation}
C_{\mathrm{R}}(\epsilon,\alpha)=\max_{Q_{U,X}}\Big[I(U;Y)-(1-\alpha)I(U;X)\Big],\label{EQ:CR_capacity_example_simp2}
\end{equation}
which is calculated with respect to $Q_{S_1}(s_1)Q_{U,X}(u,x)W_{s_1}(y|x)$. Now, since $S_1$ does not appear in any of the mutual information terms, their value remains unchanged if the above joint distribution is replaced with $Q_{U,X}(u,x)W_{Q_1}(y|x)$, where $W_{Q_1}:\mathcal{X}\to\mathcal{P}(\mathcal{Y})$ is the average DMC $W_{Q_1}(y|x)=\sum_{s_1\in\mathcal{S}_1}Q_{S_1}(s_1)W_{s_1}(y|x)$ (see Remark \ref{REM:CR_capacity_interpretation}). Noting that the DMC $W_{Q_1}$ is a binary symmetric channel with crossover probability $\epsilon$ ($\mbox{BSC}(\epsilon)$), we have that $C_{\mathrm{R}}(\epsilon,\alpha)$ is the secrecy-capacity of BS-BE WTC with a $\mbox{BSC}(\epsilon)$ between the legitimate users and a binary erasure channel with erasure probability $\alpha$ ($\mbox{BEC}(\alpha)$) to the eavesdropper \cite{Ulukus_BSCBEC_WTC2011}.
\begin{figure*}[!t]
\begin{equation}
R_{\mathrm{S}}^*(\mathfrak{W},\mathfrak{V})\triangleq\lim_{k\to\infty}\frac{1}{k}\sup_{P_{\bar{V}_k,\bar{X}^k}\in\mathcal{P}(\mathcal{V}_k\times\mathcal{X}^k)}\left[\min_{Q\in\mathcal{P}(\mathcal{S})}I_{P^{(Q)}}\big(\bar{V}_k;\bar{Y}^k_Q\big)-\max_{s^k\in\mathcal{S}^k}I_{P^{(Q)}}\big(\bar{V}_k;\bar{Z}_{s^k}^k\big)\right]\label{EQ:multiletter_capacity}    
\end{equation}
\hrulefill
\begin{equation}
P^{(Q)}_{\bar{V}_k,\bar{X}^k,\bar{Y}_Q^k,\bar{Z}_{s^k}^k}(v_k,x^k,y^k,z^k)=P_{\bar{V}_k,\bar{X}^k}\big(v_k,x^k\big)\prod_{i=1}^k\left(\sum_{s\in\mathcal{S}}Q(s)W_s(y_i|x_i)\right)V_{s_i}(z_i|x_i)\label{EQ:multiletter_PMF}
\end{equation}
\hrulefill
\begin{equation}
R_{\mathrm{S}}^*(\mathfrak{W},\mathfrak{V},Q_S)\triangleq\lim_{n\to\infty}\frac{1}{n\ell}\sup_{P_{\bar{V}_{n\ell},\bar{X}^{n\ell}}\in\mathcal{P}(\mathcal{V}_{n\ell}\times\mathcal{X}^{n\ell})}\left[I_P\big(\bar{V}_{n\ell};\bar{Y}^{n\ell}_{Q_S}\big)-\max_{s^{n\ell}\in\mathcal{T}^{n\ell}_{Q_S}}I_P\big(\bar{V}_{n\ell};\bar{Z}_{s^{n\ell}}^{n\ell}\big)\right],\label{EQ:multiletter_capacity_type}
\end{equation}
\hrulefill
\end{figure*}

\begin{remark}
Interestingly, \eqref{EQ:CR_capacity_example_simp2} is also the SS-capacity of the WTC of type II (WTCII) with a $\mbox{BSC}(\epsilon)$ to the legitimate user and an eavesdropper who can actively choose any $\lfloor n(1-\alpha)\rfloor$ of the transmitted symbols to observe.  \cite{Goldfeld_WTCII_semantic2015}. This is not surprising since the WTCII with a noisy main channel is a particular instance of a type constrained AVWTC. Consequently, its SS-capacity (stated in Theorem 3 of \cite{Goldfeld_WTCII_semantic2015}) is recovered from Theorem \ref{TM:AVWTC_CR_capacity} by steps similar to those presented between Equations \eqref{EQ:CR_capacity_simp1}-\eqref{EQ:CR_capacity_example_simp2}. Namely, this is done by letting the AVC between the legitimate users be a DMC and treating the eavesdropper's AVC as in \eqref{EQ:CR_capacity_example_eave}. In fact, the type constrained AVWTC captures as a special case also the generalization of the WTCII from \cite{Yener_WTCII_general2015}, where the subset of symbols chosen by the eavesdropper is further corrupted by noise (i.e., by passing it through another DMC). The only actual difference between these WTCII models and the AVWTC is that the main channel in the formers is a DMC (and not an AVC), which makes CR-assisted codes unnecessary. However, this is the case for any AVWTC with a main DMC. 
\end{remark}


\begin{figure}[t!]
    \begin{center}
        \begin{psfrags}
            \psfragscanon
            \psfrag{A}[][][1]{$\mspace{-43mu}\alpha$}
            \psfrag{B}[][][1]{$C_\mathrm{R}(\epsilon,\alpha)$}
            \psfrag{C}[][][1]{\ \ \ \ \ \ \ \ \ \  $\alpha=4\epsilon(1-\epsilon)$}
            \psfrag{D}[][][1]{$\mspace{-43mu}\epsilon$}
            \psfrag{E}[][][1]{$C_\mathrm{R}(\epsilon,\alpha)$}
            \psfrag{F}[][][1]{\ \ \ \ \ \ \ \ \ $\epsilon_1=\frac{1-\sqrt{1-\alpha}}{2}$}
            \psfrag{G}[][][1]{\ \ \ \ \ \ \ \ \ \ \ \ \ \ $\epsilon_2=\frac{1+\sqrt{1-\alpha}}{2}$}
            \subfloat[]{\includegraphics[scale = .45]{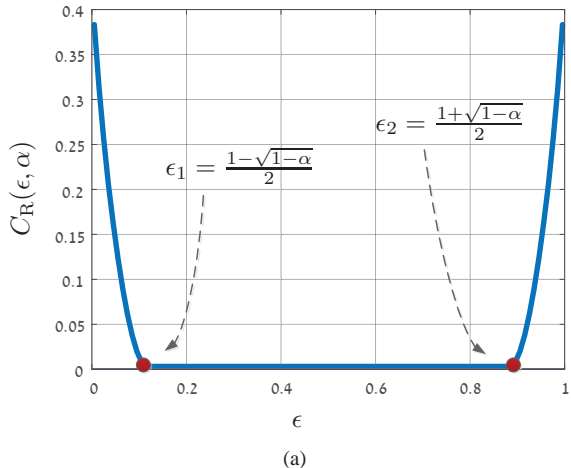}}\\ \subfloat[]{\includegraphics[scale = .45]{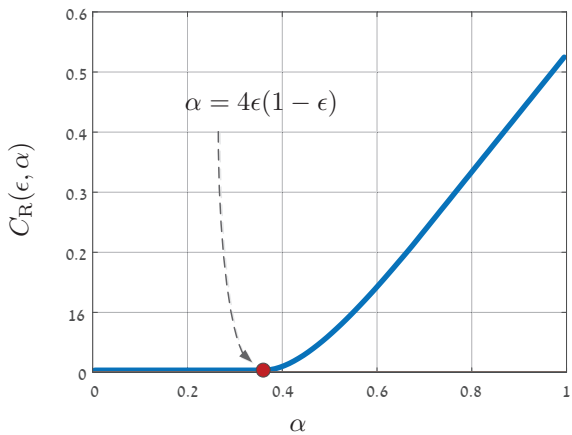}}
            \caption{CR-assisted SS-capacity $C_\mathrm{R}(\epsilon,\alpha)$ versus: (a) the proscribed type for the main channel $Q_{S_1}(1)=\epsilon$, which corresponds to the portion of flipped symbols in the BS-BE WTC; (b) the proscribed type for the eavesdropper's channel $Q_{S_2}(1)=\alpha$, which corresponds to the portion of erasures in the BS-BE WTC.} \label{FIG:BSCBECWTC}
            \psfragscanoff
        \end{psfrags}
     \end{center}
 \end{figure}

Fig. \ref{FIG:BSCBECWTC} depicts the CR-assisted SS-capacity of the considered AVWTC as a function of type constraints on the main and on the eavesdropper's channels. The variation of $C_\mathrm{R}(\epsilon,\alpha)$ as a function of $Q_{S_1}(1)=\epsilon$ for a fixed $\alpha=0.4$ is shown in Fig. \ref{FIG:BSCBECWTC}(a), while Fig. \ref{FIG:BSCBECWTC}(b) presents the SS-capacity as a function of $Q_{S_2}(1)=\alpha$ when $\epsilon=0.1$ is fixed. The curves are plotted by parametrizing the joint PMF of the binary random variables $U$ and $X$ and spanning over the possible probability values. As mentioned before, $C_\mathrm{R}(\epsilon,\alpha)$ is also the secrecy-capacity of a BS-BE WTC.

This WTC was studied in \cite{Ulukus_BSCBEC_WTC2011}, where it was shown that the secrecy-capacity is zero if $\alpha<4\epsilon(1-\epsilon)$. When $\epsilon=0.1$ the threshold value of $\alpha$ is $0.36$. Indeed, Fig. \ref{FIG:BSCBECWTC}(b) reveals that $C_\mathrm{R}(0.1,\alpha)=0$ for any $\alpha<0.36$. Beyond 0.36, the SS-capacity monotonically increases with $\alpha$, since the larger the probability of an erasure, the worse the channel to the eavesdropper is. From the opposite perspective, a fixed $\alpha=0.4$ induces two real solutions to the equation $0.4=4\epsilon(1-\epsilon)$, which are $\epsilon_1\approx 0.1127$ and $\epsilon_2\approx0.8872$. The condition $0.4<4\epsilon(1-\epsilon)$ is then satisfied for any $\epsilon\in(\epsilon_1,\epsilon_2)$, which gives a zero SS-capacity in that region in Fig. \ref{FIG:BSCBECWTC}(b). Also observe that as a function of $\epsilon$, $C_\mathrm{R}(\epsilon,0.4)$ grows as the crossover probability approaches the extreme values of 0 or 1.

\subsection{Relation Between Theorem \ref{TM:AVWTC_CR_capacity} and the Multi-Letter CR-Assisted Secrecy-Capacity Characterization} \label{SUBSUBSEC:multi_letter_comparison}

We compare the single-letter result of Theorem \ref{TM:AVWTC_CR_capacity} to the multi-letter description of the secrecy-capacity as derived in \cite{Boche_AVWTC_multiletter_capacity2015}. Reciting the result of \cite{Boche_AVWTC_multiletter_capacity2015}, the (strong) secrecy-capacity of an AVWTC $(\mathfrak{W},\mathfrak{V})$ with unconstrained states is given in \eqref{EQ:multiletter_capacity} at the top of this page\footnote{We use macrons in the following notation of random variables not because they are multi-dimensions, but to distinguish them from other random variable that have yet to be introduced.}, where the subscript $P$ indicates that the mutual information terms are calculated with respect to a joint distribution that for each $k\in\mathbb{N}$, $Q\in\mathcal{P}(\mathcal{S})$ and $s^k\in\mathcal{S}^k$, is given by \eqref{EQ:multiletter_PMF} and the cardinality of $\mathcal{V}_k$ is bounded above as $|\mathcal{V}_k|\leq|\mathcal{X}|^k$.

\begin{figure*}[!t]
\setcounter{equation}{67}
\begin{equation}
R_{\mathrm{S}}^*(\mathfrak{W},\mathfrak{V},Q_S)=\lim_{\substack{k\to\infty:\\k\in\mathbb{N}_\ell}}\frac{1}{k}\sup_{P_{\bar{V}_k,\bar{X}^k}\in\mathcal{P}(\mathcal{V}_k\times\mathcal{X}^k)}\left[I_{\hat{P}}\big(V_k;Y^k\big)-\max_{s^k\in\mathcal{T}^k_{Q_S}}I_{\hat{P}}\big(V_k;Z^k\big|S^k=s^k\big)\right].\label{EQ:multiletter_capacity_type_alt}
\end{equation}
\hrulefill
\end{figure*}
\setcounter{equation}{61}

To adapt $R_{\mathrm{S}}^*(\mathfrak{W},\mathfrak{V})$ to the type constrained AVWTC $(\mathfrak{W},\mathfrak{V},Q_S)$, for $Q_S\in\mathcal{P}_\mathbb{Q}(\mathcal{S})$ (see \eqref{EQ:rational_PMFs_set}), we first account for the fact that $\mathcal{T}_{Q_S}^k$ is empty for several values of $k\in\mathbb{N}$. Denote the non-zero entries of $Q_S$ by $Q_S(s)=\frac{a_s}{b_s}$, where $s\in\supp(Q_S)$, and define $\ell=\lcm\{b_s\}_{s\in\supp(Q_S)}$, where $\lcm\mathcal{A}$, for $\mathcal{A}\subset\mathbb{N}$ with $|\mathcal{A}|<\infty$, is the \emph{least common multiple} of the elements in $\mathcal{A}$. Denote $\mathbb{N}_\ell \triangleq \big\{n\cdot\ell \big|n\in\mathbb{N}\big\}$ and henceforth consider only blocklengths that belong to $\mathbb{N}_\ell$. Having this, the adaptation of $R_{\mathrm{S}}^*(\mathfrak{W},\mathfrak{V})$ to the AVWTC with a type constraint $Q_S\in\mathcal{P}_\mathbb{Q}(\mathcal{S})$, is given in \eqref{EQ:multiletter_capacity_type} at the top of this page, where each $P_{\bar{V}_{n\ell},\bar{X}^{n\ell}}\in\mathcal{P}(\mathcal{V}_{n\ell}\times\mathcal{X}^{n\ell})$ induces the joint distribution $P\triangleq P^{(Q_S)}_{\bar{V}_{n\ell},\bar{X}^{n\ell},\bar{Y}_{Q_S}^{n\ell},\bar{Z}_{s^{n\ell}}^{n\ell}}$. As a disclaimer, we remark that the we did not directly prove\footnote{By adapting the proof steps from \cite{Boche_AVWTC_multiletter_capacity2015}.} that $R_{\mathrm{S}}^*(\mathfrak{W},\mathfrak{V},Q_S)$ is indeed a multi-letter description of the type constrained AVWTC secrecy-capacity, and we first state it merely as an educated guess. Nonetheless, as the following proposition shows, this is actually the case since $R_{\mathrm{S}}^*(\mathfrak{W},\mathfrak{V},Q_S)$ is equal to the CR-assisted SS-capacity formula from Theorem~\ref{TM:AVWTC_CR_capacity}.
\begin{proposition}[Multi-Letter CR-Assisted SS-Capacity]
For any $Q_S\in\mathcal{P}_\mathbb{Q}(\mathcal{S})$ it holds that
\begin{equation}
R_{\mathrm{S}}^*(\mathfrak{W},\mathfrak{V},Q_S)=C_{\mathrm{R}}(\mathfrak{W},\mathfrak{V},Q_S).
\end{equation}
\end{proposition}

\begin{IEEEproof}
We first show that $R_{\mathrm{S}}^*(\mathfrak{W},\mathfrak{V},Q_S)\geq C_{\mathrm{R}}(\mathfrak{W},\mathfrak{V},Q_S)$. Let $\bar{Q}_{U,X}\in\mathcal{P}(\mathcal{U}\times\mathcal{X})$ be the extremum achieving distributions in $C_\mathrm{R}(\mathfrak{W},\mathfrak{V},Q_S)$, and denote $\bar{Q}\triangleq \bar{Q}_{U,X}Q_SQ_{Y|X,S}Q_{Z|X,S}$, where $Q_{Y|X,S}$ and $Q_{Z|X,S}$ are defined in Theorem \ref{TM:AVWTC_CR_capacity}. For each $n\in\mathbb{N}$ and $s^{n\ell}\in\mathcal{T}_{Q_S}^{n\ell}$, denote $k=n\ell$ and define 
\begin{align*}
\bar{P}&_{\bar{U}^k,\bar{X}^k,\bar{Y}_{Q_S}^k,\bar{Z}_{s^k}^k}(u^k,x^k,y^k,z^k)\\
&\mspace{55mu}\triangleq\prod_{i=1}^k\bar{Q}_{U,X}(u_i,x_i)W_{Q_S}(y_i|x_i)V_{s_i}(z_i|x_i)\numberthis,\label{EQ:multiletter_PMF_product}
\end{align*}
where $W_{Q_S}:\mathcal{X}\to\mathcal{P}(\mathcal{Y})$ is given by $W_{Q_S}(y|x)=\sum_{s\in\mathcal{S}}Q_S(s)W_s(y|x)$. We abbreviate $\bar{P}_{\bar{U}^k,\bar{X}^k,\bar{Y}_{Q_S}^k,\bar{Z}_{s^k}^k}$ as $\bar{P}$ and note that it corresponds to taking $\mathcal{V}_k=\mathcal{U}^k$ and setting $P_{\bar{V}_k,\bar{X}^k}=\bar{Q}^k_{U,X}$ in $P^{(Q_S)}_{\bar{V}_k,\bar{X}^k,\bar{Y}_{Q_S}^k,\bar{Z}_{s^k}^k}$. As a last preliminary technical step we define $(U^k,X^k,S^k,Y^k,Z^k)$ as independent copies of $(U,X,S,Y,Z)\sim \bar{Q}$, i.e., $(U^k,X^k,S^k,Y^k,Z^k)\sim \bar{Q}^k$, and note that $\bar{P}_{\bar{U}^k,\bar{X}^k,\bar{Y}_{Q_S}^k}=\bar{Q}^k_{U,X,Y}$ and that $\bar{P}_{\bar{U}^k,\bar{X}^k,\bar{Z}_{s^k}^k}=\bar{Q}^k_{U,X,Z|S=s^k}$, for each $s^k\in\mathcal{S}^k$.

Now, evaluating the first mutual information term from \eqref{EQ:multiletter_capacity_type} under $\bar{P}$ gives
\begin{equation}
I_{\bar{P}}\big(\bar{U}^k;\bar{Y}_{Q_S}^k\big)=I_{\bar{Q}^k}(U^k;Y^k)\stackrel{(a)}=k\cdot I_{\bar{Q}}(U;Y),\label{EQ:multiletter_derivation_pre}
\end{equation}
where (a) is because $\big\{(U_i,Y_i)\big\}_{i=1}^k$ are a sequence of i.i.d. pairs according to the marginal distribution of $U$ and $Y$ with respect to $\bar{Q}$. For the subtracted term from \eqref{EQ:multiletter_capacity_type} when calculated with respect to $\bar{P}$, we have
\begin{align*}
\max_{s^k\in\mathcal{T}_{Q_S}^k}I_{\bar{P}}\big(\bar{U}_k;\bar{Z}_{s^k}^k\big)&\stackrel{(a)}=\max_{s^k\in\mathcal{T}_{Q_S}^k}I_{\bar{Q}^k}(U^k;Z^k|S^k=s^k)\\
&\stackrel{(b)}=\max_{s^k\in\mathcal{T}_{Q_S}^k}\sum_{i=1}^kI_{\bar{Q}^k}(U_i;Z_i|S_i=s_i)\\
&\stackrel{(c)}=\max_{s^k\in\mathcal{T}_{Q_S}^k}k\cdot\sum_{s\in\mathcal{S}}Q_S(s)I_{\bar{Q}}(U;Z|S=s)\\
&=k\cdot I_{\bar{Q}}(U;Z|S)\numberthis\label{EQ:multiletter_derivation2}
\end{align*}
where (a) is because $\bar{P}_{\bar{U}^k,\bar{X}^k,\bar{Z}_{s^k}^k}=\bar{Q}^k_{U,X,Z|S=s^k}$, for every $s^k\in\mathcal{S}^k$, (b) uses the product structure of $\bar{Q}^k$, while (c) follows since $\nu_{s^k}=Q_S$ for every $s^k\in\mathcal{T}_{Q_S}^k$.

Based on \eqref{EQ:multiletter_derivation_pre} and \eqref{EQ:multiletter_derivation2} we get the desired inequality since
\begin{align*}
R_{\mathrm{S}}^*(&\mathfrak{W},\mathfrak{V},Q_S)\\
&\geq\lim_{\substack{k\to\infty:\\ k\in\mathbb{N}_\ell}}\frac{1}{k}\left[I_{\bar{P}}\big(\bar{U}^k;\bar{Y}^k_{Q_S}\big)-\max_{s^k\in\mathcal{T}^k_{Q_S}}I_{\bar{P}}\big(\bar{U}^k;\bar{Z}_{s^k}^k\big)\right]\\
&=I_{\bar{Q}}(U;Y)-I_{\bar{Q}}(U;Z|S)\\
&=C_{\mathrm{R}}(\mathfrak{W},\mathfrak{V},Q_S).\numberthis
\end{align*}

The opposite inequality, that is $R_{\mathrm{S}}^*(\mathfrak{W},\mathfrak{V},Q_S)\leq C_{\mathrm{R}}(\mathfrak{W},\mathfrak{V},Q_S)$, follows by repeating the steps from the proof of Theorem \ref{TM:AVWTC_CR_converse_general} (with the proper update of notation). To avoid verbatim repetition of the same arguments, we give only an outline of the proof while describing the required adjustments. To this end, for each $k\in\mathbb{N}_\ell$ and $P_{\bar{V}_k,\bar{X}^k}\in\mathcal{P}(\mathcal{V}_k\times\mathcal{X}^k)$, it is convenient to define a new distribution $\hat{P}_{V_k,X^k,S^k,Y^k,Z^k}$ given~by 
\begin{align*}
&\mspace{-5mu}\hat{P}_{V_k,X^k,S^k,Y^k,Z^k}\big(v_k,x^k,s^k,y^k,z^k\big)\\
&\mspace{15mu}\triangleq P_{\bar{V}_k,\bar{X}^k}\big(v_k,x^k\big)Q_S^k\big(s^k\big)W_{s^k}^k\big(y^k\big|x^k\big)V_{s^k}^k\big(z^k\big|x^k\big).\numberthis
\end{align*}
Noting that $\hat{P}_{V_k,X^k,Y^k,Z^k}=P_{\bar{V}_k,\bar{X}^k,\bar{Y}_{Q_S}^k}$ and $\hat{P}_{V_k,X^k,Z^k|S^k=s^k}=P_{\bar{V}_k,\bar{X}^k,\bar{Z}_{s^k}^k}$, for every $s^k\in\mathcal{S}^k$, one may rewrite \eqref{EQ:multiletter_capacity_type} as \eqref{EQ:multiletter_capacity_type_alt} from the top of this page. 
\setcounter{equation}{68}

Having this, to establish $C_{\mathrm{R}}(\mathfrak{W},\mathfrak{V},Q_S)$ as an upper bound, fix $k\in\mathbb{N}_\ell$ and $P_{\bar{V}_k,\bar{X}^k}\in\mathcal{P}(\mathcal{V}_k\times\mathcal{X}^k)$ and consider the following. Due to the structure of $\hat{P}$ one may treat $V_k$ as the message $M$ in the proof of Theorem \ref{TM:AVWTC_CR_converse_general} and invoke simple adaptations of Lemmas \ref{LEMMA:typical_type} and \ref{LEMMA:Upsilon_security} to get 
\begin{equation}
\max_{s^k\in\mathcal{T}^k_{Q_S}}I_{\hat{P}}\big(V_k;Z^k\big|S^k=s^k\big)\geq I_{\hat{P}}(V_k;Z^k|S^k)-k\xi_{k,\alpha},\label{EQ:multi_letter_converse_UB1}
\end{equation}
where $\xi_{k,\alpha}=\log|\mathcal{Z}|\left(\alpha+2|\mathcal{S}|e^{-2k\frac{\alpha^2}{|\mathcal{S}|^2}}\right)$ and $\alpha$ is any number in $(0,1]$. Using \eqref{EQ:multi_letter_converse_UB1}, for any $k\in\mathbb{N}_\ell$ and $P_{\bar{V}_k,\bar{X}^k}\in\mathcal{P}(\mathcal{V}_k\times\mathcal{X}^k)$, we have
\begin{align*}
I_{\hat{P}}\big(V_k;Y&^k\big)-\max_{s^k\in\mathcal{T}^k_{Q_S}}I_{\hat{P}}\big(V_k;Z^k\big|S^k=s^k\big)\\
           &\leq I_{\hat{P}}\big(V_k;Y^k\big)-I_{\hat{P}}\big(V_k;Z^k\big|S^k\big)+k\xi_{k,\alpha}.\numberthis\label{EQ:multi_letter_converse_UB2}
\end{align*}
Applying standard manipulations on the RHS of \eqref{EQ:multi_letter_converse_UB2} (similar to those in the derivation of \eqref{EQ:UB_summation}), further shows that
\begin{align*}
&I_{\hat{P}}\big(V_k;Y^k\big)-\max_{s^k\in\mathcal{T}^k_{Q_S}}I_{\hat{P}}\big(V_k;Z^k\big|S^k=s^k\big)\\
&\leq k\cdot \max_{Q_{U,X}\in\mathcal{P}(\mathcal{U}\times\mathcal{X})}\Big[I_Q(U;Y)-I_Q(U;Z|S)\Big]+k\xi_{k,\alpha},\numberthis\label{EQ:multi_letter_converse_final}
\end{align*}
where the mutual information terms are taken with respect to the joint distribution from Theorem \ref{TM:AVWTC_CR_capacity}. The derivation of \eqref{EQ:multi_letter_converse_final} also relies on claims similar to those from Lemmas \ref{LEMMA:upsilon_independence} and \ref{LEMMA:Q_PMF_converse} from Section \ref{SEC:AVWTC_CR_converse_general_proof}. Dividing both sides by $k$, taking the supremum of the LHS over all $P_{\bar{V}_k,\bar{X}^k}\in\mathcal{P}(\mathcal{V}_k\times\mathcal{X}^k)$ and replacing $\alpha\in(0,1]$ with a sequence $\big\{\alpha_k\big\}_{k\in\mathbb{N}_\ell}$ that decays sufficiently slowly to zero, the proof is completed by letting $k\to\infty$.
\end{IEEEproof}

\ \\

\subsection{Proof of Theorem~\ref{TM:AVWTC_CR_capacity} from Theorems \ref{TM:AVWTC_CR_achievability_general} and \ref{TM:AVWTC_CR_converse_general}}\label{SUBSEC:AVWTC_CR_capacity_proof}

\textbf{Achievability:} For the direct part, denote the RHS of \eqref{EQ:AVWTC_CR_capacity} by $C^\star_{\mathrm{R}}(\mathfrak{W},\mathfrak{V},Q_S)$ and assume that $R<C^\star_{\mathrm{R}}(\mathfrak{W},\mathfrak{V},Q_S)$. We show that there exists $\delta_0>0$, such that for any $\epsilon>0$ there is a CR $(n,M_n,K_n)$-code $\mathsf{C}_n$ that satisfies \eqref{EQ:CR_achievability_type}. 

For any $\delta>0$, define
\begin{equation}
\mathcal{P}_\delta(Q_S)\mspace{-2mu}\triangleq\mspace{-2mu}\Big\{P\in\mathcal{P}(\mathcal{S})\Big|\big|P(s)-Q_S(s)\big|\leq\delta Q_S(s),\ \forall s\in\mathcal{S}\Big\}\mspace{-2mu},\label{EQ:P-delta_PMFs_def}
\end{equation}
and note that $\mathcal{P}_\delta(Q_S)$ is convex and closed. Applying Theorem \ref{TM:AVWTC_CR_achievability_general} with $\mathcal{Q}=\mathcal{P}_\delta(Q_S)$ and recalling Definition \ref{DEF:CR_achievability} of $\mathcal{Q}$-constrained achievability, yields that if\footnote{For simplicity of notation, throughout the proof of Theorem \ref{TM:AVWTC_CR_capacity} we write $Q_j$ instead of $Q_S^{(j)}$, for $j=1,2$.}
\begin{equation}
R<\max_{Q_{U,X}}\left[\min_{Q_1\in\mathcal{P}_\delta(Q_S)}\mspace{-2mu}I_{Q_1}(U;Y)\mspace{-2mu}-\mspace{-2mu}\max_{Q_2\in\mathcal{P}_\delta(Q_S)}\mspace{-2mu}I_{Q_2}(U;Z|S)\right]\mspace{-2mu},\label{EQ:achievability_rate_assumption}
\end{equation}
then there exists a CR $(n,M_n,K_n)$-code $\mathsf{C}_n$ that satisfies \eqref{EQ:CR_achievability_type}. Thus, to establish the achievability of $R$ it suffices to show that there exists $\delta_0>0$ for which \eqref{EQ:achievability_rate_assumption} holds. The following lemma, that is proven in Appendix \ref{APPEN:delta_exists_proof} using continuity, fills that gap.
\begin{lemma}\label{LEMMA:delta_exists}
The following limiting relation holds:
\begin{align*}
\max_{Q_{U,X}}&\left[\min_{Q_1\in\mathcal{P}_{\delta}(Q_S)}I_{Q_1}(U;Y)-\max_{Q_2\in\mathcal{P}_{\delta}(Q_S)}I_{Q_2}(U;Z|S)\right]\\&\mspace{220mu}\nearrow
C^\star_{\mathrm{R}}(\mathfrak{W},\mathfrak{V},Q_S),\numberthis\label{EQ:delta_exists}
\end{align*}
as $\delta\searrow 0$.
\end{lemma}



\textbf{Converse:} Assume that $R$ is an achievable CR-assisted SS-rate for the type constrained AVWTC $(\mathfrak{W}^n,\mathfrak{V}^n,Q_S)$. Then, there exists $\delta>0$, such that for all $\epsilon>0$ and sufficiently large $n$, there exists a CR $(n,M_n,K_n)$-code $\mathsf{C}_n$ that satisfies \eqref{EQ:CR_achievability_type}. Define 
\begin{equation}
    \mathcal{R}_\delta(Q_S)\triangleq \mathcal{P}_\delta(Q_S)\cap\mathcal{P}_\mathbb{Q}(\mathcal{S}),\label{EQ:all_raqtionals_delta}
\end{equation}
(see \eqref{EQ:rational_PMFs_set} and \eqref{EQ:P-delta_PMFs_def}) which is the set of all rational PMFs on $\mathcal{S}$ that are element-wise $\delta$-close to $Q_S$. Recall the definition of $\mathcal{S}_\mathcal{Q}^n$ from \eqref{EQ:Q_PMFS}, where $\mathcal{Q}\subseteq\mathcal{P}(\mathcal{S})$ is any subset of PMFs. Taking $\mathcal{Q}=\mathcal{R}_\delta(Q_S)$ gives $\mathcal{S}_\mathcal{Q}^n=\mathcal{T}_\delta^n(Q_S)$, for any $n\in\mathbb{N}$.

Using Theorem \ref{TM:AVWTC_CR_converse_general} with $\mathcal{Q}=\mathcal{R}_\delta(Q_S)$ gives the following upper bound:
\begin{equation}
R\leq \max_{Q_{V,U,X}}\inf_{\hat{Q}_S\in\mathcal{R}_\delta(Q_S)}\Big[I_{\hat{Q}}(U;Y|V)-I_{\hat{Q}}(U;S,Z|V)\Big],\label{EQ:type_converse_new}
\end{equation}
where $I_{\hat{Q}}$ denotes that the mutual information terms are calculated with respect to the marginals of $Q_{V,U,X}\hat{Q}_SQ_{Y|X,S}Q_{Z|X,S}$ for some $\hat{Q}_S\in\mathcal{R}_\delta(Q_S)$, where $Q_{Y|X,S}(y|x,s)=W_s(y|x)$ and $Q_{Z|X,S}(z|x,s)=V_s(z|x)$, for all $(s,x,y,z)\in\mathcal{S}\times\mathcal{X}\times\mathcal{Y}\times\mathcal{Z}$.

We first show that any $\hat{Q}_S\in\mathcal{R}_\delta(Q_S)$ in the joint distribution of $(V,U,X,S,Y,Z)$ can be replaced with the $Q_S$ while causing only a small change in the value of the mutual information terms. Let $Q_{V,U,X}^\star\in\mathcal{P}(\mathcal{V}\times\mathcal{U}\times\mathcal{X})$ be the maximizer of the RHS of \eqref{EQ:type_converse_new}. With some abuse of notation we denote by $I_{\hat{Q}}$ and $I_Q$ a mutual information term calculated with respect to $Q^\star_{V,U,X}\hat{Q}_SQ_{Y|X,S}Q_{Z|X,S}$ or $Q^\star_{V,U,X}Q_SQ_{Y|X,S}Q_{Z|X,S}$, respectively. By the definition of $\mathcal{R}_\delta(Q_S)$, for any $\hat{Q}_S\in\mathcal{R}_\delta(Q_S)$ we have
\begin{equation}
\big|\hat{Q}_S(s)-Q_S(s)\big|\leq \frac{\delta}{|\mathcal{S}|},\quad\forall s\in\mathcal{S}.
\end{equation}
The continuity of the mutual information implies that there exists a function $f(\delta)$, such that $\lim_{\delta\to 0}f(\delta)= 0$ is independent of $Q^\star_{V,U,X}$ and
\begin{align*}
I_{\hat{Q}}(U;Y|V&)-I_{\hat{Q}}(U;S,Z|V)\\
&\leq I_Q(U;Y|V)-I_Q(U;S,Z|V)+f(\delta).\numberthis\label{EQ:type_typical_converse}
\end{align*}
Further notice that Definition \ref{DEF:CR_achievability_type} of the type constrained achievability gives
\begin{subequations}
\begin{align}
\mathcal{E}\big(\mathfrak{W}^n,\mathcal{Q}_{\delta'}(Q_S),\mathsf{C}_n\big)&\leq\epsilon\label{EQ:CR_achievability_reliability_delta'}\\
\mathcal{L}_\mathrm{Sem}\big(\mathfrak{V}^n,\mathcal{Q}_{\delta'}(Q_S),\mathsf{C}_n\big)&\leq\epsilon.\label{EQ:CR_achievability_security_delta'}
\end{align}
\end{subequations}
for any $\delta'\in(0,\delta)$ (see Remark \ref{REM:smaller_delta_AVWTC_achievability}). Taking $\delta'\to 0$ in \eqref{EQ:type_typical_converse}, while noting that the rational distributions are dense in $\mathcal{P}(\mathcal{S})$ and that $I_{\hat{Q}}(U;Y|V)-I_{\hat{Q}}(U;S,Z|V)$ is continuous in $\hat{Q}_S$ gives
\begin{equation}
R\leq I_Q(U;Y|V)-I_Q(U;S,Z|V).\label{EQ:type_converse_VUB}
\end{equation}

Our last step is to remove the conditioning on $V$. The structure of the joint distribution $Q_{V,U,X,S,Y,Z}=Q_{V,U,X}Q_SQ_{Y|X,S}Q_{Z|X,S}$ implies that for any $v\in\mathcal{V}$, the conditional distribution of $(U,X,S,Y,Z)$ given $V=v$ factors as
$Q_{U,X,S,Y,Z|V=v}=Q_{U,X|V=v}Q_SQ_{Y|X,S}Q_{Z|X,S}$. Denoting by $I_{Q_v}$ a mutual information term taken with respect to $Q_{U,X|V=v}Q_SQ_{Y|X,S}Q_{Z|X,S}$, we further upper bound $R$ from \eqref{EQ:type_converse_VUB} as
\begin{align*}
  R&\leq I_Q(U;Y|V)-I_Q(U;S,Z|V)\\
   &=\sum_{v\in\mathcal{V}}Q_V(v)\Big[I_Q(U;Y|V=v)-I_Q(U;S,Z|V=v)\Big]\\
   &\leq\max_{v\in\mathcal{V}}\Big[I_Q(U;Y|V=v)-I_Q(U;S,Z|V=v)\Big]\\
   &=\max_{v\in\mathcal{V}}\Big[I_{Q_v}(U;Y)-I_{Q_v}(U;S,Z)\Big]\\
   &\stackrel{(a)}=\max_{v\in\mathcal{V}}\Big[I_{Q_v}(U;Y)-I_{Q_v}(U;Z|S)\Big]\\
   &\leq\max_{Q_{U,X}}\Big[I_{Q}(U;Y)-I_{Q}(U;Z|S)\Big],\numberthis\label{EQ:from_UV_to_U}
\end{align*}
where (a) is because $U$ and $S$ are independent under $Q_v$, for every $v\in\mathcal{V}$. This completes the proof.

\section{Proof of Theorem~\ref{TM:AVWTC_CR_achievability_general}}\label{SEC:AVWTC_CR_achievability_general_proof}
     
The proof first constructs a reliable CR-code over a family of doubly-exponentially many uncorrelated codes. Specifically, the family consists of all realizations of a random wiretap code with i.i.d. codewords. Reliability is then established via a simple adaptation of the standard AVC error probability analysis \cite{Csiszar_AVWTC_constrained_random1988}. Being double-exponential in size, however, the original family of codes is too large to derive SS in the sense of \eqref{EQ:security_CR_codes}. Therefore, a Chernoff bound is used to show that only a polynomial sub-family of codes is sufficient for reliability, and having that, the double-exponential decay that Lemma \ref{LEMMA:soft_covering} provides is leveraged to prove SS over the reduced CR-code.

Without loss of generality, we assume that $\mathcal{Q}$ contains at least one rational distribution; otherwise, there is nothing to prove (see Remark \ref{REM:no_rational_PMF}). Consequently, we henceforth refer only to blocklengths $n\in\mathbb{N}$ for which $\mathcal{S}_\mathcal{Q}^n\neq\emptyset$. We show that \eqref{EQ:AVWTC_CR_unconstrained_achievability} is achievable when $U=X$. Then, using standard channel prefixing arguments, the RHS of \eqref{EQ:AVWTC_CR_unconstrained_achievability} is achieved.

Fix $\epsilon>0$, a PMF $Q_{U,X}\in\mathcal{P}(\mathcal{U}\times\mathcal{X})$ and let $Q_X$ be its marginal on $\mathcal{X}$. For any $Q_S\in\mathcal{Q}$, the joint distribution of $(X,S,Y,Z)$ is $Q_{X,S,Y,Z}=Q_XQ_SQ_{Y|X,S}Q_{Z|X,S}$, where $Q_{Y|X,S}$ and $Q_{Z|X,S}$ are given in Theorem \ref{TM:AVWTC_CR_achievability_general}. Since $Q_X$, $Q_{Y|X,S}$ and $Q_{Z|X,S}$ stay fixed throughout the proof, we use $I_{Q_S}$ to denote a mutual information term taken with respect to $Q_XQ_SQ_{Y|X,S}Q_{Z|X,S}$. Let $W$ be a random variable uniformly distributed over $\mathcal{W}_n\triangleq\big[1:2^{n\tilde{R}}\big]$, where $\tilde{R}\in\mathbb{R}_+$, that is chosen independently of the message $M$; $W$ stands for the stochastic part of the encoder.



\par\textbf{Random Codebook $\bm{\mathsf{B}_n}$:} A random codebook is a collection of independent random vectors $\mathsf{B}_n=\big\{\mathbf{X}(m,w)\big\}_{(m,w)\in\mathcal{M}_n\times\mathcal{W}_n}$, each distributed according to $Q_X^n$. Set $\tilde{K}_n=|\mathcal{X}|^{n|\mathcal{M}_n||\mathcal{W}_n|}$ and index (with respect to some arbitrary order) all possible realizations of $\mathsf{B}_n$ by $\tilde{\Gamma}_n=[1:\tilde{K}_n]$ to obtain the set of codebooks $\mathfrak{B}_n=\left\{\mathcal{B}_n^{(\gamma)}\right\}_{\gamma\in\tilde{\Gamma}_n}$, where $\mathcal{B}_n^{(\gamma)}=\big\{\mathbf{x}(m,w,\gamma)\big\}_{(m,w)\in\mathcal{M}_n\times\mathcal{W}_n}$ is the $\gamma$-th element of $\mathfrak{B}_n$. Further, we define the measure $\tilde{\mu}_n$ on $\tilde{\Gamma}_n$ as
\begin{equation}
\tilde{\mu}_n(\gamma)=\prod_{(m,w)\in\mathcal{M}_n\times\mathcal{W}_n}Q_{X}^n\big(\mathbf{x}(m,w,\gamma)\big),\quad \forall\gamma\in\tilde{\Gamma}_n.\label{EQ:mu_gamma}
\end{equation}


\par\textbf{Stochastic Encoder $\bm{f^{(n)}_\gamma}$:} Fix $\gamma\in\tilde{\Gamma}_n$. To send the message $m\in\mathcal{M}_n$ the encoder randomly and uniformly chooses $w$ from $\mathcal{W}_n$ and feeds $\mathbf{x}(m,w,\gamma)\in\mathcal{B}_n^{(\gamma)}$ into the AVWTC. The stochastic encoder $f^{(n)}_\gamma:\mathcal{M}_n\to\mathcal{P}(\mathcal{X}^n)$ is thus defined by
\begin{equation}
f^{(n)}_\gamma(\mathbf{x}|m)=\sum_{w\in\mathcal{W}_n}2^{-n\tilde{R}}\mathds{1}_{\big\{\mathbf{x}(m,w,\gamma)=\mathbf{x}\big\}}.\label{EQ:stochastic_encoder_gamma}
\end{equation}


\par\textbf{Decoder $\bm{\phi^{(n)}_\gamma}$:} Both $m$ and $w$ are decoded by the legitimate user. With some abuse of notation, for every $\gamma\in\tilde{\Gamma}_n$ we consider a decoding rule $\phi^{(n)}_\gamma:\mathcal{Y}^n\to\big(\mathcal{M}_n\times\mathcal{W}_n\big)\cup\{e\}$ that is defined in terms of a non-negative-valued function $d:\mathcal{X}^n\times\mathcal{Y}^n\to\mathbb{R}_+$ as
\begin{equation}
\phi^{(n)}_\gamma(\mathbf{y})=\begin{cases}(m,w),\quad \begin{array}{c}
     \max\limits_{\substack{(m',w')\in\mathcal{M}_n\times\mathcal{W}_n:\\(m',w')\neq(m,w)}}d\big(\mathbf{x}(m',w',\gamma),\mathbf{y}\big)\\\mspace{114mu}<d\big(\mathbf{x}(m,w,\gamma),\mathbf{y}\big)
\end{array}\\
e,\quad \mbox{no $(m,w)$ as above exists}\end{cases}.\label{EQ:decoder_gamma}
\end{equation}
Although the decoding rule is given in terms of an arbitrary function $d$, we soon limit ourselves to a specific choice with respect to which we establish reliability. We use an arbitrary $d$ for now to state Lemma \ref{LEMMA:decoding} in its most general form, thus emphasizing the generality of this decoding rule.

For every $\gamma\in\tilde{\Gamma}_n$ denote $c_n^{(\gamma)}\triangleq\left(f^{(n)}_\gamma,\phi^{(n)}_\gamma\right)$ as the associated uncorrelated $(n,M_n)$-code, and let $\tilde{\mathcal{C}}_n\triangleq\big\{c_n^{(\gamma)}\big\}_{\gamma\in\tilde{\Gamma}_n}$. The CR $(n,M_n,\tilde{K}_n)$-code $\tilde{\mathsf{C}}_n$ is thus defined by the family $\tilde{\mathcal{C}}_n$, the index set $\tilde{\Gamma}_n$ of size $\tilde{K}_n$, and the measure $\tilde{\mu}_n\in\mathcal{P}(\tilde{\Gamma}_n)$ from \eqref{EQ:mu_gamma}. Note that $\tilde{\mathsf{C}}_n$ is a CR-code over a family of doubly-exponentially many $(n,M_n)$-codes. 
\ \\
\par \textbf{Error Probability Analysis:} We first show the $\tilde{\mathsf{C}}_n$ is reliable. Having that, we use a Chernoff bound to reduce our CR-code to be only polynomial (with the blocklength) in size and then establish SS. The reliability of $\tilde{\mathsf{C}}_n$ relies on the following lemma, which is effectively an adaptation of \cite[Lemma 12.9]{Csiszar_Korner_Book2011}. For completeness, we prove the lemma is given in Appendix \ref{APPEN:decoding_proof}.

\begin{lemma}\label{LEMMA:decoding}
If $\mathbb{E}_{Q_X^n}d(\mathbf{X},\mathbf{y})\leq 1$, for all $\mathbf{y}\in\mathcal{Y}^n$, then for every channel $W_n:\mathcal{X}^n\to\mathcal{P}(\mathcal{Y}^n)$, $\eta>0$ and $(m,w)\in\mathcal{M}_n\times\mathcal{W}_n$, we have
\begin{equation}
\mathcal{E}_{m,w}(W_n,\tilde{\mathsf{C}}_n)\leq \mathbb{P}_{Q_X^nW_n}\left(d(\mathbf{X},\mathbf{Y})<\frac{|\mathcal{M}_n||\mathcal{W}_n|}{\eta}\right)+\eta,\label{EQ:error_probability_bound}
\end{equation}
where 
\begin{equation}
\mathcal{E}_{m,w}(W_n,\tilde{\mathsf{C}}_n)=\sum_{\gamma\in\tilde{\Gamma}_n}\tilde{\mu}_n(\gamma)\mspace{-8mu}\sum_{\substack{\mathbf{y}\in\mathcal{Y}^n:\\\phi^{(n)}_\gamma(\mathbf{y})\neq (m,w)}}\mspace{-5mu}W_n\big(\mathbf{y}\big|\mathbf{x}(m,w,\gamma)\big)
\end{equation}
is the expected error probability in decoding $(m,w)$ over the ensemble $\tilde{\mathcal{C}}_n$.
\end{lemma}

Assume without loss of generality that all entries of the matrices $W\in\mathfrak{W}$ are bounded below by a positive constant $\upsilon>0$. Indeed, consider a modified family of channels $\mathfrak{W}_\upsilon$, formally defined by $W_\upsilon(y|x)=(1-\upsilon|\mathcal{Y}|)W(y|x)+\upsilon$, where $W\in\mathfrak{W}$. Replacing $\mathfrak{W}$ with $\mathfrak{W}_\upsilon$ causes a negligible change in $I(U;Y)-I(U;Z|S)$ if $\upsilon$ is small, for any $Q_SQ_{U,X}$. Further, any sequence of CR codes that is reliable with respect to $\mathfrak{W}_\upsilon$ is trivially modified at the decoder to give the same maximal error probability under $\mathfrak{W}$ as the original one did for $\mathfrak{W}_\upsilon$. The modified decoder simulates the output of the AVC $\mathfrak{W}$ to look as if it was generated by $\mathfrak{W}_\upsilon$. Specifically, upon observing an output sequence $\mathbf{y}\in\mathcal{Y}$ generated by $\mathfrak{W}$, for each time instance $i\in[1:n]$, the modified decoder draws a $\mbox{Bernoulli}\big(\upsilon|\cal{Y}|\big)$ distributed random variable: If the outcome is 0, then the observed $y_i$ is preserved. If, on the other hand, the outcome is 1, the new decoder draws a symbol $y$ uniformly from $\mathcal{Y}$ and replaces the observed $y_i$ with the uniformly chosen symbol. This makes the new $y$-sequence look like it was generated by the AVC $\mathfrak{W}_\upsilon$, and the original decoder is then used. 


Our next steps up until Lemma 5 (included) follow close resemblance to \cite[Lemma 12.10]{Csiszar_Korner_Book2011}. For any $Q\in\mathcal{P}(\mathcal{S})$ define $W_Q:\mathcal{X}\to\mathcal{P}(\mathcal{Y})$, the averaged DMC under $Q$, as 
\begin{equation}
W_Q(y|x)\triangleq\sum_{s\in\mathcal{S}}Q(s)W_s(y|x),\quad\forall(x,y)\in\mathcal{X}\times\mathcal{Y}.\label{EQ:reliability_analysis_WQ}
\end{equation}
Let $\tilde{Q}\in\mathcal{Q}$ be a minimizer of $\min_{Q\in\mathcal{Q}}I_Q(X;Y)$ and set $d:\mathcal{X}^n\times\mathcal{Y}^n\to\mathbb{R}_+$ as
\begin{equation}
d(\mathbf{x},\mathbf{y})=\frac{W^n_{\tilde{Q}}(\mathbf{y}|\mathbf{x})}{\tilde{Q}^n_Y(\mathbf{y})}=\prod_{i=1}^n\frac{W_{\tilde{Q}}(y_i|x_i)}{\tilde{Q}_Y(y_i)},
\label{EQ:reliability_analysis_d}
\end{equation}
for all $(\mathbf{x},\mathbf{y})\in\mathcal{X}^n\times\mathcal{Y}^n$ with $\tilde{Q}^n_Y(\mathbf{y})\neq0$, where
\begin{equation}
\tilde{Q}_Y(y)=\sum_{x\in\mathcal{X}}Q_X(x)W_{\tilde{Q}}(y|x).\label{EQ:reliability_analysis_QY}
\end{equation}
If $\tilde{Q}^n_Y(\mathbf{y})=0$, set $d(\mathbf{x},\mathbf{y})=1$, for all $\mathbf{x}\in\mathcal{X}^n$. Clearly
\begin{equation}
\mathbb{E}_{Q_X^n}d(\mathbf{X},\mathbf{y})=1,\quad\forall\mathbf{y}\in\mathcal{Y}^n.
\end{equation}
Lemma \ref{LEMMA:decoding} implies that for every $\eta>0$, $\mathbf{s}\in\mathcal{S}^n$ and $(m,w)\in\mathcal{M}\times\mathcal{W}$, 
\begin{equation}
\mathcal{E}_{m,w}(W^n_\mathbf{s},\mspace{-2mu}\tilde{\mathsf{C}}_n)\mspace{-2.5mu}\leq\mspace{-2mu} \mathbb{P}_{Q_X^nW_\mathbf{s}^n}\mspace{-5mu}\left(\frac{W^n_{\tilde{Q}}(\mathbf{Y}_\mathbf{s}|\mathbf{X})}{\tilde{Q}^n_Y(\mathbf{Y}_\mathbf{s})}\mspace{-3mu}<\mspace{-3mu}\frac{2|\mathcal{M}_n||\mathcal{W}_n|}{\eta}\right)\mspace{-2mu}+\mspace{-2mu}\frac{\eta}{2},\label{EQ:reliability_analysis_error_prob_UB}
\end{equation}
where $(\mathbf{X},\mathbf{Y}_\mathbf{s})\sim Q_X^nW^n_\mathbf{s}$.

Fix $\mathbf{s}\in\mathcal{S}^n$ and define the random variable $L_n(\mathbf{X},\mathbf{Y}_\mathbf{s})\triangleq \log\frac{W_{\tilde{Q}}^n(\mathbf{Y}_\mathbf{s}|\mathbf{X})}{\tilde{Q}_Y^n(\mathbf{Y}_\mathbf{s})}$. Observe that 
\begin{align*}
&\mathbb{E}_{Q_X^nW^n_\mathbf{s}}L_n(\mathbf{X},\mathbf{Y}_\mathbf{s})\\
&=\sum_{(\mathbf{x},\mathbf{y})\in\mathcal{X}^n\times\mathcal{Y}^n}Q_X^n(\mathbf{x})W^n_\mathbf{s}(\mathbf{y}|\mathbf{x})\log\left(\prod_{i=1}^n\frac{W_{\tilde{Q}}(y_i|x_i)}{\tilde{Q}_Y(y_i)}\right)\\
&=\sum_{i=1}^n\sum_{(x,y)\in\mathcal{X}\times\mathcal{Y}}Q_X(x)W_{s_i}(y|x)\log\left(\frac{W_{\tilde{Q}}(y|x)}{\tilde{Q}_Y(y)}\right).\numberthis
\end{align*}
For any $\mathbf{s}\in\mathcal{S}^n$ and $(x,y)\in\mathcal{X}\times\mathcal{Y}$, denote
\begin{equation}
\sum_{i=1}^nW_{s_i}(y|x)=n\sum_{s\in\mathcal{S}}\nu_\mathbf{s}(s)W_s(y|x)\triangleq nW_{\nu_\mathbf{s}}(y|x), \label{EQ:empirical_averaged_channel}
\end{equation}
where $\nu_\mathbf{s}$ is the empirical PMF of $\mathbf{s}$, as defined in \eqref{EQ:empirical_PMF}.
Consequently, we have
\begin{align*}
\mathbb{E}&_{Q_X^nW^n_\mathbf{s}}L_n(\mathbf{X},\mathbf{Y}_\mathbf{s})\\
&\mspace{20mu}=n\cdot\mspace{-15mu}\sum_{(x,y)\in\mathcal{X}\times\mathcal{Y}}\mspace{-8mu}Q_X(x)W_{\nu_\mathbf{s}}(y|x)\log\left(\frac{W_{\tilde{Q}}(y|x)}{\tilde{Q}_Y(y)}\right).\numberthis\label{EQ:expected_L1}
\end{align*}
We next show that if $\mathbf{s}\in\mathcal{S}_\mathcal{Q}^n$, then the RHS of \eqref{EQ:expected_L1} is lower bounded by $I_{\tilde{Q}}(X;Y)$.

\begin{lemma}\label{LEMMA:pseudo_MI_geq_minimal_MI}
For any $Q\in\mathcal{Q}$, the following relation holds
\begin{equation}
\sum_{(x,y)\in\mathcal{X}\times\mathcal{Y}}Q_X(x)W_Q(y|x)\log\left(\frac{W_{\tilde{Q}}(y|x)}{\tilde{Q}_Y(y)}\right)\geq I_{\tilde{Q}}(X;Y).\label{EQ:pseudo_MI_geq_minimal_MI}
\end{equation}
\end{lemma}
Lemma \ref{LEMMA:pseudo_MI_geq_minimal_MI} is proven in Appendix \ref{APPEN:pseudo_MI_geq_minimal_MI_proof}. Now, if $\mathbf{s}\in\mathcal{S}_\mathcal{Q}^n$, then $\nu_\mathbf{s}\in\mathcal{Q}$. By Lemma \ref{LEMMA:pseudo_MI_geq_minimal_MI} and \eqref{EQ:expected_L1} this gives
\begin{equation}
\mathbb{E}_{Q_X^nW^n_\mathbf{s}}L_n(\mathbf{X},\mathbf{Y}_\mathbf{s})\geq nI_{\tilde{Q}}(X;Y),\quad\forall\mathbf{s}\in\mathcal{S}_\mathcal{Q}^n.\label{EQ:expected_l_mutual_info}
\end{equation}

Furthermore, since $W_s(y|x)>\upsilon>0$ for every $(s,x,y)\in\mathcal{S}\times\mathcal{X}\times\mathcal{Y}$, it holds that
\begin{equation}
\left|\log\frac{W_{\tilde{Q}}(y|x)}{\tilde{Q}_Y(y)}\right|\leq\left|\log\frac{1}{\upsilon}\right|=-\log\upsilon,\label{EQ:log_LB}
\end{equation}
which implies that
\begin{equation}
\left|\log\frac{W_{\tilde{Q}}(Y|X)}{\tilde{Q}_Y(Y)}\right|\leq-\log\upsilon\label{EQ:log_LB_wp1} 
\end{equation}
is true with probability 1, for any $(X,Y)\sim P_{X,Y}\in\mathcal{P}(\mathcal{X}\times\mathcal{Y})$. This yields
\begin{equation}
\mbox{var}\left(\log\frac{W_{\tilde{Q}}(Y|X)}{\tilde{Q}_Y(Y)}\right)\leq\log^2\upsilon.\label{EQ:var_LB_wp1} 
\end{equation}

Having \eqref{EQ:reliability_analysis_error_prob_UB}, \eqref{EQ:expected_l_mutual_info} and \eqref{EQ:var_LB_wp1}, the proof of reliability is concluded as follows. Recall that $\mathcal{M}_n=[1:M_n]$ and set
\begin{equation}
M_n = \left\lfloor 2^{\left(\min_{Q\in\mathcal{Q}}I_Q(X;Y)-\tilde{R}-\frac{\delta}{2}\right)}\right\rfloor,\label{EQ:reliability_rate}
\end{equation}
for some $\delta>0$ to be specified later. Using \eqref{EQ:reliability_analysis_error_prob_UB} with $\eta_n=\frac{1}{n}$ in the role of $\eta$, we have that for $n$ sufficiently large and all $(m,w)\in\mathcal{M}_n\times\mathcal{W}_n$ and $\mathbf{s}\in\mathcal{S}_\mathcal{Q}^n$, we have
\begin{align*}
&\mathcal{E}_{m,w}(W^n_\mathbf{s},\tilde{\mathsf{C}}_n)\\
&\leq \mathbb{P}_{Q_X^nW_\mathbf{s}^n}\left(\frac{W^n_{\tilde{Q}}(\mathbf{Y}_\mathbf{s}|\mathbf{X})}{\tilde{Q}^n_Y(\mathbf{Y}_\mathbf{s})}<\frac{2|\mathcal{M}_n||\mathcal{W}_n|}{\eta_n}\right)+\frac{\eta_n}{2}\\
&=\mathbb{P}_{Q_X^nW_\mathbf{s}^n}\bigg(L_n(\mathbf{X},\mathbf{Y}_\mathbf{s})<\log|\mathcal{M}_n||\mathcal{W}_n|-\log\frac{\eta_n}{2}\bigg)+\frac{\eta_n}{2}\\
&\stackrel{(a)}\leq\mathbb{P}_{Q_X^nW_\mathbf{s}^n}\bigg(L_n(\mathbf{X},\mathbf{Y}_\mathbf{s})\mspace{-3mu}<\mspace{-3mu}nI_{\tilde{Q}}(X;Y)\mspace{-2mu}-\mspace{-2mu}\frac{n\delta}{2}\mspace{-2mu}-\mspace{-2mu}\log\frac{\eta_n}{2}\bigg)\mspace{-3mu}+\mspace{-3mu}\frac{\eta_n}{2}\\
&\stackrel{(b)}\leq\mathbb{P}_{Q_X^nW_\mathbf{s}^n}\bigg(\big|\mathbb{E}L_n(\mathbf{X},\mathbf{Y}_\mathbf{s})-L_n(\mathbf{X},\mathbf{Y}_\mathbf{s})\big|>\frac{n\delta}{2}\bigg)+\frac{\eta_n}{2}\\
&\stackrel{(c)}\leq\frac{4\mbox{var}\big(L_n(\mathbf{X},\mathbf{Y}_\mathbf{s})\big)}{n^2\delta^2}+\frac{\eta_n}{2}\\
&\stackrel{(d)}\leq\frac{4\log^2\upsilon}{n\delta^2}+\frac{\eta_n}{2}\\
&\stackrel{(e)}=\frac{c_{\delta,\nu}}{n}\numberthis\label{EQ:mw_reliability_established}
\end{align*}
where (a) and (b) use \eqref{EQ:reliability_rate} and \eqref{EQ:expected_l_mutual_info}, respectively, (c) is Chebyshev's inequality, (d) follows by the pairwise independence of $(\mathbf{X},\mathbf{Y}_\mathbf{s})$ across time and \eqref{EQ:var_LB_wp1}, while (e) is by setting $c_{\delta,\nu}=\frac{1}{2}+\frac{4\log^2\upsilon}{\delta^2}$. Concluding, \eqref{EQ:mw_reliability_established} yields
\begin{equation}
\max_{\substack{\mathbf{s}\in\mathcal{S}_\mathcal{Q}^n,\\(m,w)\in\mathcal{M}_n\times\mathcal{W}_n}}\mathcal{E}_{m,w}(W_\mathbf{s}^n,\tilde{\mathsf{C}}_n)\leq \frac{c_{\delta,\nu}}{n},\label{EQ:mw_reliability_established_final}
\end{equation}
which implies the reliability of CR-code $\tilde{\mathsf{C}}_n$.

\textbf{CR Reduction for Reliability:} Our next step is to reduce the CR-code $\tilde{\mathsf{C}}_n$ over the family $\tilde{\mathcal{C}}_n$ of size $\tilde{K}_n=|\mathcal{X}|^{n|\mathcal{M}_n||\mathcal{W}_n|}$, to be over a family of codes that is no more than polynomial in size (see \cite{Ahlsweded_Elimination1978} for Ahlswede's original CR elimination argument for the classic AVC). This reduction is crucial for the subsequent security analysis. To do so, let $\big\{G_k\big\}_{k=1}^{K_n}$ be a collection of $K_n\in\mathbb{N}$ i.i.d. random variables with values in $\tilde{\Gamma}_n$ and a common distribution $\tilde{\mu}_n$. Each realization $\gamma_k\in\tilde{\Gamma}_n$ of $G_k$, $k\in[1:K_n]$, corresponds to a codebook $\mathcal{B}_n^{(\gamma)}\in\mathfrak{B}_n$ which, in turn, induces a code $c_n^{(\gamma)}$. 

We show that averaging the error probabilities associated with each random code $C_n(k)\triangleq c_n^{(G_k)}$ results in a vanishing term, with arbitrarily high probability. In a later stage, we extract a realization of $\big\{C_n(k)\big\}_{k=1}^{K_n}$ that is both reliable and semantically-secure, and define our CR-code to be uniformly distributed over the codes in the realization.


Thus, for each $\mathbf{s}\in\mathcal{S}^n$ and $(m,w)\in\mathcal{M}_n\times\mathcal{W}_n$, consider the random variable
\begin{equation}
\frac{1}{K_n}\sum_{k=1}^{K_n} e_{m,w}\big(W_\mathbf{s}^n,C_n(k)\big),\label{EQ:reduced_CR_averaged_errors}
\end{equation}
where a possible value of each $e_{m,w}\big(W_\mathbf{s}^n,C_n(k)\big)$, $k\in[1:K_n]$ and $\gamma_k\in\tilde{\Gamma}_n$, is 
\begin{align*}
e_{m,w}\big(W_\mathbf{s}^n,c_n(k)\big)&\triangleq e_{m,w}\big(W_\mathbf{s}^n,c_n^{(\gamma_k)}\big)\\
&=\mspace{-8mu}\sum_{\substack{\mathbf{y}\in\mathcal{Y}^n:\\\phi^{(n)}_{\gamma_k}(\mathbf{y})\neq (m,w)}}\mspace{-18mu}W_\mathbf{s}^n\big(\mathbf{y}\big|\mathbf{x}(m,w,\gamma_k)\big).\numberthis
\end{align*}
This is an average of $K_n$ independent random variables, each bounded between 0 and 1. Furthermore, the expected value of each equals to $\mathcal{E}_{m,w}(W^n_\mathbf{s},\tilde{\mathsf{C}}_n)$, and therefore, \eqref{EQ:mw_reliability_established_final} implies that
\begin{equation}
\mathbb{E}_{\tilde{\mu}_n}e_{m,w}\big(W_\mathbf{s}^n,C_n(k)\big)\leq\frac{c_{\delta,\nu}}{n},
\end{equation}
for each $\mathbf{s}\in\mathcal{S}_\mathcal{Q}^n$, $(m,w)\in\mathcal{M}_n\times\mathcal{W}_n$ and $k\in[1:K_n]$. The probability of \eqref{EQ:reduced_CR_averaged_errors} not decaying to zero under any $\mathbf{s}\in\mathcal{S}_\mathcal{Q}^n$ and $(m,w)\in\mathcal{M}_n\times\mathcal{W}_n$ is upper bounded using the version of the Chernoff bound from \cite[Lemma 4]{Goldfeld_WTCII_semantic2015}.
\begin{lemma}[Chernoff Bound \cite{Goldfeld_WTCII_semantic2015}]\label{LEMMA:Chernoff}
Let $\big\{X_\ell\big\}_{\ell=1}^L$ be a collection of i.i.d. random variables with common distribution $P$, such that $\supp(P)\subseteq[0,B]$ and $\mathbb{E}X_\ell\leq\mu\neq 0$, for all $\ell\in[1:L]$. Then for any $c$ with $\frac{c}{\mu} \in [1,2]$,
\begin{equation}
    \mathbb{P}_{P^L} \left( \frac{1}{L} \sum_{\ell=1}^L X_\ell \geq c \right) \leq e^{-\frac{L \mu}{3 B} \left( \frac{c}{\mu} - 1 \right)^2}.\label{EQ:Chernoff}
\end{equation}
\end{lemma}
Setting $K_n=n^3$ and using \eqref{EQ:Chernoff} with $L=K_n$, $\mu = \frac{c_{\delta,\nu}}{n}$, $B=1$, and $\frac{c}{\mu} = 2$, assures that $\frac{1}{K_n}\sum_{k=1}^{K_n} e_{m,w}\big(W_\mathbf{s}^n,C_n(k)\big)$ is arbitrarily small with probability super-exponentially close to 1. That is, for each $\mathbf{s}\in\mathcal{S}_\mathcal{Q}^n$ and $(m,w)\in\mathcal{M}_n\times\mathcal{W}_n$, we have
\begin{align*}
    \mathbb{P}_{\tilde{\mu}_n^n} \left(\frac{1}{K_n}\sum_{k=1}^{K_n} e_{m,w}\big(W_\mathbf{s}^n,C_n(k)\big)\geq \frac{2c_{\delta,\nu}}{n}  \right)&\leq e^{-\frac{1}{3}\frac{c_{\delta,\nu}K_n}{n}}\\
    &=e^{-\frac{c_{\delta,\nu}n^2}{3}}.\numberthis\label{EQ:reduced_CR_state_bad_probability_bound}
\end{align*}
By \eqref{EQ:reduced_CR_state_bad_probability_bound} and the union bound, we have
\begin{align*}
&\mathbb{P}_{\tilde{\mu}_n^n}\left(\left\{\max_{\substack{\mathbf{s}\in\mathcal{S}_\mathcal{Q}^n,\\(m,w)\\\in\mathcal{M}_n\times\mathcal{W}_n}}\frac{1}{K_n}\sum_{k=1}^{K_n} e_{m,w}\big(W_\mathbf{s}^n,C_n(k)\big)<\frac{2c_{\delta,\nu}}{n}\right\}^c\mspace{3mu}\right)\\
&=\mathbb{P}_{\tilde{\mu}_n^n} \left(\max_{\substack{\mathbf{s}\in\mathcal{S}_\mathcal{Q}^n,\\(m,w)\\\in\mathcal{M}_n\times\mathcal{W}_n}}\frac{1}{K_n}\sum_{k=1}^{K_n} e_{m,w}\big(W_\mathbf{s}^n,C_n(k)\big)\geq\frac{2c_{\delta,\nu}}{n}\right)\\
&\leq \sum_{\mathbf{s}\in\mathcal{S}_\mathcal{Q}^n}\sum_{\substack{(m,w)\\\in\mathcal{M}_n\times\mathcal{W}_n}}\mspace{-5mu}\mathbb{P}_{\tilde{\mu}_n^n} \Bigg(\frac{1}{K_n}\sum_{k=1}^{K_n}e_{m,w}\big(W_\mathbf{s}^n,C_n(k)\big)\geq\frac{2c_{\delta,\nu}}{n}\Bigg)\\
&\leq \big|\mathcal{S}_\mathcal{Q}^n\big|\cdot|\mathcal{M}_n|\cdot|\mathcal{W}_n|\cdot e^{-\frac{c_{\delta,\nu}n^2}{3}}\\
&\stackrel{(a)}\leq |S|^n\cdot 2^{n\big(\min_{Q\in\mathcal{Q}}I_Q(X;Y)-\frac{\delta}{2}\big)}\cdot e^{-\frac{c_{\delta,\nu}n^2}{3}}\\
&\triangleq \kappa_n^{(1)},\numberthis\label{EQ:reduced_CR_bad_probability_bound}
\end{align*}
where (a) uses \eqref{EQ:reliability_rate} and $\big|\mathcal{S}_\mathcal{Q}^n\big|\leq|S|^n$. Note that $\kappa_n^{(1)}\to 0$ as $n\to\infty$.

\par \textbf{Security Analysis:} We show that the probability of $\big\{C_n(k)\big\}_{k=1}^{K_n}$ violating the SS requirement is arbitrarily small. First, for any $\gamma\in\tilde{\Gamma}_n$, $P_M\in\mathcal{P}(\mathcal{M}_n)$ and $\mathbf{s}\in\mathcal{S}^n$, let $P^{(\gamma,\mathbf{s})}_{M,W,\mathbf{X},\mathbf{Z}_\mathbf{s}}$ be the induced joint distribution over $\mathcal{M}_n\times\mathcal{Z}^n$, which is given by (see \eqref{EQ:stochastic_encoder_gamma})
\begin{subequations}
\begin{equation}
P^{(\gamma,\mathbf{s})}_{M,\mathbf{Z}_\mathbf{s}}(m,\mathbf{z})=P_M(m)\frac{1}{|\mathcal{W}_n|}\sum_{w\in\mathcal{W}_n}V_\mathbf{s}^n\big(\mathbf{z}\big|\mathbf{x}(m,w,\gamma)\big).
\end{equation}
Accounting also for the random codebook construction, we define $G_n\sim\tilde{\mu}_n$ as a random variable taking values in $\tilde{\Gamma}_n$ and set
\begin{equation}
P^{(\mathbf{s})}_{G_n,M,\mathbf{Z}_\mathbf{s}}(\gamma,m,\mathbf{z})=\tilde{\mu}_n(\gamma)P^{(\gamma,\mathbf{s})}_{M,\mathbf{Z}_\mathbf{s}}(m,\mathbf{z}).
\end{equation}\label{EQ:achievabiliy_induced_PMF}%
\end{subequations}

For any $\mathbf{s}\in\mathcal{S}^n$ and $\gamma\in\Gamma_n$, have
\begin{align*}
&\max_{\substack{\mathbf{s}\in\mathcal{S}_\mathcal{Q}^n,\\P_M\in\mathcal{P}(\mathcal{M}_n)}}D\Big(P^{(\mathbf{s})}_{\mathbf{Z}_\mathbf{s}|M,G_n=\gamma}\Big|\Big|P^{(\mathbf{s})}_{\mathbf{Z}_\mathbf{s}|G_n=\gamma}\Big|P_M\Big)\\
&\stackrel{(a)}\leq \max_{\substack{\mathbf{s}\in\mathcal{S}_\mathcal{Q}^n,\\P_M\in\mathcal{P}(\mathcal{M}_n)}}D\Big(P^{(\mathbf{s})}_{\mathbf{Z}_\mathbf{s}|M,G_n=\gamma}\Big|\Big|Q^n_{Z|S=\mathbf{s}}\Big|P_M\Big)\\
&=\max_{\substack{\mathbf{s}\in\mathcal{S}_\mathcal{Q}^n,\\P_M\in\mathcal{P}(\mathcal{M}_n)}}\sum_{m\in\mathcal{M}_n}P_M(m)D\Big(P^{(\mathbf{s})}_{\mathbf{Z}_\mathbf{s}|M=m,G_n=\gamma}\Big|\Big|Q^n_{Z|S=\mathbf{s}}\Big)\\
&\leq \max_{\substack{\mathbf{s}\in\mathcal{S}_\mathcal{Q}^n,\\m\in\mathcal{M}_n}}D\Big(P^{(\mathbf{s})}_{\mathbf{Z}_\mathbf{s}|M=m,G_n=\gamma}\Big|\Big|Q^n_{Z|S=\mathbf{s}}\Big),\numberthis\label{EQ:ACWTC_security_prop1}
\end{align*}
where (a) is because for any $\mathbf{s}\in\mathcal{S}_\mathcal{Q}^n$ and $P_M\in\mathcal{P}(\mathcal{M}_n)$
\begin{align*}
&D\left(P^{(\mathbf{s})}_{\mathbf{Z}_\mathbf{s}|M,G_n=\gamma}\Big|\Big|P^{(\mathbf{s})}_{\mathbf{Z}_\mathbf{s}|G_n=\gamma}\Big|P_M\right)\\
&=D\mspace{-1mu}\Big(P^{(\mathbf{s})}_{\mathbf{Z}_\mathbf{s}|M,G_n=\gamma}\Big|\Big|Q^n_{Z|S=\mathbf{s}}\Big|P_M\Big)\mspace{-2mu}-\mspace{-2mu}D\mspace{-1mu}\Big(\mspace{-1mu}P^{(\mathbf{s})}_{\mathbf{Z}_\mathbf{s}|G_n=\gamma}\Big|\Big|Q^n_{Z|S=\mathbf{s}}\Big)\\
&\leq D\left(P^{(\mathbf{s})}_{\mathbf{Z}_\mathbf{s}|M,G_n=\gamma}\Big|\Big|Q^n_{Z|S=\mathbf{s}}\Big|P_M\right).\numberthis\label{EQ:AVWTC_security_divergence_grows}
\end{align*}

Furthermore, by Lemma \ref{LEMMA:soft_covering}, for any $m\in\mathcal{M}_n$ and $\mathbf{s}\in\mathcal{S}^n$ with empirical PMF $\nu_\mathbf{s}$, taking $\tilde{R}> I_{\nu_\mathbf{s}Q}(X;Z|S)+\zeta$ for any $\zeta>0$, implies that there exist $\gamma_1,\gamma_2>0$ (uniform in $\mathbf{s}$), such that for $n$ large enough
\begin{equation}
        \mathbb{P}_{\tilde{\mu}_n}\bigg(D\Big(P^{(\mathbf{s})}_{\mathbf{Z}_\mathbf{s}|M=m,G_n}\Big|\Big|Q^n_{Z|S=\mathbf{s}}\Big)> e^{-n\gamma_1}\bigg)\leq e^{- e^{n\gamma_2}}.\label{EQ:AVWTC_security_soft_covering}
\end{equation}
As all $\mathbf{s}\in\mathcal{S}_\mathcal{Q}^n$ have $\nu_\mathbf{s}\in\mathcal{Q}$, setting
\begin{equation}
\tilde{R}=\max_{Q\in\mathcal{Q}}I_Q(X;Z|S)+\frac{\delta}{2}\label{EQ:secrecy_rate}
\end{equation}
gives \eqref{EQ:AVWTC_security_soft_covering} for every $\mathbf{s}\in\mathcal{S}_\mathcal{Q}^n$.

Now, with respect to the collection of the i.i.d. random variables $\big\{G_n(k)\big\}_{k=1}^{K_n}$ from before (which are, in fact, i.i.d. copies of $G_n$), with $K_n=n^3$, we have that for $n$ sufficiently large\footnote{In the following chain of inequalities, we consider conditional marginal distributions of a joint distribution $P^{(\mathbf{s})}_{G_n(k),M,W,\mathbf{X},\mathbf{Z}_\mathbf{s}}$, where $k\in[1:K_n]$, each defined exactly like in \eqref{EQ:achievabiliy_induced_PMF}, but with $G_n(k)$ in the role of $G_n$.}
\begin{align*}
&\mathbb{P}_{\tilde{\mu}_n^n}\Vast(\mspace{-2mu}\max_{\substack{k\in[1:K_n],\\\mathbf{s}\in\mathcal{S}_\mathcal{Q}^n,\\P_M\in\mathcal{P}(\mathcal{M}_n)}}\mspace{-2mu}D\Big(P^{(\mathbf{s})}_{\mathbf{Z}_\mathbf{s}|M,G_n(k)}\Big|\Big|P^{(\mathbf{s})}_{\mathbf{Z}_\mathbf{s}|G_n(k)}\Big|P_M\Big)\mspace{-1mu}>\mspace{-1mu}e^{-n\gamma_1}\mspace{-2mu}\Vast)\\
&\stackrel{(a)}\leq\mathbb{P}_{\tilde{\mu}_n^n}\left(\max_{\substack{k\in[1:K_n],\\\mathbf{s}\in\mathcal{S}_\mathcal{Q}^n,\\m\in\mathcal{M}_n}}D\Big(P^{(\mathbf{s})}_{\mathbf{Z}_\mathbf{s}|M=m,G_n(k)}\Big|\Big|Q^n_{Z|S=\mathbf{s}}\Big)>e^{-n\gamma_1}\right)\\
    &\stackrel{(b)}\leq\sum_{k=1}^{K_n}\sum_{\mathbf{s}\in\mathcal{S}_\mathcal{Q}^n}\sum_{m\in\mathcal{M}_n}\mspace{-5mu}\mathbb{P}_{\tilde{\mu}_n}\mspace{-2mu}\bigg(\mspace{-2mu}D\Big(P^{(\mathbf{s})}_{\mathbf{Z}_\mathbf{s}|M=m,G_n}\Big|\Big|Q^n_{Z|S=\mathbf{s}}\Big)\mspace{-3mu}>\mspace{-3mu}e^{-n\gamma_1}\mspace{-3mu}\bigg)\\
    &\stackrel{(c)}\leq n^3\cdot |\mathcal{S}|^n\cdot2^{nR}\cdot e^{-e^{n\gamma_2}}\\
    &\triangleq\kappa_n^{(2)},\numberthis\label{EQ:AVWTC_security_prob_UB}
\end{align*}
where (a) uses \eqref{EQ:ACWTC_security_prop1}, (b) follows by the union bound and because $\big\{G_n(k)\big\}_{k=1}^{K_n}$ being i.i.d. copies of $G_n\sim\tilde{\mu}_n$, which implies
\begin{align*}
&\mathbb{P}_{\tilde{\mu}_n^n}\mspace{-3mu}\bigg(\mspace{-2mu}D\Big(P^{(\mathbf{s})}_{\mathbf{Z}_\mathbf{s}|M=m,C_n(k)}\Big|\Big|Q^n_{Z|S=\mathbf{s}}\Big)>e^{-n\gamma_1}\bigg)\\
&=\mathbb{P}_{\tilde{\mu}_n}\mspace{-3mu}\bigg(\mspace{-2mu}D\Big(P^{(\mathbf{s})}_{\mathbf{Z}_\mathbf{s}|M=m,C_n}\Big|\Big|Q^n_{Z|S=\mathbf{s}}\Big)\mspace{-3mu}>\mspace{-3mu}e^{-n\gamma_1}\mspace{-4mu}\bigg),\ \ \forall k\mspace{-2mu}\in\mspace{-2mu}[1\mspace{-3mu}:\mspace{-3mu}K_n],
\end{align*}
while (c) is by \eqref{EQ:AVWTC_security_soft_covering}-\eqref{EQ:secrecy_rate}. The double-exponential decay of probability that Lemma \ref{LEMMA:soft_covering} provides yields $\kappa_n^{(2)}\to 0$ as $n\to\infty$.


\textbf{Realization Extraction and CR-code Construction:} As long as the rate constraints in \eqref{EQ:reliability_rate} and \eqref{EQ:secrecy_rate} hold, Equations \eqref{EQ:reduced_CR_bad_probability_bound} and \eqref{EQ:AVWTC_security_prob_UB} along with the Selection Lemma from \cite[Lemma 5]{Goldfeld_WTCII_semantic2015}, imply the existence of a realization of $\big\{G_n(k)\big\}_{k=1}^{K_n}$, denoted by $\big\{\gamma_k\big\}_{k=1}^{K_n}$, that for any $n$ sufficiently large satisfies
\begin{subequations}
\begin{align}
\max_{\substack{\mathbf{s}\in\mathcal{S}_\mathcal{Q}^n,\\(m,w)\in\mathcal{M}_n\times\mathcal{W}_n}}\frac{1}{K_n}\sum_{k=1}^{K_n} e_{m,w}\big(W_\mathbf{s}^n,c_n(k)\big)&\leq \frac{2c_{\delta,\nu}}{n}\label{EQ:reduced_CR_realization1_rel}\\
\max_{\substack{k\in[1:K_n],\\\mathbf{s}\in\mathcal{S}_\mathcal{Q}^n,\\P_M\in\mathcal{P}(\mathcal{M}_n)}}D\Big(P^{(\gamma_k,\mathbf{s})}_{\mathbf{Z}_\mathbf{s}|M}\Big|\Big|P^{(\gamma_k,\mathbf{s})}_{\mathbf{Z}_\mathbf{s}}\Big|P_M\Big)&\leq e^{-n\gamma_1}.\label{EQ:reduced_CR_realization1_sec}
\end{align}\label{EQ:reduced_CR_realization1}%
\end{subequations}
where, as defined in the CR-reduction part of the proof, $c_n(k)\triangleq c_n^{(\gamma_k)}$. 

Set $\Gamma_n=[1:K_n]$, $\mathcal{C}_n\triangleq\big\{c_n(k)\big\}_{k\in\Gamma_n}$ and $\mu_n(k)=K_n^{-1}$. Associating a CR $(n,M_n,K_n)$-code $\mathsf{C}_n$ with $\Gamma_n$, $\mathcal{C}_n$ and $\mu_n$, \eqref{EQ:reduced_CR_realization1_sec} is clearly equivalent to 
\begin{equation}
\mathcal{L}_{\mathrm{Sem}}(\mathfrak{V}^n,\mathcal{Q},\mathsf{C}_n)\leq e^{-n\gamma_1}.\label{EQ:AVWTC_security_established_reduced}
\end{equation}
Next, since for every $\mathbf{s}\in\mathcal{S}_\mathcal{Q}^n$ and $(m,w)\in\mathcal{M}_n\times\mathcal{W}_n$, we have
\begin{align*}
\mathcal{E}_{m,w}(W_\mathbf{s}^n,\mathsf{C}_n)&\geq\sum_{k\in\Gamma_n}\mspace{-3mu}\mu_n(k)\mspace{-4mu}\sum_{\mathbf{x}\in\mathcal{X}^n}\mspace{-3mu}f_{\gamma_k}(\mathbf{x}|m)\mspace{-11mu}\sum_{\substack{\mathbf{y}\in\mathcal{Y}^n:\\\phi^{(n)}_{\gamma_k}(\mathbf{y})\neq m}}\mspace{-9mu}W_\mathbf{s}^n(\mathbf{y}|\mathbf{x})\\
&=\mathcal{E}_{m}(W_\mathbf{s}^n,\mathsf{C}_n),\numberthis
\end{align*}
\eqref{EQ:reduced_CR_realization1_rel} implies
\begin{equation}
\mathcal{E}(\mathfrak{W}^n,\mathcal{Q},\mathsf{C}_n)=\max_{\substack{\mathbf{s}\in\mathcal{S}_\mathcal{Q}^n,\\m\in\mathcal{M}_n}}\mathcal{E}_m(W_\mathbf{s}^n,\mathsf{C}_n)\leq\frac{2c_{\delta,\nu}}{n}.\label{EQ:m_reliability_established_reduced}
\end{equation}
The proof is concluded by combining \eqref{EQ:reliability_rate} with \eqref{EQ:secrecy_rate} to eliminate $\tilde{R}$, which leaves us with 
\begin{equation}
\frac{1}{n}\log M_n\leq \min_{Q_1\in\mathcal{Q}}I_{Q_1}(X;Y)-\max_{Q_2\in\mathcal{Q}}I_{Q_2}(X;Z|S)-\delta,
\end{equation}
\eqref{EQ:AVWTC_security_established_reduced} and \eqref{EQ:m_reliability_established_reduced}. As $\delta>0$ was arbitrary, this implies the existence of a sufficiently large $n$ for which \eqref{EQ:CR_achievability} is satisfied.



\begin{remark}[Relation to Uncorrelated SS-Capacity] Recall that $K_n=n^3$, i.e., our reduced CR-code $\mathsf{C}_n$ is only polynomial in size. This has implication to the uncorrelated scenario because if the uncorrelated SS-capacity is strictly positive, one may replace the shared randomness between the legitimate parties with local randomness at the transmitter (which is always available in WTC scenarios). In a CR-code, the shared randomness is used for selecting which code $c_n(\gamma)$, where $\gamma\in\Gamma_n$, from $\mathcal{C}_n$ will be employed thorough the transmission. Instead, the transmitter may select $\gamma\in\Gamma_n$ and communicate it to the receiver as a prefix. Since $K_n=n^3$, the positivity of the uncorrelated capacity ensures the reliable transmission of $\gamma$ with a vanishing rate. A condition that differentiates between $\mathcal{Q}$-constrained AVWTCs with zero and non-zero uncorrelated capacities and a dichotomy result (stating that the uncorrelated capacity is either zero or equal to the CR-assisted capacity) are thus the missing pieces in telling whether the RHS of \eqref{EQ:AVWTC_CR_unconstrained_achievability} lower bounds the uncorrelated SS-capacity of a given AVWTC. Such a dichotomy result \cite{MolavianJazi_AVWTC_thesis2009} based on a certain threshold property (namely, the symmetrizability of the main AVC) \cite{Boche_superactive_AVWTC2016} is known for the scenario with unconstrained states. Therefore, Theorem \ref{TM:AVWTC_CR_achievability_general} holds for uncorrelated codes when $\mathcal{Q}=\mathcal{P}(\mathcal{S})$ and the considered AVWTC satisfies the condition from \cite{Boche_superactive_AVWTC2016} for having a positive uncorrelated SS-capacity.
\end{remark}


\section{Proof for Theorem~\ref{TM:AVWTC_CR_converse_general}}\label{SEC:AVWTC_CR_converse_general_proof}

Fix $\emptyset\neq\mathcal{Q}\subset\mathcal{P}(\mathcal{S})$ (if $\mathcal{Q}=\emptyset$ there is nothing to prove) and assume without loss of generality that $\mathcal{Q}\subseteq\mathcal{P}_\mathbb{Q}(\mathcal{S})$, i.e., that it contains only rational PMFs. Otherwise, the CR-assisted SS-capacity being a monotone non-increasing function of the constraint set (see Remark \ref{REM:no_rational_PMF}), implies
\begin{equation}
C_\mathrm{R}(\mathfrak{W},\mathfrak{V},\mathcal{Q})\leq C_\mathrm{R}\big(\mathfrak{W},\mathfrak{V},\mathcal{Q}\cap\mathcal{P}_\mathbb{Q}(\mathcal{S})\big). 
\end{equation}
Let $R\in\mathbb{R}_+$ be an achievable CR-assisted SS-rate for the $\mathcal{Q}$-constrained AVWTC $(\mathfrak{W}^n,\mathfrak{V}^n,\mathcal{Q})$. Then, for all $\epsilon>0$ and sufficiently large $n$, there exists a CR $(n,M_n,K_n)$-code $\mathsf{C}_n$ that satisfies \eqref{EQ:CR_achievability}. To get the max-inf upper bound of Theorem \ref{TM:AVWTC_CR_converse_general}, we derive an upper bound on $C_\mathrm{R}(\mathfrak{W},\mathfrak{V},\mathcal{Q})$ that is uniform in $Q_S\in\mathcal{Q}$.


Fix $\epsilon>0$ and let $\mathsf{C}_n$ be the corresponding CR $(n,M_n,K_n)$-code that satisfies \eqref{EQ:CR_achievability} for some sufficiently large $n$. Further let $\mathcal{C}_n=\big\{c_n(\gamma)\big\}_{\gamma\in\Gamma_n}$, where $|\Gamma_n|=K_n$, and $\mu_n\in\mathcal{P}(\Gamma_n)$ be the associated family of $(n,M_n)$-codes and the PMF over this family, respectively. For any $Q_S\in\mathcal{Q}$ the reliability and the security constrains for achievability stated in \eqref{EQ:CR_achievability_reliability}-\eqref{EQ:CR_achievability_security} continue to hold when restricting the state sequences to $\mathcal{T}^n_{Q_S}$ (instead of allowing any $\mathbf{s}\in\mathcal{S}^n$ with empirical PMF $\nu_\mathbf{s}\in\mathcal{Q}$). Thus, for any $Q_S\in\mathcal{Q}$ and  sufficiently large $n$ we have
\begin{subequations}
\begin{align}
\mspace{-8mu}\max_{\substack{\mathbf{s}\in\mathcal{T}^n_{Q_S},\\m\in\mathcal{M}_n}}\mathcal{E}_m(W_\mathbf{s}^n,\mathsf{C}_n)&\leq\mathcal{E}(\mathfrak{W}^n,\mathcal{Q},\mathsf{C}_n)\leq\epsilon\label{EQ:reliability_typical_type}\\
\mspace{-8mu}\max_{\substack{\gamma\in\Gamma_n,\\\mathbf{s}\in\mathcal{T}^n_{Q_S},\\P_M\in\mathcal{P}(\mathcal{M}_n)}}\mspace{-17mu}\ell\big(V_\mathbf{s}^n,P_M,c_n(\gamma)\big)&\leq\mathcal{L}_\mathrm{Sem}(\mathfrak{V}^n,\mathcal{Q},\mathsf{C}_n)\leq\epsilon.\label{EQ:security_typical_type}
\end{align}\label{EQ:reliability_security_typical_type}%
\end{subequations}
Although the value of $n$ beyond which \eqref{EQ:reliability_security_typical_type} becomes valid may depend on $\mathcal{Q}$, it is independent of any certain $Q_S\in\mathcal{Q}$.

Fix $Q_S\in\mathcal{Q}$ and recall that if $n\in\mathbb{N}$ is a blocklength for which $\mathcal{T}^n_{Q_S}=\emptyset$, then \eqref{EQ:CR_achievability_reliability}-\eqref{EQ:CR_achievability_security} are trivially satisfied. We avoid these trivial blocklengths by henceforth only considering values of $n$ that belong to $\mathbb{N}_\ell$, as defined in Section \ref{SUBSUBSEC:multi_letter_comparison}. To remind the reader, $\mathbb{N}_\ell \triangleq \big\{n\cdot\ell \big|n\in\mathbb{N}\big\}$, where $\ell$ is the least common denominator of all the non-zero entries of $Q_S$.


Since for any $Q_S\in\mathcal{Q}$, \eqref{EQ:reliability_typical_type} ensures that $\mathcal{E}_m\big(W_\mathbf{s}^n,\mathsf{C}_n\big)\leq\epsilon$, for all $\mathbf{s}\in\mathcal{T}^n_{Q_S}$ and $m\in\mathcal{M}_n$, we have
\begin{align*}
\bar{\mathcal{E}}(\mathfrak{W}^n,Q_S,\mathsf{C}_n)\mspace{-2mu}&\triangleq\mspace{-2mu}\max_{\mathbf{s}\in\mathcal{T}^n_{Q_S}}\sum_{\gamma\in\Gamma_n}\mspace{-4mu}\mu_n(\gamma)\frac{1}{M_n}\mspace{-5mu}\sum_{m\in\mathcal{M}_n}\mspace{-8mu}e_m\big(W_\mathbf{s}^n,c_n(\gamma)\big)\\
&\leq\epsilon,\numberthis\label{EQ:CR_achievability_average_error}
\end{align*}
for $n$ large enough. Similarly, by \eqref{EQ:security_typical_type} it also holds that
\begin{equation}
\mathcal{L}(\mathfrak{V}^n,Q_S,\mathsf{C}_n)\triangleq \max_{\mathbf{s}\in\mathcal{T}^n_{Q_S}}\sum_{\gamma\in\Gamma_n}\mu_n(\gamma)\ell\big(V_\mathbf{s}^n,P^{(U)}_M,c_n(\gamma)\big)\leq \epsilon,\label{EQ:CR_achievability_strong_sec}
\end{equation}
where $P^{(U)}_M$ is the uniform PMF on $\mathcal{M}_n$. In other words, a small maximal error probability and SS (for all the codes in $\mathcal{C}_n$) imply small average error probability and strong secrecy (when taking the expectation over the ensemble $\mathcal{C}_n$).

For any $\gamma\in\Gamma_n$ let $\Upsilon^{(\gamma)}$ be a PMF on $\mathcal{S}^n\times\mathcal{M}_n\times\mathcal{X}^n\times\mathcal{Y}^n\times\mathcal{Z}^n\times\hat{\mathcal{M}_n}$ defined by 
\begin{align*}
\Upsilon^{(\gamma)}_{\mathbf{S},M,\mathbf{X},\mathbf{Y},\mathbf{Z},\hat{M}}&(\mathbf{s},m,\mathbf{x},\mathbf{y},\mathbf{z},\hat{m})\\&\mspace{-20mu}\triangleq Q_S^n(\mathbf{s})P^{(\gamma,\mathbf{s})}_{M,\mathbf{X},\mathbf{Y}_\mathbf{s},\mathbf{Z}_\mathbf{s},\hat{M}}(m,\mathbf{x},\mathbf{y},\mathbf{z},\hat{m}),\numberthis\label{EQ:iid_state_PMF}
\end{align*}
where $P^{(\gamma,\mathbf{s})}$ is an abbreviation of $P^{(c_n(\gamma),\mathbf{s})}$ from \eqref{EQ:AVWTC_induced_PMF} and we set $P^{(\gamma,\mathbf{s})}_M=P^{(U)}_M$, for all $\gamma\in\Gamma_n$ and $\mathbf{s}\in\mathcal{S}^n$. Thus, for every $\mathbf{s}\in\mathcal{S}^n$ with $Q_S^n(\mathbf{s})>0$ and any $\gamma\in\Gamma_n$, the conditional PMF of $\Upsilon^{(\gamma)}$ given $\mathbf{S}=\mathbf{s}$ equals the corresponding induced PMF $P^{(\gamma,\mathbf{s})}$. 
Furthermore, let $C_n$ be a random variable that describes the choice of an $(n,M_n)$-code $c_n(\gamma)$, $\gamma\in\Gamma_n$, from the family $\mathcal{C}_n$ according to the distribution $\mu_n$. We now set
\begin{subequations}
\begin{align*}
&\Upsilon_{C_n,\mathbf{S},M,\mathbf{X},\mathbf{Y},\mathbf{Z},\hat{M}}\big(c_n(\gamma),\mathbf{s},m,\mathbf{x},\mathbf{y},\mathbf{z},\hat{m}\big)\\&\mspace{70mu}\triangleq \mu_n(\gamma)\Upsilon^{(\gamma)}_{\mathbf{S},M,\mathbf{X},\mathbf{Y},\mathbf{Z},\hat{M}}(\mathbf{s},m,\mathbf{x},\mathbf{y},\mathbf{z},\hat{m})\numberthis\label{EQ:averaged_upsilon}\\
&P^{(\mathbf{s})}_{C_n,M,\mathbf{X},\mathbf{Y}_\mathds{s},\mathbf{Z}_\mathds{s},\hat{M}}\big(c_n(\gamma),m,\mathbf{x},\mathbf{y},\mathbf{z},\hat{m})\\&\mspace{100mu}\triangleq \mu_n(\gamma)P^{(\gamma,\mathbf{s})}_{M,\mathbf{X},\mathbf{Y}_\mathds{s},\mathbf{Z}_\mathds{s},\hat{M}}(m,\mathbf{x},\mathbf{y},\mathbf{z},\hat{m}\big).\numberthis\label{EQ:averaged_P}
\end{align*}\label{EQ:averaged_upsilon_P}%
\end{subequations}
Henceforth, we use $I_\Upsilon(\cdot)$ and $I_P(\cdot)$ to indicate that a mutual information term is calculated with respect to $\Upsilon$ or $P^{(\mathbf{s})}$ from \eqref{EQ:averaged_upsilon_P}. We now present three technical lemmas that are essential in establishing the result of Theorem \ref{TM:AVWTC_CR_converse_general}. For the proofs of Lemmas \ref{LEMMA:typical_type}, \ref{LEMMA:Upsilon_security} and \ref{LEMMA:average_channel_reliability} see Appendices \ref{APPEN:typical_type_proof}, \ref{APPEN:Upsilon_security_proof} and \ref{APPEN:average_channel_reliability_proof}, respectively. 

\begin{lemma}[Leakage under Typical State Sequence]\label{LEMMA:typical_type} For any $Q_S\in\mathcal{P}(\mathcal{S})$, $\alpha\in (0,1]$, $n\in\mathbb{N}_\ell$ and $\mathbf{s}_1\in\mathcal{T}^n_\alpha(Q_S)$, there exists $\mathbf{s}_2\in\mathcal{T}^n_{Q_S}$, such that
\begin{align*}
\Big|I_\Upsilon(M;Z^n|S^n=\mathbf{s}_1,C_n)-I_\Upsilon(M;Z^n&|S^n=\mathbf{s}_2,C_n)\Big|\\
&\leq n\alpha\log|\mathcal{Z}|.\numberthis\label{EQ:information_difference_typical_type}
\end{align*}
\end{lemma}

\begin{lemma}[Average Leakage under $\bm{\Upsilon}$]\label{LEMMA:Upsilon_security}
For any $Q_S\in\mathcal{Q}$, $\alpha\in (0,1]$ and $n\in\mathbb{N}_\ell$ sufficiently large that is independent of $Q_S$ and $\alpha$, the following relation holds
\begin{equation}
I_\Upsilon(M;Z^n|S^n,C_n)\leq n\eta_{n,\alpha}^{(1)},\label{EQ:secrecy_averaged_channel}
\end{equation}
where $\eta_{n,\alpha}^{(1)}\triangleq \frac{\epsilon}{n}+\log|\mathcal{Z}|\left(\alpha+2|\mathcal{S}|e^{-2n\frac{\alpha^2}{|\mathcal{S}|^2}}\right)$.
\end{lemma}

Recall the definition of the averaged DMC $W_{Q}:\mathcal{X}\to\mathcal{P}(\mathcal{Y})$ from \eqref{EQ:reliability_analysis_WQ}, given by
\begin{equation}
W_{Q}(y|x)=\sum_{s\in\mathcal{S}}Q_S(s)W_s(y|x),\quad\forall(x,y)\in\mathcal{X}\times\mathcal{Y}.
\end{equation}
The $n$-fold extension of $W_{Q}$ satisfies
\begin{align*}
W_{Q}^n(\mathbf{y}|\mathbf{x})&=\prod_{i=1}^n\sum_{s\in\mathcal{S}}Q_S(s)W_s(y_i|x_i)\\
&=\sum_{\mathbf{s}\in\mathcal{S}^n}Q_S^n(\mathbf{s})W^n_\mathbf{s}(\mathbf{y}|\mathbf{x})\\
&=\Upsilon(\mathbf{y}|\mathbf{x}),\quad\forall(\mathbf{x},\mathbf{y})\in\mathcal{X}^n\times\mathcal{Y}^n.\numberthis
\end{align*}
Thus, the conditional marginal PMF $\Upsilon_{\mathbf{Y}|\mathbf{X}}$ of $\Upsilon$ from \eqref{EQ:averaged_upsilon_P} describes an $n$-length block transmission over the average channel $W_{Q}^n$. As subsequently shown, the derivation of our single-letter upper bound relies on the normalized equivocation of the message $M$ given an output sequence $Y^n$ of the average DMC $W_Q^n$ being small. Commonly, Fano's inequality implies that quantities such as $\frac{1}{n}H(M|Y^n)$ can be made arbitrarily small with $n$. Here, however, this equivocation term is not directly related to the performance criteria defining CR-assisted achievability. A brute force application of Fano's inequality based on  \eqref{EQ:reliability_typical_type} gives
\begin{equation}
\max_{\mathbf{s}\in\mathcal{T}^n_{Q_S}}H_P(M|Y^n_\mathbf{s},C_n)\leq 1+\epsilon\log M_n.\label{EQ:Fano_insufficient}
\end{equation}
However, it remains to be shown that \eqref{EQ:Fano_insufficient} implies that $\frac{1}{n}H_\Upsilon(M|Y^n,C_n)$ is small. In general, for any $\mathbf{s}\in\mathcal{T}^n_{Q_S}$, the channel $W_\mathbf{s}\in\mathfrak{W}$ is at least as good as $W_Q$, meaning that the averaged channel induces a possibly larger equivocation. Nonetheless, the equivocation of the message given the output sequence of $W_Q^n$ is upper bounded in Lemma \ref{LEMMA:average_channel_reliability}.

\begin{lemma}[Equivocation under Averaged Channel]\label{LEMMA:average_channel_reliability}
For any $Q_S\in\mathcal{Q}$,  $\alpha\in\left(0,\frac{1}{2}\right]$ and $n\in\mathbb{N}_\ell$ sufficiently large that is independent of $Q_S$ and $\alpha$, the following relation holds
\begin{equation}
H_\Upsilon(M|Y^n,C_n)\leq n\eta_{n,\alpha}^{(2)},
\end{equation}
where $\eta_{n,\alpha}^{(2)}=\frac{1}{n}+\frac{1}{n}\log M_n\left(\epsilon+2|\mathcal{S}|e^{-2n\frac{\alpha^2}{|\mathcal{S}|^2}}\right)+\alpha\log|\mathcal{Y}|+2h(\alpha)+|\mathcal{S}|\frac{\log(n+1)}{n}$ and $h$ is the binary entropy function.
\end{lemma}

Fix $d\in\left(0,\frac{1}{2}\right)$ and let $\alpha_n=n^{-\left(\frac{1}{2}-d\right)}$, for $n\in\mathbb{N}_\ell$. Accordingly, $\big\{\alpha_n\big\}_{n\in\mathbb{N}_\ell}$ vanishes to 0 slower than $\frac{1}{\sqrt{n}}$, which means that for sufficiently large $n$, random noise is typical with respect to $\alpha_n$ with arbitrarily high probability. Furthermore, $\alpha_n\in\left(0,\frac{1}{2}\right]$, for every $n\geq 2^{\frac{2}{1-2d}}$, so replacing $\alpha$ from Lemmas \ref{LEMMA:Upsilon_security} and \ref{LEMMA:average_channel_reliability} with $\alpha_n$, for sufficiently large $n$ we have
\begin{subequations}
\begin{align}
I_\Upsilon(M;Z^n|S^n,C_n)&\leq n\eta_{n}^{(1)}\\
H_\Upsilon(M|Y^n,C_n)&\leq n\eta_{n}^{(2)},
\end{align}\label{EQ:final_equivocation_SS_UB}%
\end{subequations}
where 
\begin{subequations}
\begin{align}
\eta_n^{(1)}&\triangleq \eta_{n,\alpha_n}^{(1)}=\frac{\epsilon}{n}+\log|\mathcal{Z}|\left(\alpha_n+2|\mathcal{S}|e^{-\frac{2n^{2d}}{|S|^2}}\right)\\
\eta_n^{(2)}&\triangleq \eta_{n,\alpha_n}^{(2)}=\frac{1}{n}\mspace{-3mu}+\mspace{-3mu}\frac{1}{n}\log M_n \left(\mspace{-2mu}\epsilon\mspace{-3mu}+\mspace{-3mu}2|\mathcal{S}|e^{-\frac{2n^{2d}}{|S|^2}}\mspace{-2mu}\right)\mspace{-3mu}+\mspace{-3mu}\alpha_n\log|\mathcal{Y}|\nonumber\\
&\mspace{140mu}+2h(\alpha_n)+|\mathcal{S}|\frac{\log(n+1)}{n},
\end{align}\label{EQ:final_equivocation_SS_etas}%
\end{subequations}
and consequently $\lim_{n\to\infty}\eta^{(j)}_n=0$, for $j=1,2$. Note that the independence of $n$ and $\alpha$ is essential for applying the Lemmas with the vanishing sequence $\big\{\alpha_n\big\}_{n\in\mathbb{N}_\ell}$. Furthermore, \eqref{EQ:final_equivocation_SS_UB} uniformly hold for all $Q_S\in\mathcal{Q}$.

Having \eqref{EQ:final_equivocation_SS_UB}-\eqref{EQ:final_equivocation_SS_etas}, we proceed with upper bounding the achievable rate $R$. Unless explicitly stated otherwise, all subsequent information measures are taken with respect to $\Upsilon$, which is therefore omitted from the notation of mutual information. For any $Q_S$ and  $n\in\mathbb{N}_\ell$ sufficiently large (in particular, larger than $2^{\frac{2}{1-2d}}$), we have
\begin{align*}
\log& M_n\\
&\stackrel{(a)}\leq I(M;Y^n|C_n)-I(M;S^n,Z^n|C_n)+n\eta_n\\
&\begin{multlined}[b][.43\textwidth]\stackrel{(b)}=\sum_{i=1}^n\Big[I(M;Y^i,S_{i+1}^n,Z_{i+1}^n|C_n)\\-I(M;Y^{i-1},S_i^n,Z_i^n|C_n)\Big]+n\eta_n\end{multlined}\\
&\begin{multlined}[b][.43\textwidth]=\sum_{i=1}^n\Big[I(M;Y_i|Y^{i-1},S_{i+1}^n,Z_{i+1}^n,C_n)\\-I(M;S_i,Z_i|Y^{i-1},S_{i+1}^n,Z_{i+1}^n,C_n)\Big]+n\eta_n\end{multlined}\\   
&\stackrel{(c)}=\sum_{i=1}^n\Big[I(M;Y_i|V_i)-I(M;S_i,Z_i|V_i)\Big]+n\eta_n,\numberthis\label{EQ:UB_summation}
  \end{align*}
where:\\
(a) uses \eqref{EQ:final_equivocation_SS_UB} and the independence of $(M,S^n,C_n)$, while defining $\eta_n\triangleq \eta_n^{(1)}+\eta_n^{(2)}$;\\
(b) follows by a telescoping identity \cite[Equations (9) and (11)]{Kramer_telescopic2011} and the independence of $C_n$, $M$ and $S^n$;\\
(c) is by defining $V_i\triangleq(Y^{i-1},S_{i+1}^n,Z_{i+1}^n,C_n)$, for all $i\in[1:n]$. The identification of $V_i$ is uniform in $Q_S\in\mathcal{Q}$.

The bound in \eqref{EQ:UB_summation} is rewritten by introducing a time-sharing random variable $T$ that is uniformly distributed over the set $[1:n]$ and is independent of $(S^n,M,X^n,Y^n,Z^n)$:
\begin{align*}
&\frac{1}{n}\log M_n\\
    &\leq\frac{1}{n}\sum_{t=1}^n\Big[I(M;Y_t|V_t)-I(M;S_t,Z_t|V_t)\Big]+\eta_n\\
    &=\sum_{t=1}^n\mathbb{P}\big(T=t\big)\Big[I(M;Y_t|V_t)-I(M;S_t,Z_t|V_t)\Big]+\eta_n\\
    &= I(M;Y_T|V_T,T)-I(M;S_T,Z_T|V_T,T)+\eta_n.\numberthis\label{EQ:AVWTC_time_sharing_UB}
\end{align*}
Denoting $S_T\triangleq S$, $V\triangleq (V_T,T)$, $U\triangleq (M,V)$, $X\triangleq X_T$,  $Y\triangleq Y_T$ and $Z\triangleq Z_T$. Lemma \ref{LEMMA:upsilon_independence} establishes an independence property under $\Upsilon$, which is key in deriving the factorization property of the distribution of $(S,V,U,X,Y,Z)$ (induced by $\Upsilon$), stated in Lemma \ref{LEMMA:Q_PMF_converse}. Both lemmas are proven in Appendix \ref{APPEN:upsilon_independence_proof}.

\begin{lemma}\label{LEMMA:upsilon_independence}
For any $Q_S\in\mathcal{Q}$, $S_i$ and $(C_n,S^{n\backslash i},M,X^n,Y^{n\backslash i},Z^{n\backslash i})$ are independent under $\Upsilon$ from \eqref{EQ:averaged_upsilon}.
\end{lemma}


\begin{lemma}\label{LEMMA:Q_PMF_converse}
For any $Q_S\in\mathcal{Q}$ and $(s,v,u,x,y,z)\in\mathcal{S}\times\mathcal{V}\times\mathcal{U}\times\mathcal{X}\times\mathcal{Y}\times\mathcal{Z}$, where $\mathcal{V}$ and $\mathcal{U}$ are the alphabets that correspond to the definitions of $V$ and $U$ stated above, the following factorization holds
\begin{align*}
\mathbb{P}_\Upsilon(V&=v,U=u,X=x,S=s,Y=y,Z=z)\\
&=\mathbb{P}_\Upsilon(V=v,U=u,X=x)Q_S(s)W_s(y|x)V_s(z|x).\numberthis\label{EQ:Q_PMF_converse}
\end{align*}
\end{lemma}
Denoting $\mathbb{P}_\Upsilon(V=v,U=u,X=x)\triangleq Q_{V,U,X}(v,u,x)$, for all $(v,u,x)\in\mathcal{V}\times\mathcal{U}\times\mathcal{X}$, Lemma \ref{LEMMA:Q_PMF_converse} shows that the joint distribution of $(S,V,U,X,Y,Z)$ factors as stated in Theorem \ref{TM:AVWTC_CR_converse_general}.

Finally, we substitute $\eta_n=\eta_n^{(1)}+\eta_n^{(2)}$, while using the definition of $\eta_n^{(2)}$ and \eqref{EQ:CR_achievability_rate}, to get that for any $Q_S\in\mathcal{Q}$ and $n$ sufficiently large
\begin{align*}
R&<\frac{I_{Q}(U;Y|V)-I_{Q}(U;S,Z|V)}{1-\epsilon-2|\mathcal{S}|e^{-2n^{2d}}}\\
&+\frac{\eta^{(1)}_n+\epsilon+\frac{1}{n}+\alpha_n\log|\mathcal{Y}|+2h(\alpha_n)+|\mathcal{S}|\frac{\log(n+1)}{n}}{1-\epsilon-2|\mathcal{S}|e^{-\frac{2n^{2d}}{|S|^2}}}+\epsilon,\numberthis
\end{align*}
where $I_Q$ denotes that the underlying distribution of the mutual information terms is $Q_{V,U,X}Q_SQ_{Y|X,S}Q_{Z|X,S}$, where $Q_{Y|X,S}(y|x,s)=W_s(y|x)$ and $Q_{Z|X,S}(z|x,s)=V_s(z|x)$, for all $(s,x,y,z)\in\mathcal{S}\times\mathcal{X}\times\mathcal{Y}\times\mathcal{Z}$. Letting $n\to\infty$ (which takes $\alpha_n$ and $\eta_n^{(1)}$ to 0) and $\epsilon\to 0$ gives
\begin{equation}
R\leq I_{Q}(U;Y|V)-I_{Q}(U;S,Z|V),\quad\forall Q_S\in\mathcal{Q}.\label{EQ:type_converse}
\end{equation}
Taking an infimum of the RHS \eqref{EQ:type_converse} over all $Q_S\in\mathcal{Q}$ further gives
\begin{equation}
R\leq\inf_{Q_S\in\mathcal{Q}} \Big[I_{Q}(U;Y|V)-I_{Q}(U;S,Z|V)\Big].\label{EQ:type_converse_inf}
\end{equation}
Further upper bounding the RHS of \eqref{EQ:type_converse_inf} by maximizing it over all $Q_{V,U,X}\in\mathcal{P}(\mathcal{V}\times\mathcal{U}\times\mathcal{X})$
concludes the proof.

\section{Summary and Concluding Remarks}\label{SEC:summary}

We derived the CR-assisted SS-capacity of the AVWTC with type constrained states. The constraint allows only the state sequences whose empirical distribution is within a small gap from the prescribed type. Achievability relies on a general single-letter lower bound on the capacity of the $\mathcal{Q}$-constrained AVWTC that does not assume the existence of a best channel to the eavesdropper. To establish SS under each of the exponentially many possible states, the mutual information between the message and the eavesdropper's observations was shown to be negligible even when maximized over all message distributions, choices of state sequences and realizations of the CR-code. The SS analysis was based on a heterogeneous version of the strong soft-covering lemma that was recently presented in \cite{Goldfeld_WTCII_semantic2015}. The lemma showed that the probability (with respect to a randomly generated codebook) of the soft-covering phenomenon happening is doubly-exponentially close to one, when transmitting over a state-dependent channel with a certain state sequence realization. The condition for the above is that the rate of the codebook is above the conditional mutual information between the input and output given the state. An application of the union bound combined with a CR-code reduction argument (based on a Chernoff bound) then establishes SS. The resulting reliable and semantically-secure reduced CR-code is over a family of (uncorrelated) codes that is only polynomial in size.


The converse for the type constrained scenario used a general upper bound on $C_{\mathrm{R}}(\mathfrak{W},\mathfrak{V},\mathcal{Q})$. Derived uniformly over the constraint set, the upper bound has a max-inf form, and when specialized to a compound WTC over a corresponding constraint set, it improves upon the previously best known single-letter upper bound for that problem \cite[Theorem 2]{Liang_compoundWTC_Journal2007}. The proof of the upper bound showed that reliability and SS under all state sequences in any type-class imply similar performance when the state sequence is i.i.d. according to the type. The main challenge was in proving that the normalized equivocation of the message given the output sequence is negligible for outputs generated by the averaged main channel. This step required a continuity property that was derived via a novel distribution coupling argument. Combining our upper and lower bounds with some continuity arguments established the SS-capacity of the type constrained AVWTC. The formula has the structure of two subtracted mutual information terms. The first term suggests that the legitimate users effectively transmit over the averaged DMC, which is in general no better than any of the main channels associated with each state. The second (subtracted) mutual information term corresponds to ensuring secrecy versus an eavesdropper with perfect CSI.


Our main goal was to find a single-letter description of the admissible secrecy-rate in an AVWTC scenario while accounting for each of its exponential number of security constraints (instead of relying on assumptions that degenerate the scenario to a single dominating constraint). The heterogeneous strong soft-covering lemma allowed us to do just that, while upgrading to SS. Our achievability proof showed the existence of CR-assisted SS-capacity achieving CR-code of polynomial size. Consequently, combining our code construction with a condition that identifies whether a given type constrained AVWTC has zero or non-zero uncorrelated capacity, will suffice for characterizing the uncorrelated SS-capacity. Such a differentiating condition being currently unknown, we pose it as a question for future research.

\appendices


\section{Proof of Lemma \ref{LEMMA:delta_exists}}\label{APPEN:delta_exists_proof}

Let $Q_{U,X}\in\mathcal{P}(\mathcal{U}\times\mathcal{X})$ and denote 
\begin{subequations}
\begin{align}
\mathcal{I}(Q_S,Q_{U,X})&\triangleq I_{Q_S}(U;Y)-I_{Q_S}(U;Z|S),\\
\mathcal{I}_{\delta}(Q_S,Q_{U,X})&\triangleq \min_{Q_1\in\mathcal{P}_\delta(Q_S)}I_{Q_1}(U;Y)\nonumber\\
&\mspace{60mu}-\max_{Q_2\in\mathcal{P}_\delta(Q_S)}I_{Q_2}(U;Z|S),
\end{align}
\end{subequations}
where $I_Q$ stands for the mutual information term being calculated with respect to $Q$ as the state distribution. Note that for any $Q_{U,X}$ and $\delta>0$ we have
\begin{equation}
\mathcal{I}_{\delta}(Q_S,Q_{U,X})\leq \mathcal{I}(Q_S,Q_{U,X}),\label{EQ:achievability_rate_assumption_needed_ineq1}
\end{equation}
and therefore 
\begin{align*}
\mathcal{I}_{\delta}^*(Q_S)&\triangleq \max_{Q_{U,X}}\mathcal{I}_{\delta}(Q_S,Q_{U,X})\\
                           &\leq\max_{Q_{U,X}}\mathcal{I}(Q_S,Q_{U,X})\\
                           &=C^\star_{\mathrm{R}}(\mathfrak{W},\mathfrak{V},Q_S).\numberthis\label{EQ:achievability_rate_assumption_needed_ineq11}
\end{align*}

Fix $Q_{U,X}\in\mathcal{P}(\mathcal{U}\times\mathcal{X})$. The continuity of mutual information implies that for every $Q_1,Q_2\in\mathcal{P}_\delta(Q_S)$, we have
\begin{subequations}
\begin{align}
\Big|I_{Q_S}(U;Y)-I_{Q_1}(U;Y)\Big|&\leq f_1(\delta)\label{EQ:delta_continuity_y}\\
\Big|I_{Q_S}(U;Z|S)-I_{Q_2}(U;Z|S)\Big|&\leq f_2(\delta),\label{EQ:delta_continuity_z}
\end{align}\label{EQ:delta_continuity}%
\end{subequations}
where $\lim_{\delta\to 0}f_j(\delta)=0$, for $j=1,2$, uniformly in $Q_{U,X}$ (i.e., $f_1$ and $f_2$ are independent of $Q_{U,X}$). For any $\delta>0$, if $Q_1^\star\in\mathcal{P}_\delta(Q_S)$ achieves $\min_{Q_1\in\mathcal{P}_\delta(Q_S)}I_{Q_1}(U;Y)$, then \eqref{EQ:delta_continuity_y} implies
\begin{align*}
I_{Q_S}(U;Y)&\leq I_{Q^\star_1}(U;Y)+f_1(\delta)\\                                 
            &=\min_{Q_1\in\mathcal{P}_\delta(Q_S)}I_{Q_1}(U;Y)+f_1(\delta).\numberthis\label{EQ:Q_delta_relation_y}
\end{align*}
Similarly, we also have
\begin{equation}
I_{Q_S}(U;Z|S)\geq \max_{Q_2\in\mathcal{P}_\delta(Q_S)}I_{Q_2}(U;Z|S)-f_2(\delta).\label{EQ:Q_delta_relation_z}
\end{equation}

Combining \eqref{EQ:Q_delta_relation_y} and \eqref{EQ:Q_delta_relation_z} shows that
\begin{equation}
\mathcal{I}(Q_S,Q_{U,X})\leq \mathcal{I}_{\delta}(Q_S,Q_{U,X})+f_1(\delta)+f_2(\delta),
\end{equation}
for every $Q_{U,X}\in\mathcal{P}(\mathcal{U}\times\mathcal{X})$ and $\delta>0$, which, in turn, implies
\begin{equation}
C^\star_{\mathrm{R}}(\mathfrak{W},\mathfrak{V},Q_S)\leq \mathcal{I}_{\delta}^*(Q_S)+f_1(\delta)+f_2(\delta),\quad\forall\delta>0.\label{EQ:IQ_IQdelta_UB}
\end{equation}
Since $\lim_{\delta\to 0}f_j(\delta)=0$, for $j=1,2$, \eqref{EQ:achievability_rate_assumption_needed_ineq11} and \eqref{EQ:IQ_IQdelta_UB} produce~\eqref{EQ:delta_exists}.

\section{Proof of Lemma \ref{LEMMA:decoding}}\label{APPEN:decoding_proof}

Let $\eta>0$ be arbitrary and with respect to the random codebook $\mathsf{B}_n$, for every $\mathbf{y}\in\mathcal{Y}^n$, define the random variable
\begin{equation}
\Phi_n(\mathbf{y})=\begin{cases}(m,w),\quad\begin{array}{c}
     \max\limits_{\substack{(m',w')\in\mathcal{M}_n\times\mathcal{W}_n:\\(m',w')\neq(m,w)}}d\big(\mathbf{X}(m',w',\gamma),\mathbf{y}\big)\\\mspace{114mu}<d\big(\mathbf{X}(m,w,\gamma),\mathbf{y}\big)
\end{array}\\
e,\ \mbox{no $(m,w)$ as above exists}\end{cases}\mspace{-25mu}.\label{EQ:decoder_rv}
\end{equation}
Furthermore, for every $(m,w)\in\mathcal{M}_n\times\mathcal{W}_n$ and $\mathbf{y}\in\mathcal{Y}^n$, also set $Z(\mathbf{y},m,w)=\mathds{1}_{\big\{\Phi_n(\mathbf{y})\neq(m,w)\big\}}$. With respect to the measure $\tilde{\mu}_n$ from \eqref{EQ:mu_gamma}, we have
\begin{align*}
&\mathcal{E}_{m,w}(W_n,\tilde{\mathsf{C}}_n)\\
    &=\mathbb{E}_{\tilde{\mu}_n}\sum_{\mathbf{y}\in\mathcal{Y}^n}W_n\big(\mathbf{y}\big|\mathbf{X}(m,w)\big)Z(\mathbf{y},m,w)\\
    &\begin{multlined}[b][.45\textwidth]\stackrel{(a)}=\sum_{\mathbf{x}\in\mathcal{X}^n}Q_X^n(\mathbf{x})\sum_{\mathbf{y}\in\mathcal{Y}^n}W_n(\mathbf{y}|\mathbf{x})\\\times\mathbb{E}_{\tilde{\mu}_n}\Big[Z(\mathbf{y},m,w)\Big|\mathbf{X}(m,w)=\mathbf{x}\Big]\end{multlined}\\
    &\begin{multlined}[b][.45\textwidth]=\sum_{(\mathbf{x},\mathbf{y})\in\mathcal{X}^n\times\mathcal{Y}^n}Q_X^n(\mathbf{x})W_n(\mathbf{y}|\mathbf{x})\\\times\mathbb{P}_{\tilde{\mu}_n}\Big(\Phi_n(\mathbf{y})\neq(m,w)\Big|\mathbf{X}(m,w)=\mathbf{x}\Big)\end{multlined}\\
    &\begin{multlined}[b][.45\textwidth]\stackrel{(b)}=\sum_{(\mathbf{x},\mathbf{y})\in\mathcal{X}^n\times\mathcal{Y}^n}Q_X^n(\mathbf{x})W_n(\mathbf{y}|\mathbf{x})\\\times\mathbb{P}_{\tilde{\mu}_n}\left(\max_{\substack{(m',w')\in\mathcal{M}_n\times\mathcal{W}_n:\\(m',w')\neq(m,w)}}d\big(\mathbf{X}(m',w'),\mathbf{y}\big)\geq d(\mathbf{x},\mathbf{y})\right),\end{multlined}\numberthis\label{EQ:decoding_proof_prob}
\end{align*}
where (a) is the law of total expectation (by first taking a conditional expectation on $\mathbf{X}(m,w)$), while (b) uses the definition of $\Phi_n(\mathbf{y})$ and the independence of the random vectors in the collection $\mathsf{B}_n$.

For all $(\mathbf{x},\mathbf{y})\in\mathcal{X}^n\times\mathcal{Y}^n$ with $d(\mathbf{x},\mathbf{y})\geq\frac{|\mathcal{M}_n||\mathcal{W}_n|}{\eta}$, we upper bound the probability on the RHS of \eqref{EQ:decoding_proof_prob} as
\begin{align*}
&\mathbb{P}_{\tilde{\mu}_n}\left(\max_{\substack{(m',w')\in\mathcal{M}_n\times\mathcal{W}_n:\\(m',w')\neq(m,w)}}d\big(\mathbf{X}(m',w'),\mathbf{y}\big)\geq d(\mathbf{x},\mathbf{y})\right)\\
    &\leq \mathbb{P}_{\tilde{\mu}_n}\left(\max_{\substack{(m',w')\in\mathcal{M}_n\times\mathcal{W}_n:\\(m',w')\neq(m,w)}}d\big(\mathbf{X}(m',w'),\mathbf{y}\big)\geq \frac{|\mathcal{M}_n||\mathcal{W}_n|}{\eta}\right)\\
    &=\mathbb{P}_{\tilde{\mu}_n}\left(\mspace{-5mu}\bigcup_{\substack{(m',w')\in\mathcal{M}_n\times\mathcal{W}_n:\\(m',w')\neq(m,w)}}\mspace{-5mu}\left\{d\big(\mathbf{X}(m',w'),\mathbf{y}\big)\mspace{-3mu}\geq\mspace{-3mu} \frac{|\mathcal{M}_n||\mathcal{W}_n|}{\eta}\right\}\mspace{-5mu}\right)\\
    &\stackrel{(a)}\leq \sum_{\substack{(m',w')\in\mathcal{M}_n\times\mathcal{W}_n:\\(m',w')\neq(m,w)}}\mathbb{P}_{Q_X^n}\left(d\big(\mathbf{X},\mathbf{y}\big)\geq \frac{|\mathcal{M}_n||\mathcal{W}_n|}{\eta}\right)\\
    &\stackrel{(b)}\leq \eta\sum_{\substack{(m',w')\in\mathcal{M}_n\times\mathcal{W}_n:\\(m',w')\neq(m,w)}}\frac{\mathbb{E}_{Q_X^n}d\big(\mathbf{X},\mathbf{y}\big)}{|\mathcal{M}_n||\mathcal{W}_n|}\\
    &\stackrel{(c)}\leq \eta \numberthis\label{EQ:decoding_proof_prob_UB}
\end{align*}
where (a) uses the union bound and the fact that $\mathbf{X}(m',w')\sim Q_X^n$, for all $(m',w')\in\mathcal{M}_n\times\mathcal{W}_n$, (b) is Markov's inequality and (c) follows by the assumption that $\mathbb{E}_{Q_X^n}d(\mathbf{X},\mathbf{y})\leq1$, for all $\mathbf{y}\in\mathcal{Y}^n$.

Plugging \eqref{EQ:decoding_proof_prob_UB} back into \eqref{EQ:decoding_proof_prob} completes the proof:
\begin{align*}
\mathcal{E}_{m,w}(W_n,\tilde{\mathsf{C}}_n)&\leq\mspace{-10mu}\sum_{\substack{(\mathbf{x},\mathbf{y})\in\mathcal{X}^n\times\mathcal{Y}^n:\\d(\mathbf{x},\mathbf{y})<\frac{|\mathcal{M}_n||\mathcal{W}_n|}{\eta}}}\mspace{-14mu}Q_X^n(\mathbf{x})W_n(\mathbf{y}|\mathbf{x})\cdot 1\\&\mspace{60mu}+\mspace{-10mu}\sum_{\substack{(\mathbf{x},\mathbf{y})\in\mathcal{X}^n\times\mathcal{Y}^n:\\d(\mathbf{x},\mathbf{y})\geq\frac{|\mathcal{M}_n||\mathcal{W}_n|}{\eta}}}\mspace{-14mu}Q_X^n(\mathbf{x})W_n(\mathbf{y}|\mathbf{x})\cdot\eta\\
    &\leq\mathbb{P}_{Q_X^nW_n}\left(d(\mathbf{X},\mathbf{Y})<\frac{|\mathcal{M}_n||\mathcal{W}_n|}{\eta}\right)+\eta.\numberthis
\end{align*}

\section{Proof of Lemma \ref{LEMMA:pseudo_MI_geq_minimal_MI}}\label{APPEN:pseudo_MI_geq_minimal_MI_proof}
First define
\begin{equation}
  \mathfrak{W}_{\mathcal{Q}}\triangleq\Big\{W_Q:\mathcal{X}\to\mathcal{P}(\mathcal{Y})\Big|Q\in\mathcal{Q}\Big\},
\end{equation}
and note that the convexity of $\mathcal{Q}$ implies that $\mathfrak{W}_{\mathcal{Q}}$ is also a convex set. Throughout this proof we make use of a slightly modified notation of mutual information. Specifically, we represent the mutual information between a pair of random variables in terms of their underlying joint distribution, i.e., for any $P\in\mathcal{P}(\mathcal{X})$ and $W:\mathcal{X}\to\mathcal{P}(\mathcal{Y})$, let $I(P,W)\triangleq I(X;Y)$, where $(X,Y)\sim P\cdot W$. Accordingly, we may write
\begin{equation}
\min_{Q\in\mathcal{Q}}I_Q(X;Y)=\min_{W\in\mathfrak{W}_{\mathcal{Q}}}I(Q_X,W).\label{EQ:MI_alt_representation}
\end{equation}
Let $\tilde{W}\in\mathfrak{W}_{\mathcal{Q}}$ be a channel that achieves the RHS of \eqref{EQ:MI_alt_representation}, i.e., with respect to the notation in the error probability analysis from Section \ref{SEC:AVWTC_CR_achievability_general_proof}, we have
\begin{equation}
I\big(Q_X,\tilde{W}\big)=I_{\tilde{Q}}(X;Y).
\end{equation}
The convexity of $\mathfrak{W}_{\mathcal{Q}}$ implies that for every $W\in\mathfrak{W}_{\mathcal{Q}}$ and $\alpha\in[0,1]$
\begin{equation}
I\big(Q_X,\alpha W+(1-\alpha)\tilde{W}\big)\geq I\big(Q_X,\tilde{W}\big),
\end{equation}
and therefore,
\begin{equation}
\lim_{\alpha\searrow 0}\frac{\partial}{\partial \alpha}I\big(Q_X,\alpha W+(1-\alpha)\tilde{W}\big)\geq 0.
\end{equation}
Similarly to \cite[Equation (12.19)]{Csiszar_Korner_Book2011}, since 
\begin{align*}
\frac{\partial}{\partial \alpha}&I\big(Q_X,\alpha W+(1-\alpha)\tilde{W}\big)\\
&=\sum_{(x,y)\in\mathcal{X}\times\mathcal{Y}}Q_X(x)\big(W(y|x)-\tilde{W}(y|x)\big)\\&\mspace{60mu}\times\log\left(\frac{\alpha W(y|x)+(1-\alpha)\tilde{W}(y|x)}{\alpha Q_Y(y)+(1-\alpha)\tilde{Q}_Y(y)}\right),\numberthis
\end{align*}
where $Q_Y(y)=\sum_{x\in\mathcal{X}}Q_X(x)W(y|x)$ and $\tilde{Q}_Y(y)=\sum_{x\in\mathcal{X}}Q_X(x)\tilde{W}(y|x)$, it follows that
\begin{align*}
\sum_{(x,y)\in\mathcal{X}\times\mathcal{Y}}\mspace{-15mu}Q_X(x)W(y|x)\log\left(\frac{\tilde{W}(y|x)}{\tilde{Q}_Y(y)}\right)&\geq I\big(Q_X,\tilde{W}\big)\\
    &=I_{\tilde{Q}}(X;Y).\numberthis
\end{align*}

\ \\


\section{Proof of Lemma \ref{LEMMA:typical_type}}\label{APPEN:typical_type_proof}

Fix $Q_S\in\mathcal{Q}$, $\alpha\in(0,1]$, $n\in\mathbb{N}_\ell$ and $\mathbf{s}_1\in\mathcal{T}^n_\alpha(Q_S)$. Clearly, there exists an $\mathbf{s}_2\in\mathcal{T}^n_{Q_S}$, such that 
\begin{equation}
d_H(\mathbf{s}_1,\mathbf{s}_2)\leq n\alpha,\label{EQ:Hamming_distance}
\end{equation}
where $d_H:\mathcal{S}^n\times\mathcal{S}^n\to [0:n]$ is the Hamming distance function. Let $\mathcal{A}$ be the set of indices for which the components of $\mathbf{s}_1$ and $\mathbf{s}_2$ coincide, i.e.,
\begin{equation}
\mathcal{A}=\big\{i\in[1:n]\big|s_{1,i}=s_{2,i}\big\}.\label{EQ:set_s1s2_coincide}
\end{equation}
Note that \eqref{EQ:Hamming_distance} implies that $|\mathcal{A}^c|\leq n\alpha$.

Recall that for any subset $\emptyset\neq\mathcal{A}\subseteq[1:n]$ and any $n$-dimensional vector $\mathbf{x}\in\mathcal{X}^n$, we denote the vector of elements from $\mathbf{x}$ with indices in $\mathcal{A}$ by $\mathbf{x}^\mathcal{A}$, that is, $\mathbf{x}^\mathcal{A}=(x_i)_{i\in\mathcal{A}}$. Similar convention is used for random vectors, while using uppercase letters. By the mutual information chain rule, the absolute value of the difference of mutual information terms from \eqref{EQ:information_difference_typical_type} is upper bounded as follows:
\begin{align*}
&\big|I_\Upsilon(M;Z^n|S^n=\mathbf{s}_1,C_n)-I_\Upsilon(M;Z^n|S^n=\mathbf{s}_2,C_n)\big|\\
&\begin{multlined}[b][.39\textwidth]\leq\Big|I_\Upsilon\big(M;\mathbf{Z}^{\mathcal{A}}\big|S^n=\mathbf{s}_1,C_n\big)-I_\Upsilon\big(M;\mathbf{Z}^{\mathcal{A}}\big|S^n=\mathbf{s}_2,C_n\big)\Big|\\+\Big|I_\Upsilon\big(M;\mathbf{Z}^{\mathcal{A}^c}\big|\mathbf{Z}^{\mathcal{A}},S^n=\mathbf{s}_1,C_n\big)\\-I_\Upsilon\big(M;\mathbf{Z}^{\mathcal{A}^c}\big|\mathbf{Z}^{\mathcal{A}},S^n=\mathbf{s}_2,C_n\big)\Big|\end{multlined}\\
&\stackrel{(a)}\leq\Big|I_\Upsilon\big(M;\mathbf{Z}^{\mathcal{A}}\big|\mathbf{S}^{\mathcal{A}}=\mathbf{s}_1^{\mathcal{A}},C_n\big)\mspace{-3mu}-\mspace{-3mu}I_\Upsilon\big(M;\mathbf{Z}^{\mathcal{A}}\big|\mathbf{S}^{\mathcal{A}}=\mathbf{s}_2^{\mathcal{A}},C_n\big)\Big|\\
&\mspace{127mu}+\max\left\{\begin{array}{l}
     I_\Upsilon\big(M;\mathbf{Z}^{\mathcal{A}^c}\big|\mathbf{Z}^{\mathcal{A}},\mathbf{S}=\mathbf{s}_1,C_n\big), \\
     I_\Upsilon\big(M;\mathbf{Z}^{\mathcal{A}^c}\big|\mathbf{Z}^{\mathcal{A}},\mathbf{S}=\mathbf{s}_2,C_n\big)\end{array}\right\}\\
&\stackrel{(b)}\leq \max\Big\{H_\Upsilon\big(\mathbf{Z}^{\mathcal{A}^c}\big|S^n=\mathbf{s}_1\big),H_\Upsilon\big(\mathbf{Z}^{\mathcal{A}^c}\big|S^n=\mathbf{s}_2\big)\Big\}\\
&\stackrel{(c)}\leq n\alpha\log|\mathcal{Z}|\numberthis\label{EQ:MI_typical_type_differnece_UB}
\end{align*}
where:\\
(a) is because for any $m\in\mathcal{M}_n$, $\mathbf{z}^\mathcal{A}\in\mathcal{Z}^{|\mathcal{A}|}$ and $\mathbf{s}_j$, where $j=1,2$, \eqref{EQ:iid_state_PMF} implies
\begin{align*}
\Upsilon\big(m,\mathbf{z}^\mathcal{A}\big|\mathbf{s}_j\big)&=\sum_{\gamma\in\Gamma_n}\mu_n(\gamma)\frac{1}{M_n}\sum_{\mathbf{x}\in\mathcal{X}^n}f_\gamma(\mathbf{x}|m)V_{\mathbf{s}_j^{\mathcal{A}}}^{|\mathcal{A}|}\big(\mathbf{z}^\mathcal{A}\big|\mathbf{x}^\mathcal{A}\big)\\
&=\Upsilon\big(m,\mathbf{z}^\mathcal{A}\big|\mathbf{s}^\mathcal{A}_j\big)\numberthis\label{EQ:Upsilon_memoryless};
\end{align*}
(b) is since $\mathbf{s}_1^{\mathcal{A}}=\mathbf{s}_2^{\mathcal{A}}$, because conditioning cannot increase entropy and the memoryless property from \eqref{EQ:Upsilon_memoryless};\\
(c) holds since entropy is maximized by the uniform distribution and because $|\mathcal{A}^c|\leq n\alpha$.


\section{Proof of Lemma \ref{LEMMA:Upsilon_security}}\label{APPEN:Upsilon_security_proof}

Fix $Q_S\in\mathcal{Q}$ and $\alpha\in (0,1]$. First observe that for any $n\in\mathbb{N}_\ell$, we have
\begin{align*}
\max_{\mathbf{s}\in\mathcal{T}_\alpha^n(Q_S)}&I_\Upsilon(M;Z^n|S^n=\mathbf{s},C_n)\\
&\stackrel{(a)}\leq \max_{\mathbf{s}\in\mathcal{T}^n_{Q_S}}I_\Upsilon(M;Z^n|S^n=\mathbf{s},C_n)+n\alpha\log|\mathcal{Z}|\\
&\stackrel{(b)}=\mathcal{L}(\mathfrak{V}^n,Q_S,\mathsf{C}_n)+n\alpha\log|\mathcal{Z}|,\numberthis\label{EQ:typical_type_UB}
\end{align*}
where (a) follows from Lemma \ref{LEMMA:typical_type}, while (b) is because $\Upsilon_{M,\mathbf{Z}|\mathbf{S}=\mathbf{s}}=P^{(\mathbf{s})}_{M,\mathbf{Z}}$, for every $\mathbf{s}\in\mathcal{S}^n$, and the notation in \eqref{EQ:CR_achievability_strong_sec}. On account of \eqref{EQ:CR_achievability_strong_sec}, for sufficiently large values of $n$ that are independent of $Q_S\in\mathcal{Q}$, it holds that
\begin{equation}
\mathcal{L}(\mathfrak{V}^n,Q_S,\mathsf{C}_n)\leq \epsilon.\label{EQ:strong_secrecy_lemma_proof}
\end{equation}

Consequently, for those values of $n$, we have\footnote{Only the last step relies on $n$ being sufficiently large; all other steps are valid for every $n\in\mathbb{N}_\ell$}
\begin{align*}
&I_\Upsilon(M;Z^n|S^n,C_n)\\
    &=\sum_{\mathbf{s}\in\mathcal{S}^n}Q_S^n(\mathbf{s})I_\Upsilon(M;Z^n|S^n=\mathbf{s},C_n)\\
    &\begin{multlined}[b][.46\textwidth]\leq\sum_{\mathbf{s}\in\mathcal{T}_{\alpha}^n(Q_S)}Q_S^n(\mathbf{s})I_\Upsilon(M;Z^n|S^n=\mathbf{s},C_n)\\+\sum_{\mathbf{s}\notin\mathcal{T}_\alpha^n(Q_S)}Q_S^n(\mathbf{s})n\log|\mathcal{Z}|\end{multlined}\\
    &\begin{multlined}[b][.46\textwidth]\leq\max_{\mathbf{s}\in\mathcal{T}_\alpha^n(Q_S)}I_\Upsilon(M;Z^n|S^n=\mathbf{s},C_n)\\+ n\log|\mathcal{Z}|\cdot\mathbb{P}_{Q_S^n}\big(S^n\notin\mathcal{T}_\alpha^n(Q_S)\big)\end{multlined}\\
    &\stackrel{(a)}\leq\mathcal{L}(\mathfrak{V}^n,Q_S,\mathsf{C}_n)+n\log|\mathcal{Z}|\left(\alpha+2|\mathcal{S}|e^{-2n\frac{\alpha^2}{|\mathcal{S}|^2}}\right)\\
    &\stackrel{(b)}\leq n\eta_{n,\alpha}^{(1)},\numberthis
\end{align*}
where (a) uses \eqref{EQ:typical_type_UB} and the upper bound on the probability of drawing an atypical sequence from \eqref{EQ:uniform_atypical_bound}, while (b) follows from \eqref{EQ:strong_secrecy_lemma_proof}.


\section{Proof of Lemma \ref{LEMMA:average_channel_reliability}}\label{APPEN:average_channel_reliability_proof}

Fix $Q_S\in\mathcal{Q}$. For any $n\in\mathbb{N}_\ell$, let $\mathcal{E}_U(\mathfrak{W}^n,Q_S,\mathsf{C}_n)$ be the average error probability under the CR-code $\mathsf{C}_n$ when the adversary chooses the state sequence randomly and uniformly over $\mathcal{T}^n_{Q_S}$. Clearly, for any $n$ sufficiently large (that is independent of $Q_S$), we have
\begin{equation}
\mathcal{E}_U(\mathfrak{W}^n,Q_S,\mathsf{C}_n)\leq\bar{\mathcal{E}}(\mathfrak{W}^n,Q_S,\mathsf{C}_n)\stackrel{(a)}\leq\epsilon,\label{EQ:uniform_error_probability}
\end{equation}
where (a) is on account of \eqref{EQ:CR_achievability_average_error}. 

For each $\gamma\in\Gamma_n$, the PMF on $\mathcal{S}^n\times\mathcal{M}_n\times\mathcal{X}^n\times\mathcal{Y}^n\times\mathcal{Z}^n\times\hat{\mathcal{M}}_n$ describing the random experiment where the state sequence is uniformly drawn from $\mathcal{T}^n_{Q_S}$ is
\begin{equation}
\Lambda_1^{(\gamma)}(\mathbf{s},m,\mathbf{x},\mathbf{y},\mathbf{z},\hat{m})\triangleq \frac{\mathds{1}_{\big\{\mathbf{s}\in\mathcal{T}^n_{Q_S}\big\}}}{|\mathcal{T}^n_{Q_S}|}P^{(\gamma,\mathbf{s})}(m,\mathbf{x},\mathbf{y},\mathbf{z},\hat{m}),\label{EQ:uniform_induced_PMF_code}
\end{equation}
for all $(\mathbf{s},m,\mathbf{x},\mathbf{y},\mathbf{z},\hat{m})\in\mathcal{S}^n\times\mathcal{M}_n\times\mathcal{X}^n\times\mathcal{Y}^n\times\mathcal{Z}^n\times\hat{\mathcal{M}}_n$. As before, we set
\begin{equation}
\Lambda_1\big(c_n(\gamma),\mathbf{s},m,\mathbf{x},\mathbf{y},\mathbf{z},\hat{m}\big)\triangleq \mu_n(\gamma)\Lambda_1^{(\gamma)}(\mathbf{s},m,\mathbf{x},\mathbf{y},\mathbf{z},\hat{m}).\label{EQ:uniform_induced_PMF}
\end{equation}
By \eqref{EQ:uniform_error_probability} and Fano's inequality, we have
\begin{align*}
H_{\Lambda_1}(M|Y^n,C_n)&\leq 1+\mathcal{E}_U(\mathfrak{W}^n,Q_S,\mathsf{C}_n)\cdot\log M_n\\
&\leq 1+\epsilon\log M_n,\numberthis\label{EQ:uniform_Fano}
\end{align*}
where the last inequality holds for the aforementioned sufficiently large $n$ values.

To upper bound $H_\Upsilon(M|Y^n,C_n)$ in terms of $H_{\Lambda_1}(M|Y^n,C_n)$, we first index all the types in $\mathcal{P}_n(\mathcal{S})$ by $i\in\mathcal{B}\triangleq\big[1:|\mathcal{P}_n(\mathcal{S})|\big]$. Set $Q_1=Q_S$, and associate with every $Q_i\in\mathcal{P}_n(\mathcal{S})$, $i\neq 1$, a PMF $\Lambda_i$ on $\mathcal{C}_n\times\mathcal{S}^n\times\mathcal{M}_n\times\mathcal{X}^n\times\mathcal{Y}^n\times\mathcal{Z}^n\times\hat{\mathcal{M}}_n$, that is defined analogously to $\Lambda_1$ from \eqref{EQ:uniform_induced_PMF}. Thus, with respect to $\Lambda_i$, the state sequence is uniformly chosen from $\mathcal{T}^n_{Q_i}$.

Let $B$ be a random variable over $\mathcal{B}$ that takes the value $B=i$, for $i\in\mathcal{B}$, if an i.i.d. state sequence $S^n\sim Q_S^n$ satisfies $S^n\in\mathcal{T}^n_{Q_i}$. First note that
\begin{align*}
I_\Upsilon(M;Y^n|C_n)&=I_\Upsilon(M;B,Y^n|C_n)-I_\Upsilon(M;B|Y^n,C_n)\\
                              &\geq I_\Upsilon(M;B,Y^n|C_n)-\log|\mathcal{B}|\\
                              &\geq I_\Upsilon(M;B,Y^n|C_n)-|\mathcal{S}|\log(n+1),\numberthis\label{EQ:add_conditioning_B}
\end{align*}
which implies
\begin{equation}
H_\Upsilon(M|Y^n,C_n)\leq H_\Upsilon(M|Y^n,B,C_n)+|\mathcal{S}|\log(n+1).\label{EQ:equivocation_average_channel_B_UB}
\end{equation}

Next, we expand the conditional entropy from the RHS of \eqref{EQ:equivocation_average_channel_B_UB} with respect to $B$, while splitting it into typical and atypical realizations of $B$. Fix $\alpha\in\left(0,\frac{1}{2}\right]$ and define
\begin{equation}
\mathcal{I}(Q_S,\alpha)=\Big\{i\in\mathcal{B}\mspace{3mu}\Big|\mathcal{T}^n_{Q_i}\subset\mathcal{T}_\alpha^n(Q_S)\Big\}.\label{EQ:typical_types_indices}
\end{equation}
For any $n\in\mathbb{N}_\ell$, we have
\begin{align*}
&H_\Upsilon(M|Y^n,B,C_n)\\
&\begin{multlined}[b][.46\textwidth]\leq\sum_{i\in\mathcal{I}(Q_S,\alpha)}\mathbb{P}_{Q_S^n}\big(S^n\in\mathcal{T}^n_{Q_i}\big)H_\Upsilon(M|Y^n,B=i,C_n)\\+\mathbb{P}_{Q_S^n}\big(S^n\notin\mathcal{T}_\alpha^n(Q_S)\big)\cdot\log M_n\end{multlined}\\
&\begin{multlined}[b][.46\textwidth]\stackrel{(a)}\leq\sum_{i\in\mathcal{I}(Q_S,\alpha)}\mathbb{P}_{Q_S^n}\big(S^n\in\mathcal{T}^n_{Q_i}\big)H_\Upsilon(M|Y^n,S^n\in\mathcal{T}^n_{Q_i},C_n)\\+2|\mathcal{S}|e^{-2n\frac{\alpha^2}{|\mathcal{S}|^2}}\log M_n,\end{multlined}\numberthis\label{EQ:equivocation_conditinail_onB_UB}
\end{align*}
where (a) uses \eqref{EQ:uniform_atypical_bound} and the definition of $B$. Recall that
\begin{equation}
\mathbb{P}_{Q_S^n}\big(S^n=\mathbf{s}\big|S^n\in\mathcal{T}^n_{Q_i}\big)=\frac{\mathds{1}_{\big\{\mathbf{s}\in\mathcal{T}^n_{Q_i}\big\}}}{|\mathcal{T}^n_{Q_i}|},\quad\forall \mathbf{s}\in\mathcal{S}^n\ ,\ \forall i\in\mathcal{B},
\end{equation}
and therefore,
\begin{align*}
H_\Upsilon(M|Y^n,B=i,C_n)&=H_\Upsilon(M|Y^n,S^n\in\mathcal{T}^n_{Q_i},C_n)\\
                         &=H_{\Lambda_i}(M|Y^n,C_n).\numberthis\label{EQ:equivocation_Upsilon_Lambda_equal}
\end{align*}

Having this, we remark that the main idea in upper bounding $H_\Upsilon(M|Y^n,C_n)$ is to use \eqref{EQ:equivocation_average_channel_B_UB} and \eqref{EQ:equivocation_conditinail_onB_UB}, while arguing that the conditional entropy $H_{\Lambda_i}(M|Y^n,C_n)$ is small for any $1\neq i\in\mathcal{I}(Q_S,\alpha)$ (i.e., for any random state sequence in the typical set) as long as we know it is small for one single type in the typical set (i.e., as long as \eqref{EQ:uniform_Fano} holds). To do so, we show that $H_{\Lambda_i}(M|Y^n,C_n)$, for all $i\in\mathcal{I}(Q_S,\alpha)$, can be replaced with $H_{\Lambda_1}(M|Y^n,C_n)$ with a correction term that can be made arbitrarily small when normalized by $n$. With some abuse of notation, we henceforth denote by $\Lambda_1$ the marginal PMF of \eqref{EQ:uniform_induced_PMF} on $\mathcal{S}^n\times\mathcal{M}_n\times\mathcal{X}^n\times\mathcal{Y}^n$. Similarly, for any $i\in\mathcal{I}(Q_S,\alpha)$, $\Lambda_i$ denotes the marginal on $\mathcal{S}^n\times\mathcal{M}_n\times\mathcal{X}^n\times\mathcal{Y}^n$ of the original $\Lambda_i$.

Fix $i\in\mathcal{I}(Q_S,\alpha)$ and let $\Lambda_{1,i}$ be a coupling of $\Lambda_1$ and $\Lambda_i$, which is a PMF on $\mathcal{S}^n\times\mathcal{M}_n\times\mathcal{X}^n\times\mathcal{Y}^n\times\mathcal{S}^n\times\mathcal{Y}^n$ that is defined by the following steps:
\begin{enumerate}
\item Similarly to the definition of the set $\mathcal{A}$ from \eqref{EQ:set_s1s2_coincide}, for any $(\mathbf{s},\tilde{\mathbf{s}})\in\mathcal{S}^n\times\mathcal{S}^n$ set
\begin{equation}
\mathcal{A}(\mathbf{s},\tilde{\mathbf{s}})\triangleq \big\{j\in[1:n]\big|s_j=\tilde{s}_j\big\}.\label{EQ:set_s_tildes_coincide}
\end{equation}
\item For any $\gamma\in\Gamma_n$, define
\begin{align*}
\Lambda&_{1,i}^{(\gamma)}(\mathbf{s},m,\mathbf{x},\mathbf{y},\tilde{\mathbf{s}},\tilde{\mathbf{y}})\\
&\triangleq  \Lambda_1^{(\gamma)}(\mathbf{s},m,\mathbf{x},\mathbf{y})\Lambda_{1,i}(\tilde{\mathbf{s}}|\mathbf{s})\Lambda_{1,i}(\tilde{\mathbf{y}}|\mathbf{s},\mathbf{x},\mathbf{y},\tilde{\mathbf{s}}),\numberthis\label{EQ:coupling_PMF_code}
\end{align*}
where 
\begin{equation}
\Lambda_{1,i}(\tilde{\mathbf{y}}|\mathbf{s},\mathbf{x},\mathbf{y},\tilde{\mathbf{s}})= \mathds{1}\mspace{-8mu}_{\bigcap\limits_{j\in\mathcal{A}(\mathbf{s},\tilde{\mathbf{s}})}\mspace{-8mu}\big\{\tilde{y}_j=y_j\big\}}\mspace{-6mu}\prod_{j\in\mathcal{A}(\mathbf{s},\tilde{\mathbf{s}})^c}\mspace{-8mu}W_{\tilde{s}_j}(\tilde{y}_j|x_j),\label{EQ:coupling_PMF_tildey_marginal}
\end{equation}
and $\Lambda_{1,i}(\tilde{\mathbf{s}}|\mathbf{s})$ is defined by the following pseudo-algorithm.

\begin{algorithm}
\caption{Construction of $\Lambda_{1,i}(\tilde{\mathbf{s}}|\mathbf{s})$}
\label{ALG:conditional_Lambda}
\begin{algorithmic}[1]
\State $\mathbf{s}'\leftarrow \mathbf{s}$
\While{$\mathbf{s}'\notin\mathcal{T}^n_{Q_i}$}
\State $\mathcal{J}(\mathbf{s}'):=\big\{j\in[1:n]\big|N(s'_j|\mathbf{s}')>nQ_i(s'_j)\big\}$
\State $\mathcal{L}(\mathbf{s}'):=\big\{s\in\mathcal{S}\big|N(s|\mathbf{s}')<nQ_i(s)\big\}$
\State Draw $j\sim\mbox{Unif}\left(\mathcal{J}(\mathbf{s}')\right)$
\State Draw $s\sim\mbox{Unif}\left(\mathcal{L}(\mathbf{s}')\right)$
\State $s'_j\leftarrow s$
\EndWhile
\end{algorithmic}
\end{algorithm}

Namely, in each step the algorithm first uniformly chooses an index $j\in[1:n]$, such that the number of appearances of $s_j$ in $\mathbf{s}$ is above the quota allowed by $Q_i$. Then, $s_j$ is replaced with a symbol $s\in\mathcal{S}$ that is uniformly chosen from the set of symbols whose number of appearances in $\mathbf{s}$ is below the quote subscribed by $Q_i$. The procedure repeats itself until the modified sequence belongs to $\mathcal{T}^n_{Q_i}$. Clearly, the algorithm stops after a finite number of cycles, since in each cycle the sequence $\mathbf{s}'$ is adjusted so that its empirical PMF of $\nu_{\mathbf{s}'}$ is closer to $Q_i$. We give a formal justification for the finite running time argument subsequently.

\item As a last step, we set 
\begin{align*}
\Lambda_{1,i}\big(c_n(\gamma),\mathbf{s},m,&\mathbf{x},\mathbf{y},\tilde{\mathbf{s}},\tilde{\mathbf{y}}\big)\\
&=\mu_n(\gamma)\Lambda_{1,i}^{(\gamma)}(\mathbf{s},m,\mathbf{x},\mathbf{y},\tilde{\mathbf{s}},\tilde{\mathbf{y}}).\numberthis\label{EQ:coupling_PMF}
\end{align*}
\end{enumerate}

For any $1\neq i\in\mathcal{I}(Q_S,\alpha)$, the symmetry in constructing $\Lambda_{1,i}(\tilde{\mathbf{s}}|\mathbf{s})$, the uniformity of $\Lambda_{1,i}(\mathbf{s})$ over $\mathcal{T}^n_{Q_S}$, and the fact that sequences of the same type are merely permutations of one another, imply that the marginal PMF of $\tilde{S}^n$ with respect to $\Lambda_{1,i}$ is uniform over $\mathcal{T}^n_{Q_i}$, i.e., 
\begin{equation}
\Lambda_{1,i}(\tilde{\mathbf{s}})=\Lambda_i(\tilde{\mathbf{s}})=\frac{\mathds{1}_{\big\{\tilde{\mathbf{s}}\in\mathcal{T}^n_{Q_i}\big\}}}{|\mathcal{T}^n_{Q_i}|}.\label{EQ:coupling_PMF_tildes_marginal}
\end{equation}
Recalling the definition of $\Lambda_{1,i}(\tilde{\mathbf{y}}|\mathbf{s},\mathbf{x},\mathbf{y},\tilde{\mathbf{s}})$ from \eqref{EQ:coupling_PMF_tildey_marginal}, we thus obtain
\begin{equation}
\Lambda_{1,i}(\tilde{\mathbf{s}},m,\mathbf{x},\tilde{\mathbf{y}})=\Lambda_i(\tilde{\mathbf{s}},m,\mathbf{x},\tilde{\mathbf{y}}),
\end{equation}
which shows that $\Lambda_{1,i}$ is a valid coupling of $\Lambda_1$ and $\Lambda_i$.

Some additional properties of the algorithms output are needed. Denote by $K\in\mathbb{N}\cup\{\infty\}$ the number of cycles it takes until Algorithm \ref{ALG:conditional_Lambda} terminates (i.e., for now, $K$ may be infinite). For each $k\in[1:K]$ (if $K=\infty$ then $k\in[1:K]$ is to be understood as $k\in\mathbb{N}$), denote by $\mathbf{s}_k'$ the $\mathbf{s}'$ sequence obtained after the $k$-th cycle. Accordingly, $\mathbf{s}_0'=\mathbf{s}$, and if indeed $K<\infty$, then $\mathbf{s}_K'=\tilde{\mathbf{s}}$. To analyse the algorithm's operation, for each $k\in[1:K]$, define
\begin{subequations}
\begin{align}
\mathcal{L}_k^{(h)}&=\Big\{s\in\mathcal{S}\Big|N(s|\mathbf{s}_k')>nQ_i(s)\Big\}\\
\mathcal{L}_k^{(l)}&=\Big\{s\in\mathcal{S}\Big|N(s|\mathbf{s}_k')<nQ_i(s)\Big\},
\end{align}
\end{subequations}
and further set
\begin{subequations}
\begin{align}
N_k^{(h)}&=\sum_{s\in\mathcal{L}_k^{(h)}}\big|N(s|\mathbf{s}_k')-nQ_i(s)\big|\label{EQ:alg_analysis_La}\\
N_k^{(l)}&=\sum_{s\in\mathcal{L}_k^{(l)}}\big|N(s|\mathbf{s}_k')-nQ_i(s)\big|.
\end{align}
\end{subequations}
Note that $N_k^{(h)}=N_k^{(l)}$, for every $k\in[1:K]$, and that in each iteration both $N_k^{(h)}$ and $N_k^{(l)}$ reduce by 1, i.e., 
\begin{equation}
N_k^{(h)}=N_{k-1}^{(h)}-1,\quad k\in[1:K].\label{EQ:alg_analysis_Na}
\end{equation}
Clearly, Algorithm \ref{ALG:conditional_Lambda} terminates once $\mathcal{L}_k^{(h)}=\mathcal{L}_k^{(l)}=\emptyset$, or equivalently, once $N_k^{(h)}=N_k^{(l)}=0$. When combined with \eqref{EQ:alg_analysis_Na}, this characterizes $K=N_0^{(h)}$, thus justifying the finite running time of the algorithm. Consequently, if $\big(S^n,\tilde{S}^n\big)\sim \Lambda_{1,i}$, then 
\begin{equation}
\mathbb{P}_{\Lambda_{1,i}}\Big(\big(S^n,\tilde{S}^n\big)\in\mathcal{T}^n_{Q_S}\times\mathcal{T}^n_{Q_i}\Big)=1.\label{EQ:s_tildes_Types_wp1}
\end{equation}

Another important outcome of the algorithm's operation is that the sequences $S^n$ and $\tilde{S}^n$ jointly distributed according to $\Lambda_{1,i}$ are almost surely within a Hamming distance of at most $n\alpha$. Namely, we claim that
\begin{equation}
\mathbb{P}_{\Lambda_{1,i}}\Big(d_H\big(S^n,\tilde{S}^n\big)\leq n\alpha\Big)=1.\label{EQ:s_tildes_distance_wp1}
\end{equation}
To see the validity of \eqref{EQ:s_tildes_distance_wp1}, note that in each iteration one symbol of the current $\mathbf{s}_k'$ is altered. Furthermore, an altered symbol is never modified again in any of the succeeding iterations. As a consequence, this observation implies that
\begin{equation}
d_H(\mathbf{s}_{k-1}',\mathbf{s}_k')=1,\quad\forall k\in[1:K],
\end{equation}
and when combined with the fact the $d_H(\mathbf{s},\mathbf{s}_0')=0$, we obtain
\begin{equation}
d_H(\mathbf{s},\mathbf{s}_k')=d_H(\mathbf{s},\mathbf{s}_{k-1}')+1,\quad\forall k\in[1:K].
\end{equation}
Thus, to show that for $d_H(\mathbf{s},\tilde{\mathbf{s}})\leq n\alpha$, it suffices to show that the number of cycles $K\leq n\alpha$ (keeping in mind that $\mathbf{s}_K'=\tilde{\mathbf{s}}$). Indeed, we have
\begin{equation}
K=N_0^{(h)}\stackrel{(a)}=\mspace{-4mu}\sum_{s\in\mathcal{L}_0^{(h)}}\mspace{-4mu}\big|N(s|\mathbf{s})-nQ_i(s)\big|\stackrel{(b)}\leq n\mspace{-5mu}\sum_{s\in\mathcal{L}_0^{(h)}}\mspace{-3mu}\frac{\alpha}{|\mathcal{S}|}\stackrel{(c)}\leq n\alpha,
\end{equation}
where (a) uses \eqref{EQ:alg_analysis_La}, (b) is because $\mathcal{T}^n_{Q_i}\subset\mathcal{T}_\alpha^n(Q_S)$, while (c) follows since $\big|\mathcal{L}_0^{(h)}\big|\leq|\mathcal{S}|$.


Having \eqref{EQ:s_tildes_Types_wp1}-\eqref{EQ:s_tildes_distance_wp1}, let $\mathcal{A}_i\big(S^n,\tilde{S}^n\big)$ be a random variable defined by \eqref{EQ:set_s_tildes_coincide}, where $\big(S^n,\tilde{S}^n\big)\sim \Lambda_{1,i}$ and $1\neq i\in\mathcal{I}(Q_S,\alpha)$. We abbreviate $\mathcal{A}_i\big(S^n,\tilde{S}^n\big)$ as $\mathcal{A}_i$ (and further omit the index $i$, when it is clear from the context). Define the random variable $\mathcal{A}_i^c\triangleq [1:n]\backslash \mathcal{A}_i$, and denote its alphabet by $\mathfrak{A}^c_i$. As a consequence of \eqref{EQ:s_tildes_Types_wp1}, for any $n\in\mathbb{N}_\ell$, the cardinality of $\mathcal{A}_i^c$ is upper bounded by \cite[Section 3.1]{Gelvin_entropy_counting2014}
\begin{equation}
|\mathfrak{A}^c_i|=\sum_{j=0}^{\lfloor n\alpha \rfloor}\binom{n}{j}\leq 2^{nh(\alpha)},\quad\forall 1\neq i\in\mathcal{I}(Q_S,\alpha),\label{EQ:A_alphabet_UB}
\end{equation}
where $h$ is the binary entropy function. Since $\mathcal{A}_i$ and $\mathcal{A}^c_i$ uniquely define one another, \eqref{EQ:A_alphabet_UB} yields
\begin{equation}
H_{\Lambda_{1,i}}(\mathcal{A})=H_{\Lambda_{1,i}}(\mathcal{A}^c)\leq \log|\mathfrak{A}_i^c|\leq nh(\alpha),\ \ \forall i\mspace{-3mu}\in\mspace{-3mu}\mathcal{I}(Q_S,\alpha),\label{EQ:A_entropy_UB}
\end{equation}
\indent We are now ready to link the mutual information between the message and the legitimate user's output sequence under $\Lambda_i$, to the corresponding term 
under $\Lambda_1$. This, in turn, yields an upper bound on $H_{\Lambda_i}(M|Y^n,C_n)$ in terms of $H_{\Lambda_1}(M|Y^n,C_n)$, that when combined with \eqref{EQ:equivocation_conditinail_onB_UB}, suffices to establish Lemma \ref{LEMMA:average_channel_reliability}. In the following chain of inequalities $(\tilde{S}^n,\tilde{Y}^n)$ and $(S^n,Y^n)$ denote the state sequence and legitimate channel output when distributed according to $\Lambda_i$ and $\Lambda_1$, respectively. Fix $1\neq i\in\mathcal{I}(Q_S,\alpha)$ and for any $n\in\mathbb{N}_\ell$ consider:
\begin{align*}
&I_{\Lambda_i}(M;\tilde{Y}^n|C_n)\\
    &\stackrel{(a)}\geq I_{\Lambda_{1,i}}(M;\tilde{Y}^n|\mathcal{A},C_n)-H_{\Lambda_{1,i}}(\mathcal{A})\\
    &\geq I_{\Lambda_{1,i}}(M;\tilde{\mathbf{Y}}^\mathcal{A}|\mathcal{A},C_n)-H_{\Lambda_{1,i}}(\mathcal{A})\\
    &\stackrel{(b)}= I_{\Lambda_{1,i}}(M;\mathbf{Y}^\mathcal{A}|\mathcal{A},C_n)-H_{\Lambda_{1,i}}(\mathcal{A})\\
    &= I_{\Lambda_{1,i}}(M;\mathbf{Y}^\mathcal{A}|\mathcal{A},C_n)+I_{\Lambda_{1,i}}(M;\mathbf{Y}^{\mathcal{A}^c}|\mathbf{Y}^\mathcal{A},\mathcal{A},C_n)\\&\mspace{115mu}-I_{\Lambda_{1,i}}(M;\mathbf{Y}^{\mathcal{A}^c}|\mathbf{Y}^\mathcal{A},\mathcal{A},C_n)-H_{\Lambda_{1,i}}(\mathcal{A})\\
    &\stackrel{(c)}\geq I_{\Lambda_{1,i}}(M;Y^n|\mathcal{A},C_n)-n\alpha\log|\mathcal{Y}|-H_{\Lambda_{1,i}}(\mathcal{A})\\
    &\stackrel{(d)}\geq I_{\Lambda_1}(M;Y^n|C_n)-n\alpha\log|\mathcal{Y}|-2H_{\Lambda_{1,i}}(\mathcal{A})\\
    &\stackrel{(e)}\geq I_{\Lambda_1}(M;Y^n|C_n)-n\alpha\log|\mathcal{Y}|-2nh(\alpha),\numberthis\label{EQ:MI_Lambda_1andi_LB}
\end{align*}
where:\\
(a) and (d) follow by similar arguments as in the lower bound from \eqref{EQ:add_conditioning_B};\\
(b) is since since $\tilde{Y}_j=Y_j$ almost surely, for all $j\in\mathcal{A}$ (cf. \eqref{EQ:coupling_PMF_tildey_marginal});\\
(c) is because for every $a\in\mathfrak{A}_i$, we have 
\begin{equation*}
I_{\Lambda_{1,i}}(M;\mathbf{Y}^{a^c}|\mathbf{Y}^a,\mathcal{A}=a,C_n)\leq |a^c|\log|\mathcal{Y}|\leq n\alpha\log|\mathcal{Y}|;
\end{equation*}
(e) uses \eqref{EQ:A_entropy_UB}.

Clearly, \eqref{EQ:MI_Lambda_1andi_LB} yields
\begin{equation}
H_{\Lambda_i}(M|\tilde{Y}^n,C_n)\mspace{-2mu}\leq\mspace{-2mu} H_{\Lambda_1}(M|Y^n,C_n)\mspace{-2mu}+\mspace{-2mu}n\alpha\log|\mathcal{Y}|\mspace{-2mu}+\mspace{-2mu}2nh(\alpha),\numberthis\label{EQ:entropy_Lambda_1andi_LB}
\end{equation}
for each $1\neq i\in\mathcal{I}(Q_S,\alpha)$. Inserting this back into \eqref{EQ:equivocation_conditinail_onB_UB}, while keeping  \eqref{EQ:equivocation_average_channel_B_UB} and \eqref{EQ:equivocation_Upsilon_Lambda_equal} in mind, for any $n$ sufficiently large (that is independent of $Q_S$ and $\alpha$), we have 
\begin{align*}
&H_\Upsilon(M|Y^n,C_n)\\
&\begin{multlined}[b][.47\textwidth]\leq \sum_{i\in\mathcal{I}(Q_S,\alpha)}\mathbb{P}_{Q_S^n}\big(S^n\in\mathcal{T}^n_{Q_i}\big)H_{\Lambda_i}(M|Y^n,C_n)\\+2\log M_n|\mathcal{S}|e^{-2n\frac{\alpha^2}{|\mathcal{S}|^2}}+|\mathcal{S}|\log(n+1)\end{multlined}\\
&\begin{multlined}[b][.47\textwidth]\stackrel{(a)}\leq H_{\Lambda_1}(M|Y^n,C_n)+n\alpha\log|\mathcal{Y}|+2nh(\alpha)\\+2\log M_n|\mathcal{S}|e^{-2n\frac{\alpha^2}{|\mathcal{S}|^2}}+|\mathcal{S}|\log(n+1)\end{multlined}\\
&\stackrel{(b)}\leq n\eta_{n,\alpha}^{(2)},\numberthis
\end{align*}
where (a) uses \eqref{EQ:entropy_Lambda_1andi_LB}, while (b) follows by \eqref{EQ:uniform_Fano} and by setting $\eta_{n,\alpha}^{(2)}=\frac{1}{n}+\frac{1}{n}\log M_n\left(\epsilon+2|\mathcal{S}|e^{-2n\frac{\alpha^2}{|\mathcal{S}|^2}}\right)+\alpha\log|\mathcal{Y}|+2h(\alpha)+|\mathcal{S}|\frac{\log(n+1)}{n}$.


\section{Proof of Lemmas \ref{LEMMA:upsilon_independence} and \ref{LEMMA:Q_PMF_converse}}\label{APPEN:upsilon_independence_proof}


\subsection{Lemma \ref{LEMMA:upsilon_independence}}

Fix $Q_S\in\mathcal{Q}$, $\gamma\in\Gamma_n$ and $i\in[1:n]$. For any $s_i\in\mathcal{S}$ and $\big(c_n(\gamma),s^{n\backslash i},m,x^n,y^{n\backslash i},z^{n\backslash i}\big)\in\mathcal{C}_n\times\mathcal{S}^{n-1}\times\mathcal{M}_n\times\mathcal{X}^n\times\mathcal{Y}^{n-1}\times\mathcal{Z}^{n-1}$, for some $\gamma\in\Gamma_n$, we have
\begin{align*}
\Upsilon\big(c_n(\gamma),&s^{n\backslash i},m,x^n,y^{n\backslash i},z^{n\backslash i}\big|s_i\big)\\
&=\frac{\Upsilon\big(c_n(\gamma),s_i,s^{n\backslash i},m,x^n,y^{n\backslash i},z^{n\backslash i}\big|s_i\big)}{\Upsilon(s_i)}.\numberthis\label{EQ:independence_bayes}
\end{align*}
The marginal distribution of $S_i$ with respect to $\Upsilon$ from \eqref{EQ:averaged_upsilon},~is
\begin{equation}
\Upsilon(s_i)=Q_S(s_i),\label{EQ:independence_denominator}
\end{equation}
while for the numerator, we have
\begin{align*}
&\Upsilon\big(c_n(\gamma),s_i,s^{n\backslash i},m,x^n,y^{n\backslash i},z^{n\backslash i}\big)\\
&=\mu_n(\gamma)\Upsilon^{(\gamma)}(s_i)\Upsilon^{(\gamma)}\big(s^{n\backslash i},m,x^n,y^{n\backslash i},z^{n\backslash i}\big|s_i\big)\\
&\begin{multlined}[b][.47\textwidth]=\mu_n(\gamma)Q_S(s_i)Q_S^{n-1}\big(s^{n\backslash i}\big)\frac{1}{M_n}f_\gamma(x^n|m)\\\times W_{s^{n\backslash i}}^{n-1}\big(y^{n\backslash i}\big|x^{n\backslash i}\big)V_{s^{n\backslash i}}^{n-1}\big(z^{n\backslash i}\big|x^{n\backslash i}\big)\end{multlined}\\
&\begin{multlined}[b][.47\textwidth]=\mu_n(\gamma)Q_S(s_i)\Upsilon^{(\gamma)}\big(s^{n\backslash i}\big)\Upsilon^{(\gamma)}\big(m\big|s^{n\backslash i}\big)\\\times\Upsilon^{(\gamma)}\big(x^n\big|s^{n\backslash i},m\big)\Upsilon^{(\gamma)}\big(y^{n\backslash i},z^{n\backslash i}\big|s^{n\backslash i},m,x^n\big)\end{multlined}\\
&=\Upsilon(s_i)\Upsilon\big(c_n(\gamma),s^{n\backslash i},m,x^n,y^{n\backslash i},z^{n\backslash i}\big).\numberthis\label{EQ:independence_numerator}
\end{align*}
Inserting \eqref{EQ:independence_denominator} and \eqref{EQ:independence_numerator} back into \eqref{EQ:independence_bayes} completes the proof. 



\subsection{Lemma \ref{LEMMA:Q_PMF_converse}}

Again, fix $Q_S\in\mathcal{Q}$ and recall that $V=(V_T,T)$, where $V_T=(Y^{T-1},S_{T+1}^n,Z_{T+1}^n,C_n)$. Therefore, we represent a realization $v$ of $V$ as $v=(\tilde{v}_t,t)$, where $\tilde{v}_t\triangleq\big(y^{t-1},s_{t+1}^n,z_{t+1}^n,c_n(\gamma)\big)\in\mathcal{Y}^{t-1}\times\mathcal{S}^{n-t}\times\mathcal{Z}^{n-t}\times\mathcal{C}_n$, for some $\gamma\in\Gamma_n$, and $t\in[1:n]$. For any $(v,u,x,s,y,z)\in\mathcal{V}\times\mathcal{U}\times\mathcal{X}\times\mathcal{S}\times\mathcal{Y}\times\mathcal{Z}$, we have
\begin{align*}
\mathbb{P}_\Upsilon&(S=s|V=v,U=u,X=x)\\
    &\stackrel{(a)}=\mathbb{P}_\Upsilon\big(S_T=s\big|(V_T,T)=(\tilde{v}_t,t),M=m,X_T=x\big)\\
    &\stackrel{(c)}=\mathbb{P}_{\Upsilon^{(\gamma)}}(S_t=s)\\
    &=Q_S(s),\numberthis\label{EQ:factorization_proof_S}
\end{align*}
where (a) is because $U=(M,V_T,T)$, while (b) uses the independence of $T$ and $(M,X^n,S^n,Y^n,Z^n)$ and the independence relation from Lemma \ref{LEMMA:upsilon_independence}.

By similar steps to those in the derivation of \eqref{EQ:factorization_proof_S}, we also obtain
\begin{align*}
\mathbb{P}_\Upsilon(Y=y&,Z=z|V=v,U=u,X=x,S=s)\\
    &\stackrel{(a)}=\mathbb{P}_{\Upsilon^{(\gamma)}}(Y_t=y,Z_t=z|X_t=x,S_t=s)\\
    &=W_s(y|x)V_s(z|x),\numberthis
\end{align*}
where (a) also relies on the Markov relation induced by the channel.

\bibliographystyle{unsrt}
\bibliographystyle{IEEEtran}
\bibliography{ref}

\begin{IEEEbiographynophoto}{Ziv Goldfeld}
(S'13) received his B.Sc.\@ (summa cum laude) and M.Sc.\@ (summa cum laude) degrees in Electrical and Computer Engineering from the Ben-Gurion University, Israel, in 2012 and 2014, respectively. He is currently a 
student in the direct Ph.D. program for honor students in Electrical and Computer Engineering at that same institution.

Between 2003 and 2006, he served in the intelligence corps of the Israeli Defense Forces.

Ziv is a recipient of several awards, among them are the Dean's List Award, the Basor Fellowship, the Lev-Zion fellowship, IEEEI-2014 best student paper award, a Minerva Short-Term Research Grant (MRG), and a Feder Family Award in the national student contest for outstanding research work in the field of communications technology.
\end{IEEEbiographynophoto}

\begin{IEEEbiographynophoto}{Paul Cuff}
(S'08-M'10) received the B.S. degree in electrical engineering from Brigham Young University, Provo, UT, in 2004 and the M.S. and Ph.D. degrees in electrical engineering from Stanford University in 2006 and 2009. Since 2009 he has been an Assistant Professor of Electrical Engineering at Princeton University.

As a graduate student, Dr. Cuff was awarded the ISIT 2008 Student Paper Award for his work titled “Communication Requirements for Generating Correlated Random Variables” and was a recipient of the National Defense Science and Engineering Graduate Fellowship and the Numerical Technologies Fellowship. As faculty, he received the NSF Career Award in 2014 and the AFOSR Young Investigator Program Award in 2015.
\end{IEEEbiographynophoto}

\begin{IEEEbiographynophoto}{Haim H. Permuter}
(M'08-SM'13) received his B.Sc.\@ (summa cum laude) and M.Sc.\@ (summa cum laude) degrees in Electrical and Computer Engineering from the Ben-Gurion University, Israel, in 1997 and 2003, respectively, and the Ph.D. degree in Electrical Engineering from Stanford University, California in 2008.

Between 1997 and 2004, he was an officer at a research and development unit of the Israeli Defense Forces. Since 2009 he is with the department of Electrical and Computer Engineering at Ben-Gurion University where he is currently an associate professor.

Prof. Permuter is a recipient of several awards, among them the Fullbright Fellowship, the Stanford Graduate Fellowship (SGF), Allon Fellowship, and and the U.S.-Israel Binational Science Foundation Bergmann Memorial Award. Haim is currently serving on the editorial board of the IEEE Transactions on Information Theory.
\end{IEEEbiographynophoto}

\end{document}